\documentclass{article}

\usepackage{arxiv}
\usepackage[utf8]{inputenc} % allow utf-8 input
\usepackage[T1]{fontenc} % use 8-bit T1 fonts
\usepackage{hyperref}       % hyperlinks
\usepackage{url}            % simple URL typesetting
\usepackage{graphicx}
\usepackage{graphics}
\usepackage{physics}
\newcommand{\beq}{\begin{equation}}
\newcommand{\eeq}{\end{equation}}
\usepackage{float}
\usepackage[mathscr]{eucal}
\usepackage{titling}
\newcommand{\subtitle}[1]{%
  \posttitle{%
    \par\end{center}
    \begin{center}\large#1\end{center}
    \vskip0.5em}%
}

\title{Quantum Lissajous Figures via Projection}
\subtitle{\textit{This thesis has been submitted and accepted in partial fulfillment of the requirements for the degree Master of Science in physics at the Rochester Institute of Technology.}}

\author{Errico J. Russo\\
	School of Physics and Astronomy\\
	Rochester Institute of Technology\\
	Rochester, NY 14623 \\
    \texttt{ejr8685@rit.edu} \\}

\begin{document}\maketitle
\begin{abstract}
	\noindent We present and investigate a new category of quantum Lissajous states for a two-dimensional Harmonic Oscillator (2DHO) having commensurate angular frequencies. The states themselves result from the projection of ordinary coherent states (i.e. quasi-classical) onto a degenerate subspace of the 2DHO. In this way, new, highly non-classical quantum mechanically stationary states arise “organically” from the highly classical but non-stationary coherent states. The connection to Lissajous figures is evident in the result that our states so defined all have probability densities that are localized along the corresponding classical Lissajous figures. We further emphasize the important interplay between the probability current density and the emergence of quantum interference in the states we examine. In doing so, we are able to present a consistent discussion of a class of states known as vortex states.
\end{abstract}
\keywords{Lissajous figures \and Quantum mechanical 2-dimensional harmonic oscillator \and Vortex states \and Quantum interference}
\section{Introduction}
First discovered in 1815 by Nathaniel Bowditch and further studied in 1857 by Jules Antoine Lissajous, the Lissajous figures are a type of complex harmonic motion that describes the periodic orbits of systems of two harmonic oscillators with commensurable frequencies\cite{marion}. Many applications of Lissajous figures have been used in classical physics, such as in the study of trajectory planning\cite{borkar_application_2020} and bistable perception\cite{weilnhammer_revisiting_2014}. In the literature, there have been studies on two-mode coherent states for the isotropic and anisotropic 2-dimensional quantum harmonic oscillator (2DHO) that produce stationary states with the appearance of Lissajous figures, where the probability densities have localized along classical Lissajous orbits\cite{gorska_correspondence_2006,moran_coherent_2019,kumar_commensurate_2008}. Vortex structures are shown to appear in some cases of "quantum Lissajous figure" states and are connected to the appearance of quantum interference fringes\cite{chen_vortex_2003}. This emergence of quantum vortices has been observed to result in stable orbits\cite{barker_use_2000} and explain the flows in quantum dot phenomena\cite{bird_intrinsic_1998}. The vortices here are analogous to those found in superfluid helium\cite{onsager_statistical_1949,feynman_atomic_1954}, as well as those found in chemical physics\cite{mccullough_dynamics_1971,hirschfelder_quantized_1974,hirschfelder_quantum_1974}. While the appearance of Lissajous figures in the quantum regime is well determined, some states that result in Lissajous figure behavior have been miscategorized as SU(2) coherent states. We will show that the isotropic 2DHO does result in the SU(2) coherent state, but the anisotropic 2DHO does not. Also, with the results presented by Chen and Huang\cite{chen_vortex_2003}, we have a reworked explanation of the connection between quantum vortices and quantum interference. In this article, we propose a unique derivation of the quantum Lissajous states involving the projection of the two-mode coherent state onto a degenerate subspace of the 2DHO, which clears up the dilemma of the SU(2) coherent state categorization as well as the relationship between vortices and interference fringes. In section \ref{chap2_classicalLJ}, we review classical Lissajous figures in the 2DHO from the point of view of the complex normal coordinates. Section \ref{chap3_QLJviaProjection} introduces the full development of quantum Lissajous figures via projection. Section \ref{chap4_PrincipleResults} presents the principal results of the states derived in section \ref{chap3_QLJviaProjection}. Section \ref{chap5_Summary} summarizes the results of this research and discusses possible extensions of the theoretical development.
\section{Brief Review of Classical Lissajous Figures}
\label{chap2_classicalLJ}
The work presented in this section is not our original work, rather, it is included to create context for the motivation behind the quantum Lissajous figures discussed in section \ref{chap3_QLJviaProjection} and beyond. For completeness of the overall theoretical presentation, we have included this to justify the physically sensible formulation of quantum Lissajous states. First, in order to categorize a quantum Lissajous state, there must be some connection to the classical version, which is the emergence of the curves created by classical Lissajous figures appearing in a quantum state probability density. Second, we know from Ehrenfest's Theorem\cite{sakurai_modern_2017} that the average value of a quantum probability density moves like a classical particle subjected to a scalar potential.

\subsection{Lagrangian and Hamiltonian Mechanics}

Newton's second law describes the motion of a physical system. An equivalent formulation was created in the 1750s by Leonhard Euler and Joseph-Louis Lagrange. This formulation uses generalized positions and velocities, $(q_i,\dot{q}_i)$, making it easier to apply to the most general of geometric systems. The steps that are gone through here were made in such a way to justify our jump to quantum Mechanics. Following the development of Goldstein\cite{goldstein_classical_2008}, consider the Euler-Lagrange equation, which is given by

\beq
\frac{d}{dt}\frac{\partial \mathscr{L}}{\partial \dot{q}_i}-\frac{\partial \mathscr{L}}{\partial q_i}=0.
\label{eqn:euler lagrange eqn}
\eeq
The Lagrangian $\mathscr{L}(q_i,\dot{q}_i,t)$ is the difference between the kinetic and potential energies, $\mathscr{L}=K-U$. It provides a convenient way to derive the equations of motion by minimizing the action. Another important quantity is the generalized momenta

\beq
p_i=\frac{\partial \mathscr{L}}{\partial \dot{q}_i}.
\label{eqn:generalized momenta}
\eeq
Plugging the generalized momenta into Eq. (\ref{eqn:euler lagrange eqn}), we retrieve the equation for the force

\beq
\dot{p}_i=\frac{\partial \mathscr{L}}{\partial q_i}.
\label{eqn:force eqn lagrangian}
\eeq
The goal in this work is to identify quantum mechanical analogs to already known classical concepts, so it's convenient to introduce Hamilton's Formulation in Classical Mechanics to treat the jump from frameworks in the simplest way. The importance of this formulation comes partly from the fact that the quantum position and momenta operators $(\hat{q},\hat{p})$ are directly connected to the non-deterministic nature of quantum Mechanics through the Heisenberg Uncertainty Principle, $\sigma_x\sigma_p\geq\hbar/2$. 

We employ the use of the Legendre Transformation\cite{goldstein_classical_2008} to change variables from generalized position and velocities $(q_i,\dot{q}_i)$ to generalized position and momenta $(q_i,p_i)$. The point of generalized coordinates is to simplify calculations of a certain system, where, say, solving with Cartesian coordinates would be more difficult. Generalized coordinates could be as simple as rotating your coordinate system by an angle, or by using a different set of variables. Here, we choose to use generalized position and momenta for both, ease of calculation, and direct connection to Hamilton's formulation in quantum mechanics. The differential of $\mathscr{L}$ is

\begin{align}
\label{eq:differential of lagrangian}
\begin{split}
d\mathscr{L}(q_i,\dot{q}_i,t) & =\frac{\partial \mathscr{L}}{\partial q_i}dq_i+\frac{\partial\mathscr{L}}{\partial\dot{q}_i}d\dot{q}_i+\frac{\partial\mathscr{L}}{\partial t}dt
\\
 &=\dot{p}_idq_i+p_id\dot{q}_i+\frac{\partial\mathscr{L}}{\partial t}dt.
\end{split}
\end{align}

The Classical Hamiltonian is defined as the sum of the kinetic and potential energies, $\mathscr{H}=K+U$, and is a function of generalized position and momenta, $\mathscr{H}(q_i,p_i,t)$. The formal Legendre transformation from $\mathscr{L}$ to $\mathscr{H}$ is

\beq
\mathscr{H}(q_i,p_i,t)=p_i\dot{q}_i-\mathscr{L}(q_i,\dot{q}_i,t).
\label{eqn:L to H legendre trans}
\eeq
Taking the differential of Eq. (\ref{eqn:L to H legendre trans}) gives

\begin{align}
\label{eq:differential of H from legendre trans}
\begin{split}
d\mathscr{H} & =\dot{q}_idp_i+p_id\dot{q}_i-\frac{\partial\mathscr{L}}{\partial q_i}dq_i-\frac{\partial\mathscr{L}}{\partial \dot{q}_i}d\dot{q}_i-\frac{\partial\mathscr{L}}{\partial t}dt
\\
 & =\dot{q}_idp_i+p_id\dot{q}_i-\dot{p}_idq_i-p_id\dot{q}_i-\frac{\partial\mathscr{L}}{\partial t}dt,
\\
 & =\dot{q}_idp_i-\dot{p}_idq_i-\frac{\partial\mathscr{L}}{\partial t}dt
\end{split}
\end{align}
but $d\mathscr{H}$ can also be written as

\beq
d\mathscr{H}=\frac{\partial \mathscr{H}}{\partial p_i}dp_i+\frac{\partial \mathscr{H}}{\partial q_i}dq_i+\frac{\partial \mathscr{H}}{\partial t}dt,
\label{eqn:differential of H in terms of q,p,t}
\eeq
and comparing Eqs. (\ref{eq:differential of H from legendre trans}) and (\ref{eqn:differential of H in terms of q,p,t}) we arrive at Hamilton's equations

\begin{align}
\label{eq:hamiltons eqns}
\begin{split}
\frac{\partial \mathscr{H}}{\partial p_i} & =\dot{q}_i,
\\
\frac{\partial \mathscr{H}}{\partial q_i} & =-\dot{p}_i,
\\
\frac{\partial \mathscr{H}}{\partial t} & =-\frac{\partial\mathscr{L}}{\partial t}.
\end{split}
\end{align}
Now if we consider a function depending on the canonical position and momenta, say, $u(q_i,p_i,t)$, and take the time derivative of it, with the use of the chain rule for partial derivatives we attain

\begin{align}
\label{eq:chain rule of fcn u}
\begin{split}
\frac{du}{dt} & =\frac{\partial u}{\partial q_i}\frac{\partial q_i}{\partial t}+\frac{\partial u}{\partial p_i}\frac{\partial p_i}{\partial t}+\frac{\partial u}{\partial t}\frac{\partial t}{\partial t}
\\
 & =\frac{\partial u}{\partial q_i}\dot{q}_i+\frac{\partial u}{\partial p_i}\dot{p}_i+\frac{\partial u}{\partial t}
\\
 & =\frac{\partial u}{\partial q_i}\frac{\partial \mathscr{H}}{\partial p_i}-\frac{\partial u}{\partial p_i}\frac{\partial \mathscr{H}}{\partial q_i}+\frac{\partial u}{\partial t},
\end{split}
\end{align}
where in the last line, we plugged in Eq. 
(\ref{eq:hamiltons eqns}). Note that the first two terms of the last line are defined as the Poisson Bracket

\beq
\{u,\mathscr{H}\}=\frac{\partial u}{\partial q_i}\frac{\partial \mathscr{H}}{\partial p_i}-\frac{\partial u}{\partial p_i}\frac{\partial \mathscr{H}}{\partial q_i}.
\label{eqn:poisson bracket}
\eeq
The Poisson Bracket is canonically invariant, and is analogous to the commutator bracket commonly used in quantum Mechanics. Plugging in the Poisson Bracket into Eq. (\ref{eq:chain rule of fcn u}), we retrieve Hamilton's Equation in terms of the Poisson Bracket

\beq
 \frac{du}{dt}=\{u,\mathscr{H}\}+\frac{\partial u}{\partial t}.
\label{eqn:hamiltons eqn in poisson bracket formulation}
\eeq
It is important to note that Eq. (\ref{eqn:hamiltons eqn in poisson bracket formulation}) is analogous to the Heisenberg Equation from quantum Mechanics, when dealing with operators that are time dependent. This analogy will come in handy later. From here we can evaluate the time derivatives of $q_i$, $p_i$, and $\mathscr{H}$ in terms of the Poisson Bracket. First, setting $u(q_i,p_i,t)=q_i$ and using the fact that $\partial q_i/\partial t=0$ because $q_i$ is an independent variable

\begin{align}
\label{eq:Hamiltons q eqn}
\begin{split}
\frac{dq_i}{dt} & =\{q_i,\mathscr{H}\}
\\
\dot{q}_i & =\frac{\partial \mathscr{H}}{\partial p_i}=\{q_i,\mathscr{H}\}.
\end{split}
\end{align}
The same can be done setting $u(q_i,p_i,t)=p_i$ and using $\partial p_i/\partial t=0$

\begin{align}
\label{eq:Hamiltons p eqn}
\begin{split}
\frac{dp_i}{dt} & =\{p_i,\mathscr{H}\}
\\
\dot{p}_i & =\frac{\partial \mathscr{H}}{\partial q_i}=\{p_i,\mathscr{H}\}.
\end{split}
\end{align}
Finally, we set $u(q_i,p_i,t)=\mathscr{H}$

\begin{align}
\label{eq:Hamiltons H eqn}
\begin{split}
\frac{d\mathscr{H}}{dt} & =\{\mathscr{H},\mathscr{H}\}+\frac{\partial \mathscr{H}}{\partial t}
\\
 & =\frac{\partial \mathscr{H}}{\partial q_i}\frac{\partial \mathscr{H}}{\partial p_i}-\frac{\partial \mathscr{H}}{\partial p_i}\frac{\partial \mathscr{H}}{\partial q_i}+\frac{\partial \mathscr{H}}{\partial t}
\\
 & =\frac{\partial \mathscr{H}}{\partial t},
\end{split}
\end{align}
where we have used the fact that the Poisson Bracket between $\mathscr{H}$ and itself is zero, i.e. a function commutes with itself in the Poisson Bracket. In other words, Eq. (\ref{eq:Hamiltons H eqn}) says that the total time derivative of $\mathscr{H}$ is equal to the partial time derivative of $\mathscr{H}$. The Poisson Bracket of the position and momentum is also an important quantity;

\begin{align}
\label{eq:poisson bracket q p}
\begin{split}
\{q_i,p_i\}&=\frac{\partial q_i}{\partial q_i}\frac{\partial p_i}{\partial p_i}-\frac{\partial q_i}{\partial p_i}\frac{\partial p_i}{\partial q_i}=1,
\end{split}
\end{align}
and the Poisson bracket, $\{p_i,q_i\}$, is

\begin{align}
\label{eq:poisson bracket p q}
\begin{split}
\{p_i,q_i\}&=\frac{\partial p_i}{\partial q_i}\frac{\partial q_i}{\partial p_i}-\frac{\partial p_i}{\partial p_i}\frac{\partial q_i}{\partial q_i}=-1.
\end{split}
\end{align}
Comparing Eqs. (\ref{eq:poisson bracket q p}) and (\ref{eq:poisson bracket p q}) shows

\begin{align}
\label{eq:poisson bracket property 1}
\begin{split}
\{q_i,p_i\}&=-\{p_i,q_i\},
\end{split}
\end{align}
which says the Poisson bracket of two variables is the negative Poisson bracket of those same variables with their order reversed. Eq. (\ref{eq:poisson bracket property 1}) applies for Poisson Brackets of all variables. Eq. (\ref{eq:poisson bracket q p}) and in turn Eq. (\ref{eq:poisson bracket p q}) can expressed more generally in the case where the bracket is between variables of two different coordinates

\begin{align}
\label{eq:poisson bracket q_i p_j}
\begin{split}
\{q_k,p_j\}&=\sum^N_{i=1}\frac{\partial q_k}{\partial q_i}\frac{\partial p_j}{\partial p_i}-\frac{\partial q_k}{\partial p_i}\frac{\partial p_j}{\partial q_i}=\delta_{kj},
\end{split}
\end{align}
where $\delta_{ij}$ is the Kronecker Delta.
The system to be examined is the 2DHO. Considering the oscillations to be in the $x-$ and $y$-directions, the Lagrangian is given by

\beq
\mathscr{L}(x,\dot{x},y,\dot{y})=\frac{1}{2}m_x\dot{x}^2+\frac{1}{2}m_y\dot{y}^2-\frac{1}{2}m_x\omega_x^2x^2-\frac{1}{2}m_y\omega_y^2y^2.
\label{eqn:lagrangian 2dho}
\eeq
The Legendre transformation , Eq. (\ref{eqn:L to H legendre trans}), in 2-dimensions is

\beq
\mathscr{H}(x,p_x,y,p_y)=p_x\dot{x}+p_y\dot{y}-\mathscr{L}(x,\dot{x},y,\dot{y}).
\label{eqn:2dho legendre trans}
\eeq
To find the Hamiltonian, we must change variables from $(x,\dot{x},y,\dot{y})$ to $(x,p_x,y,p_y)$. This can be done by way of Eq. (\ref{eqn:generalized momenta}). Differentiating the Lagrangian with respect to the velocities $\dot{x}$ and $\dot{y}$ yields 

\beq
p_x=\frac{\partial \mathscr{L}}{\partial \dot{x}}=m_x\dot{x}\quad\Rightarrow\quad\dot{x}=\frac{p_x}{m_x},
\label{eqn:x momentum}
\eeq
and

\beq
p_y=\frac{\partial \mathscr{L}}{\partial \dot{y}}=m_y\dot{y}\quad\Rightarrow\quad\dot{y}=\frac{p_y}{m_y}.
\label{eqn:y momentum}
\eeq
Inputting Eq. (\ref{eqn:lagrangian 2dho}) into Eq. (\ref{eqn:2dho legendre trans}), then plugging in for $\dot{x}$ and $\dot{y}$ using Eqs. (\ref{eqn:x momentum}) and (\ref{eqn:y momentum}) gives the Hamiltonian in terms of positions and momenta

\beq
\mathscr{H}(x,p_x,y,p_y)=\frac{p_x^2}{2m_x}+\frac{p_y^2}{2m_y}+\frac{1}{2}m_x\omega_x^2x^2+\frac{1}{2}m_y\omega_y^2y^2.
\label{eqn:hamiltonian 2dho}
\eeq
Eq. (\ref{eqn:hamiltonian 2dho}) can also be found using the facts that $\mathscr{L}=K-U$ and $\mathscr{H}=K+U$. Using Eqs. (\ref{eq:Hamiltons q eqn}) and (\ref{eq:Hamiltons p eqn}), we can specify the differential equations for the case of the 2DHO;

\begin{align}
\label{eq:Hamiltons x and px eqns for 2dho}
\begin{split}
\dot{x}&=\frac{\partial \mathscr{H}}{\partial p_x}=\frac{p_x}{m_x},
\\
\dot{p_x}&=-\frac{\partial \mathscr{H}}{\partial x}=-m_x\omega_x^2x,
\end{split}
\end{align}
and

\begin{align}
\label{eq:Hamiltons y and py eqns for 2dho}
\begin{split}
\dot{y}&=\frac{\partial \mathscr{H}}{\partial p_y}=\frac{p_y}{m_y},
\\
\dot{p_y}&=-\frac{\partial \mathscr{H}}{\partial y}=-m_y\omega_y^2y.
\end{split}
\end{align}
It is useful to specify the Poisson Brackets for this system,

\begin{align}
\label{eq:poisson bracket 2dho properties}
\begin{split}
\{x,p_x\}&=\{y,p_y\}=1,
\\
\{x,y\}&=\{x,p_y\}=\{y,p_x\}=0.
\end{split}
\end{align}

\subsection{Classical Normal Modes}
It is convenient to introduce a new set of coordinates that resemble what is worked with in quantum Mechanics. Consider the complex coordinates below

\begin{align}
\label{eq:normal mode coordinates}
\begin{split}
\alpha & =Ax+iBp_x\quad\Leftrightarrow\quad \alpha^*=Ax-iBp_x,
\\
\beta & =Cy+iDp_y\quad\Leftrightarrow\quad \beta^*=Cy-iDp_y,
\end{split}
\end{align}
where $A,B,C,D$ are real constants and the $*$ represents complex conjugation. The use of these complex normal coordinates is even more justification for jumping from classical to quantum Lissajous figures. They play a similar role to that of the Boson operators in quantum mechanics, and in section \ref{chap3_QLJviaProjection}, these will become quantized by use of the Dirac Prescription. It is obvious that multiplying a function by its complex conjugate eliminates the cross terms, leaving the independent variables in a quadratic form. This is important because energies of the oscillators are also quadratic forms\cite{goldstein_classical_2008}. The analysis of $\alpha$ and $\beta$ are exactly the same, so it is only necessary to work through one of them, say $\alpha$. Multiplying $\alpha$ and its complex conjugate gives

\beq
\alpha^*\alpha=A^2x^2+B^2p_x^2.
\label{eqn:alphaalpha*}
\eeq
Our goal is to mimic the quantum Harmonic Oscillator, so we require $A^2=m_x\omega_x/2$ and $B^2=1/2m_x\omega_x$. Plugging into Eq. (\ref{eqn:alphaalpha*}),

\beq
\alpha^*\alpha=\frac{1}{2}m_x\omega_xx^2+\frac{p_x^2}{2m_x\omega_x}.
\label{eqn:alphaalpha* with A and B defined}
\eeq
In order to get the total energy of the $x$ oscillator, we multiply a factor of $\omega_x$ to Eq. (\ref{eqn:alphaalpha* with A and B defined})

\beq
\omega_x\alpha^*\alpha=\frac{1}{2}m_x\omega_x^2x^2+\frac{p_x^2}{2m_x}.
\label{eqn:omegax alphaalpha* with A and B defined}
\eeq
By observation, we have $C^2=m_y\omega_y/2$ and $D^2=1/2m_y\omega_y$, and the total energy of the $y$ oscillator is 

\beq
\omega_y\beta^*\beta=\frac{1}{2}m_y\omega_y^2y^2+\frac{p_y^2}{2m_y}.
\label{eqn:omegay betabeta* with C and D defined}
\eeq
Replacing $A$ and $B$ (and similarly for $C$ and $D$) in Eq. (\ref{eq:normal mode coordinates}) 

\begin{align}
\label{eq:normal mode coordinates 2}
\begin{split}
\alpha & =\sqrt{\frac{m_x\omega_x}{2}}x+\frac{i}{\sqrt{2m_x\omega_x}}p_x\quad\Leftrightarrow\quad \alpha^*=\sqrt{\frac{m_x\omega_x}{2}}x-\frac{i}{\sqrt{2m_x\omega_x}}p_x,
\\
\beta & =\sqrt{\frac{m_y\omega_y}{2}}y+\frac{i}{\sqrt{2m_y\omega_y}}p_y\quad\Leftrightarrow\quad \beta^*=\sqrt{\frac{m_y\omega_y}{2}}y-\frac{i}{\sqrt{2m_y\omega_y}}p_y.
\end{split}
\end{align}
Adding Eqs. (\ref{eqn:omegax alphaalpha* with A and B defined}) and (\ref{eqn:omegay betabeta* with C and D defined}) yields the Hamiltonian in terms of the normal mode coordinates $\alpha$ and $\beta$

\beq
\mathscr{H}(\alpha,\beta)=\omega_x\alpha^*\alpha+\omega_y\beta^*\beta.
\label{eqn:hamiltonian 2dho normal coords}
\eeq
Now suppose $\omega_x=q\omega_0$ and $\omega_y=p\omega_0$, then Eq. (\ref{eqn:hamiltonian 2dho normal coords}) becomes

\beq
\mathscr{H}(\alpha,\beta)=\omega_0(q\alpha^*\alpha+p\beta^*\beta).
\label{eqn:hamiltonian 2dho normal coords with q and p}
\eeq
Just as in the $x-y$ basis, the Poisson bracket of the normal coordinates, $\alpha$ and $\beta$, are of importance. $\alpha$ is a function of the $x$ position and momentum, and $\beta$, a function of the $y$ position and momentum, so the Poisson bracket between them is zero,

\begin{align}
\label{eq:poisson bracket alpha beta}
\begin{split}
\{\alpha,\beta\}=0.
\end{split}
\end{align}
The Poisson bracket between $\alpha$ and its complex conjugate, $\alpha^*$, is non-zero, and holds the same information as $\{x,p_x\}$, Eq. (\ref{eq:poisson bracket 2dho properties}):

\begin{align}
\label{eq:poisson bracket alpha alpha*}
\begin{split}
\{\alpha,\alpha^*\}&=\frac{\partial\alpha}{\partial x}\frac{\partial\alpha^*}{\partial p_x}-\frac{\partial\alpha}{\partial p_x}\frac{\partial\alpha^*}{\partial x}
\\
&=\sqrt{\frac{m_x\omega_x}{2}}\left(\frac{-i}{\sqrt{2m_x\omega_x}}\right)-\frac{i}{\sqrt{2m_x\omega_x}}\left(\sqrt{\frac{m_x\omega_x}{2}}\right)
\\
&=\frac{-i}{2}-\frac{i}{2}
\\
&=-i,
\end{split}
\end{align}
and similarly for $\beta$

\begin{align}
\label{eq:poisson bracket beta beta*}
\begin{split}
\{\beta,\beta^*\}&=-i.
\end{split}
\end{align}

it's instructive to look at the time dependence of $\alpha$ and $\beta$. Using Eq. (\ref{eqn:hamiltons eqn in poisson bracket formulation})

\begin{align}
\label{eq:time dependence of alpha}
\begin{split}
\frac{d\alpha}{dt}&=\{\alpha,\mathscr{H}\}+\frac{\partial\alpha}{\partial t}
\\
&=\frac{\partial\alpha}{\partial x}\frac{\partial\mathscr{H}}{\partial p_x}-\frac{\partial\alpha}{\partial p_x}\frac{\partial\mathscr{H}}{\partial x}+\frac{\partial\alpha}{\partial y}\frac{\partial\mathscr{H}}{\partial p_y}-\frac{\partial\alpha}{\partial p_y}\frac{\partial\mathscr{H}}{\partial y}+0
\\
&=\frac{\partial\alpha}{\partial x}\frac{\partial\mathscr{H}}{\partial p_x}-\frac{\partial\alpha}{\partial p_x}\frac{\partial\mathscr{H}}{\partial x}+0-0
\\
&=\sqrt{\frac{m_x\omega_x}{2}}\frac{p_x}{m_x}-\frac{i}{\sqrt{2m_x\omega_x}}m_x\omega_x^2x
\\
&=\sqrt{\frac{\omega_x}{2m_x}}p_x-i\sqrt{\frac{m_x}{2\omega_x}}\omega_x^2x
\\
&=-i\omega_x\left(\frac{i}{\sqrt{2m_x\omega_x}}p_x+\sqrt{\frac{m_x\omega_x}{2}}x\right)
\\
&=-i\omega_x\alpha.
\end{split}
\end{align}
The solution for this differential equation is simple,

\beq
\alpha(t)=e^{-i\omega_xt}\alpha(0).
\label{eqn:solution to alpha diff eq}
\eeq
$\beta$ is found in a similar manner, except now the partial derivatives with respect to $x$ and $p_x$ are zero, and nonzero with respect to $y$ and $p_y$,

\begin{align}
\label{eq:time dependence of beta}
\begin{split}
\frac{d\beta}{dt}&=\{\beta,\mathscr{H}\}+\frac{\partial\beta}{\partial t}
\\
\frac{d\beta}{dt}&=-i\omega_y\beta,
\end{split}
\end{align}
with solution

\beq
\beta(t)=e^{-i\omega_yt}\beta(0).
\label{eqn:solution to beta diff eq}
\eeq
The solution, Eq. (\ref{eqn:solution to alpha diff eq}) allows us to find the time dependent solutions for $x$ and $p_x$. Defining Eq. (\ref{eq:normal mode coordinates 2}) with explicit $t$ dependence,

\begin{align}
\label{eq:alpha normal mode coordinates with t dep}
\begin{split}
\alpha(t) & =\sqrt{\frac{m_x\omega_x}{2}}x(t)+\frac{i}{\sqrt{2m_x\omega_x}}p_x(t)
\\
 & =\sqrt{\frac{m_x\omega_x}{2}}\left(x(t)+\frac{i}{m_x\omega_x}p_x(t)\right).
\end{split}
\end{align}
The time independent piece of Eq. (\ref{eq:alpha normal mode coordinates with t dep}) is retrieved when setting $t=0$,

\begin{align}
\label{eq:alpha normal mode coordinates with t=0}
\begin{split}
\alpha(0) & =\sqrt{\frac{m_x\omega_x}{2}}\left(x(0)+\frac{i}{m_x\omega_x}p_x(0)\right)
\\
 & =\sqrt{\frac{m_x\omega_x}{2}}\left(x_0+\frac{i}{m_x\omega_x}p_{x0}\right).
\end{split}
\end{align}
Plugging Eqs. (\ref{eq:alpha normal mode coordinates with t dep}) and (\ref{eq:alpha normal mode coordinates with t=0}) allows for us to group the real and imaginary parts of the expression. The real part yields $x(t)$, the imaginary part is the momentum, $p_x(t)$ 

\begin{align}
\label{eq:x(t) and px(t)}
\begin{split}
x(t) & =x_0\cos(\omega_xt)+\frac{p_{x0}}{m_x\omega_x}\sin(\omega_xt),
\\
p_x(t) & = -m_x\omega_xx_0\sin(\omega_xt)+p_{x0}\cos(\omega_xt).
\end{split}
\end{align}
An alternative way to find these, without the use of the normal coordinate $\alpha$, is to solve the coupled set of differential equations, Eq. (\ref{eq:Hamiltons x and px eqns for 2dho}), and assume initial conditions for the position, momentum, velocity, and force as $x(0)=x_0$, $p_x(0)=p_{x0}$, $\dot{x}(0)=p_{x0}/m_x\omega_x$, and $\dot{p_x}(0)=-m_x\omega_x^2x_0$ respectively. The time dependencies of $y$ and $p_y$ follow simply as

\begin{align}
\label{eq:y(t) and py(t)}
\begin{split}
y(t) & =y_0\cos(\omega_yt)+\frac{p_{y0}}{m_y\omega_y}\sin(\omega_yt),
\\
p_y(t) & = -m_y\omega_yy_0\sin(\omega_yt)+p_{y0}\cos(\omega_yt).
\end{split}
\end{align}

Presented above, is our motivation for bridging the gap from classical mechanics to quantum mechanics.
\subsection{Classical Lissajous Figures}
\begin{figure}[H]
    \centering
    \includegraphics[width=\textwidth]{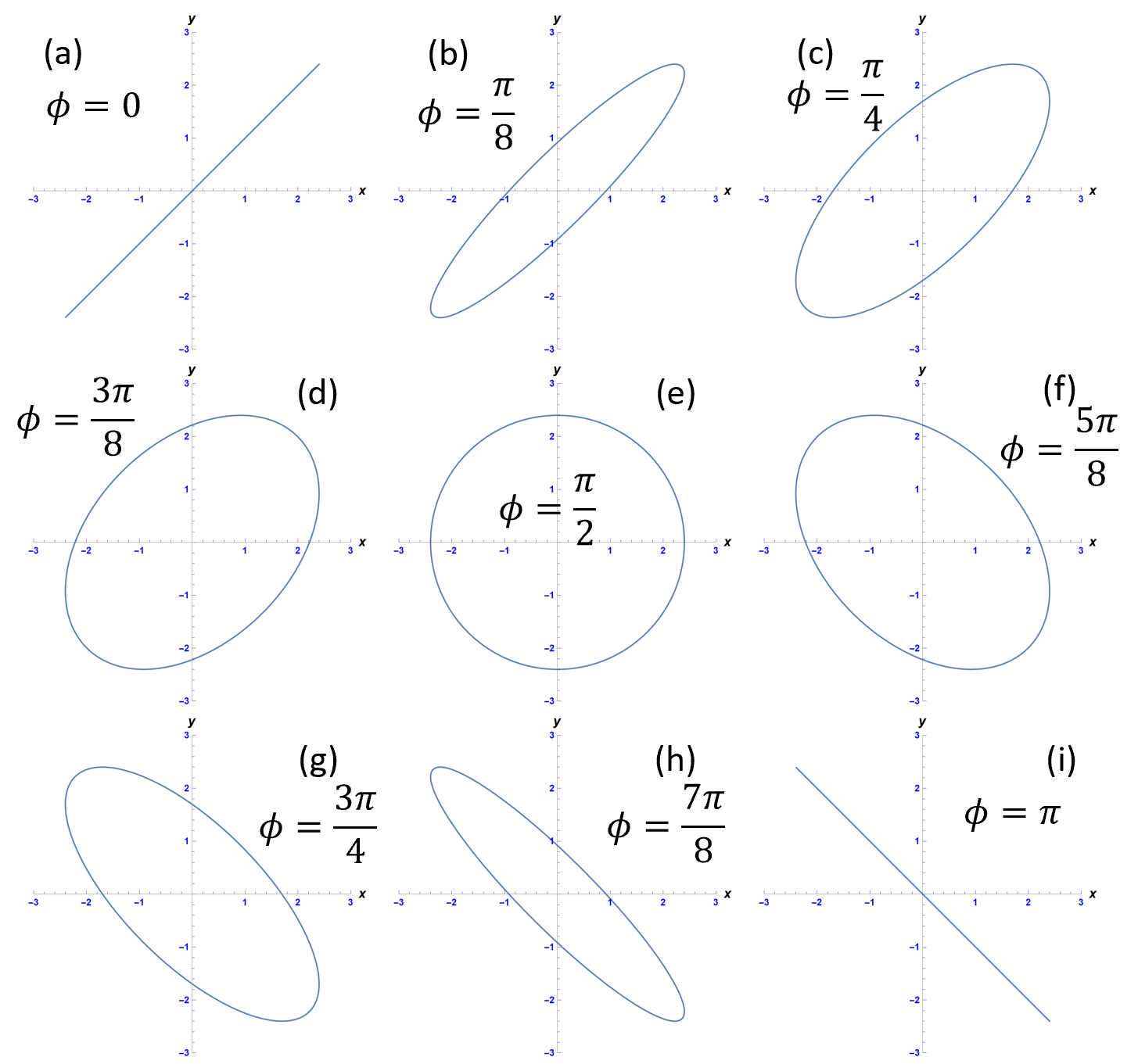}
    
    \caption{Eq. (\ref{eq:uncoupled EOMs solutions in terms of phi}) with $A=B=1$, $\omega_0=1$. Shows to evolution of the Lissajous ellipse by changing the phase, starting as a line segment, $y=x$, and morphing into ellipses until it becomes a circle. Then the major and minor axes flip until the minor axis (on $y=x$) vanishes, turning into a line segment $y=-x$. For each figure shown, there is one extreme point on the $x-$axis and one extreme point on the $y-$axis.} 
    \label{fig:CLJ-p1q1-phiVARIED}
\end{figure}

Now, we will present the concepts of Lissajous figures, as they were discovered in classical physics. Classical Lissajous figures \footnote{Other common names are Lissajous curves and Bowditch curves.} are closed orbits occurring in the complex harmonic configuration space dynamics of an nDHO, $n\geq2$, under conditions of commensurate natural angular frequencies\cite{marion}. Of the most practical importance are two- and three-dimensional Lissajous figures; in this project we focus on the two-dimensional case. The simplest examples of Lissajous figures are circles and line segments, more generally, ellipses. We can employ the 2DHO to reach the equation of the ellipse, algebraically. Following the development of Marion\cite{marion}, consider simple harmonic motion and the two degrees of freedom to be the $x-$ and $y-$directions

\begin{align}
\label{eqn:uncoupled EOMs} 
\begin{split}
\Ddot{x}+\omega_0^2x & =0,
\\
\Ddot{y}+\omega_0^2y & =0.
\end{split}
\end{align}
where $\omega_0$ is the angular frequency, and is given by $\omega_0^2=k/m$. The oscillations are uncoupled, so the solutions are sinusoidal in nature

\begin{align}
\label{eq:uncoupled EOMs solutions}
\begin{split}
x(t) & =A\cos{(\omega_0t-\delta)},
\\
y(t) & =B\cos{(\omega_0t-\mu)},
\end{split}
\end{align}
where $\delta$ and $\mu$ are phases. Adding $-\mu+\mu$ to the cosine's argument in $x(t)$ gives

\begin{align}
\label{eq:y eqn}
\begin{split}
x(t) & =A\cos{(\omega_0t-\mu+(\mu-\delta))} 
\\
& =A\cos{(\omega_0t-\mu)}\cos{(\mu-\delta))}-A\sin{(\omega_0t-\mu)}\sin{(\mu-\delta))},
\\
\end{split}
\end{align}
where we've used a trigonometric identity to break up the cosine function. From equation (\ref{eq:uncoupled EOMs solutions}) we substitute $\cos{(\omega_0t-\mu)}=y/B$ as well as defining $\phi=\mu-\delta$. After some manipulation, Eq. (\ref{eq:y eqn}) becomes the equation of an ellipse

\beq
\label{eq:general ellipse}
B^2x^2-2ABxy\cos{\phi}+A^2y^2=A^2B^2\sin^2{\phi},
\eeq
known as the general Lissajous ellipse. This is analogous to the polarization ellipse with a change of variables $x\rightarrow E_x$, $y\rightarrow E_y$, and for the constants $A\rightarrow a_1$ and $B\rightarrow a_2$  \cite{born_principles_1980}. The solutions to $x$ and $y$ are 

\begin{align}
\label{eq:uncoupled EOMs solutions in terms of phi}
\begin{split}
x(t) & =A\cos{(\omega_0t)},
\\
y(t) & =B\cos{(\omega_0t-\phi)},
\end{split}
\end{align}
as can be seen from plugging these into Eq. (\ref{eq:general ellipse}). The phase, $\phi$, and the amplitudes, $A$ and $B$, affect the eccentricity and orientation in the $x-y$ plane. Eq. (\ref{eq:uncoupled EOMs solutions in terms of phi}) is just Eq. (\ref{eq:uncoupled EOMs solutions}) in terms of $\phi$, and this set of equations is not the only set that can produce Lissajous figures. The phase difference is accounted for in $y(t)$, but can be accounted for in $x(t)$ instead, with changes to the period of evolution and to the direction in which it traces. For Eq. (\ref{eq:uncoupled EOMs solutions in terms of phi}), a positive (negative) phase, $\phi$, corresponds to a counter-clockwise (clockwise) direction of travel. What has been done to Eq. (\ref{eq:uncoupled EOMs solutions}) to get Eq. (\ref{eq:uncoupled EOMs solutions in terms of phi}) is, the phase $\alpha$ has been transferred from $x(t)$ to $y(t)$. Some special cases include: $\phi=0,\pm\pi,\pm\pi/2$. To illustrate the effect of the amplitudes $A$ and $B$, along with the phase $\phi$ on the orientation and eccentricity, we review a few basic examples. Starting with $\phi=0$, the sine function vanishes and the cosine goes to unity
\begin{align}
\label{eq:general ellipse phi=0}
\begin{split}
B^2x^2-2ABxy+A^2y^2 & =0
\\
(Bx-Ay)^2 & =0
\\
Bx & =Ay
\\
y & =\frac{B}{A}x
\end{split}
\end{align}
From the first to second line we've factored the polynomial. The plot of the last line of Eq. (\ref{eq:general ellipse phi=0}) is a straight line with slope $B/A$. Figure \ref{fig:CLJ-p1q1-phiVARIED}(a) shows a plot of this with $A=B=1$. It should be noted that the straight line has a maximal eccentricity of $\infty$. For $\phi=\pm\pi/2$
\begin{align}
\label{eq:general ellipse phi=piover2}
\begin{split}
B^2x^2+A^2y^2 & =A^2B^2
\\
\frac{x^2}{A^2}+\frac{y^2}{B^2} & =1,
\end{split}
\end{align}
where the last line is the equation of an ellipse centered at the origin of the $x-y$ plane. Figure \ref{fig:CLJ-p1q1-phiVARIED}(e) shows a plot of this with $A=B=1$, which is a circle. The eccentricity of a circle is 0, so in Fig. \ref{fig:CLJ-p1q1-phiVARIED}(a-e), the major axis is along the line $y=x$, the minor axis increases until equal to the major axis, and because of the major and minor axes being equal length, the eccentricity vanishes. Finally, for $\phi=\pm\pi$
\begin{align}
\label{eq:general ellipse phi=pi}
\begin{split}
B^2x^2+2ABxy+A^2y^2 & =0
\\
(Bx+Ay)^2 & =0
\\
Bx & =-Ay
\\
y & =-\frac{B}{A}x,
\end{split}
\end{align}
where the same steps have been taken here as in Eq. (\ref{eq:general ellipse phi=0}). The plot of the last line of Eq. (\ref{eq:general ellipse phi=pi}) is a straight line with slope $-B/A$, shown in Fig. \ref{fig:CLJ-p1q1-phiVARIED}(i). The major axis is now the line $y=-x$, so in Fig. \ref{fig:CLJ-p1q1-phiVARIED}(e-i), the minor axis (now along $y=x$) decreases to zero, and thus, the eccentricity increases back to its maximum value of $\infty$.
The whole of Fig. \ref{fig:CLJ-p1q1-phiVARIED} shows the evolution of the Lissajous ellipse through half of one period, $0<\phi\leq\pi$. Then, the back half of the evolution is completed in the range of $\pi<\phi\leq 2\pi$; Fig. \ref{fig:CLJ-p1q1-phiVARIED}(i-a) evolves backwards through the description given for Fig. \ref{fig:CLJ-p1q1-phiVARIED}(a-i). 

So far, the specific case of 1:1 Lissajous figures have been described. The ratio, 1:1, pertains to the coefficients of $\omega_0t$ in $x(t)$ and $y(t)$ in Eq. (\ref{eq:uncoupled EOMs solutions in terms of phi}). It is evident that $\omega_0/\omega_0=1$, so Figure \ref{fig:CLJ-p1q1-phiVARIED} shows the "one-to-one" ratio Lissajous curves for various values of $\phi$. The general Lissajous curve can be described using frequencies of the ratio $p:q$, where $q$ and $p$ are coprime numbers. The angular frequencies are now unique to each parametric equation; $\omega_x=q\omega_0$ and $\omega_y=p\omega_0$.
\begin{align}
\label{eq: q:p LJ}
\begin{split}
x(t) & =A\cos{(\omega_xt)},
\\
y(t) & =B\cos{(\omega_yt-\phi)},
\end{split}
\end{align}
where the ratio simplifies to $\omega_x/\omega_y=q\omega_0/p\omega_0=q/p$. In terms of $q$ and $p$, Eq. (\ref{eq: q:p LJ}) is
\begin{align}
\label{eq:q:p LJ in terms of qp}
\begin{split}
x(t) & =A\cos{(q\omega_0 t)},
\\
y(t) & =B\cos{(p\omega_0 t-\phi)}.
\end{split}
\end{align}

\begin{figure}[H]
    \centering
    \includegraphics[width=\textwidth]{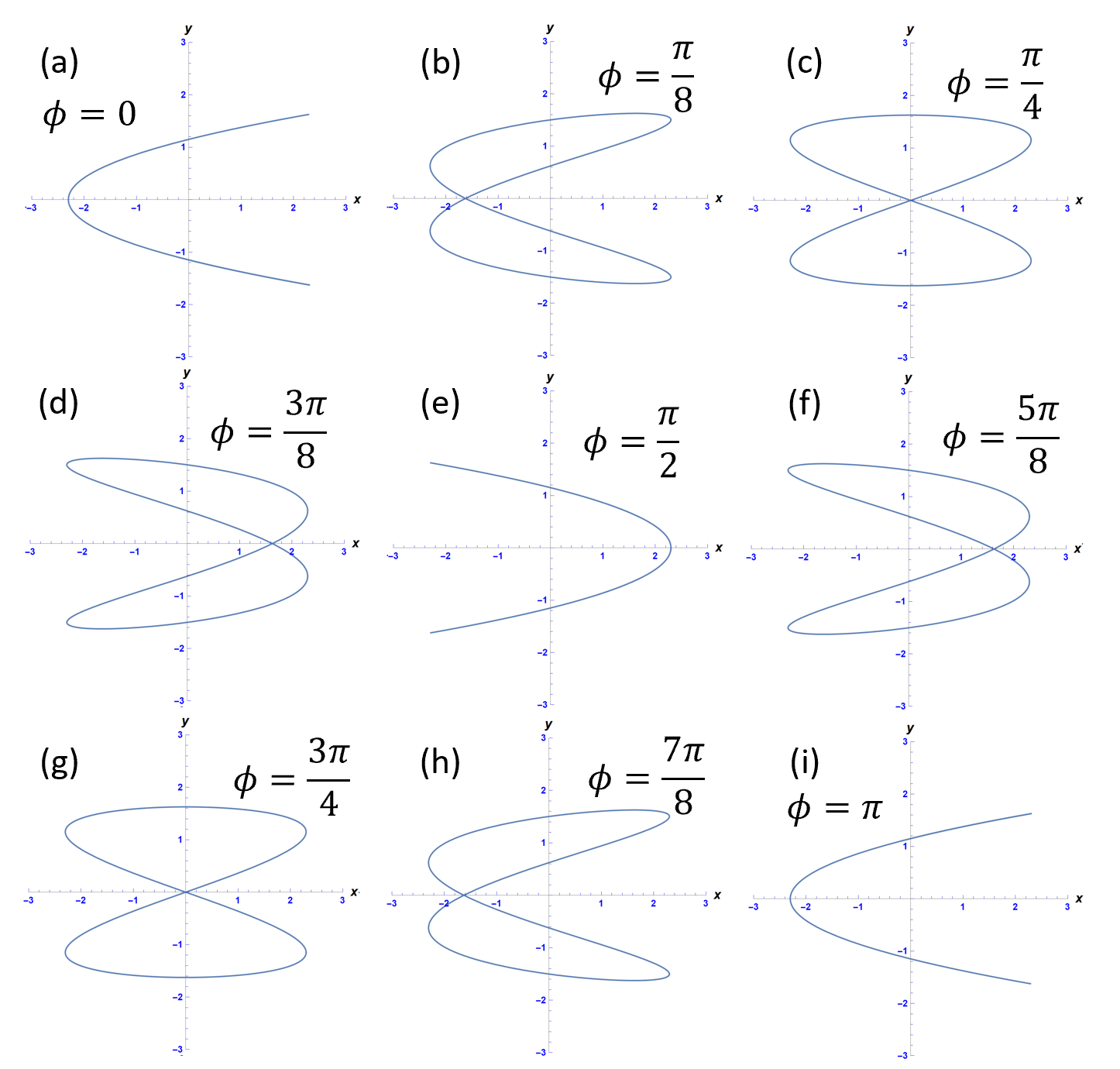}
    
    \caption{Eq. (\ref{eq:q:p LJ in terms of qp}) with $q=2$, $p=1$, $A=B=1$, $\omega_0=1$. As the phase increases the zero-point oscillates along the $x-$axis until it reaches its starting point, thus finishing one full period of the phase. Each figure shown has two extrema on the $x-$axis while only having one on the $y-$axis. For figures (a,i), the two extrema on the negative $x-$axis are at the same point, but must be counted as two due to the oscillation along the curve retracing itself. The same argument applied for the positive $x-$axis of Fig. (e).} 
    \label{fig:CLJ-p1q2-phiVARIED}
\end{figure}
Fig. \ref{fig:CLJ-p1q2-phiVARIED} shows the trajectory of the 1:2 Lissajous figure, on the range of phase, $0\leq\phi\leq\pi$. Unlike the Lissajous ellipse, the 1:2 figure completes one full period on the same phase range. A characteristic feature of Lissajous figures is the connection between $q$ and $p$ and the number of extrema on the $x-$ and $y-$axes. Fig. \ref{fig:CLJ-p1q1-phiVARIED} has $q=p=1$ and one extreme point on both the $x-$ and $y-$axes. Fig. \ref{fig:CLJ-p1q2-phiVARIED} has $q=2$ and $p=1$, and two extrema on the $x-$axis and one extreme on the $y-$axis. Therefore, there are $q$ extrema on the $x-$axis and $p$ extrema on the $y-$axis for a $p:q$ Lissajous figure.

\begin{figure}[H]
    \centering
    \includegraphics[width=\textwidth]{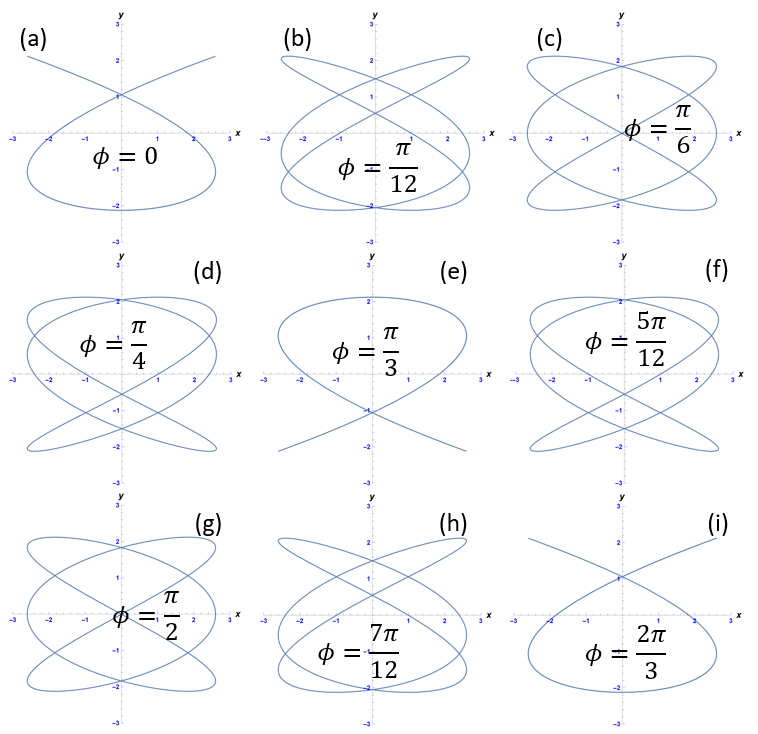}
    
    \caption{Eq. (\ref{eq:q:p LJ in terms of qp}) with $q=3$, $p=2$, $A=B=1$, $\omega_0=1$. Each figure shown has three extrema on the $x-$axis and two on the $y-$axis. Figures (a,e,i) have two overlapping extrema on the $x-$axis for the same reason as Fig. \ref{fig:CLJ-p1q2-phiVARIED}(a,e,i).} 
    \label{fig:CLJ-p2q3-phiVARIED}
\end{figure}
\begin{figure}[H]
    \centering
    \includegraphics[width=\textwidth]{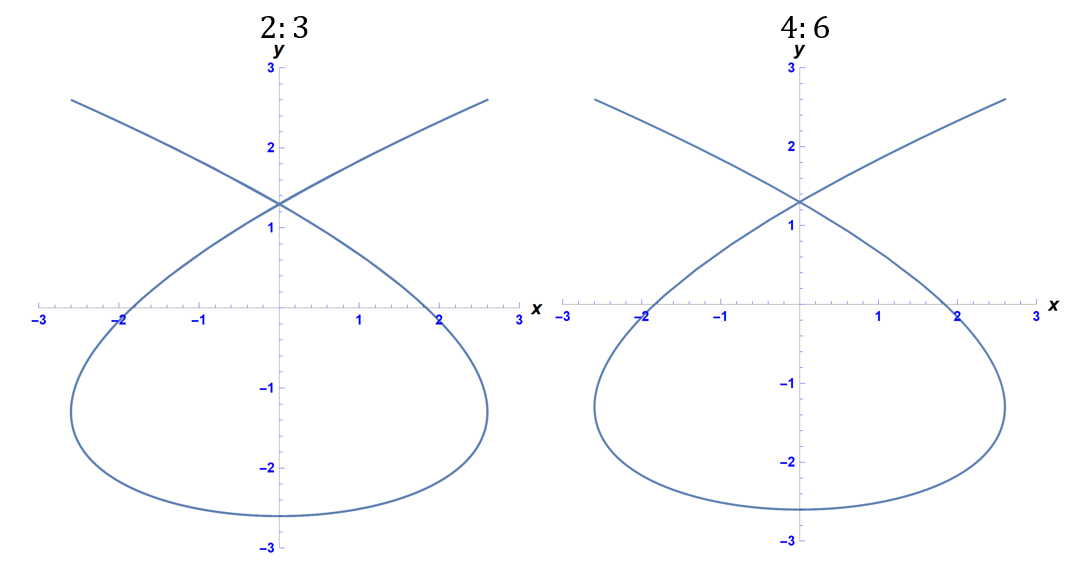}
    
    \caption{Shown is a comparison between a coprime frequency ratio $2:3$ and a higher harmonic frequency ratio $4:6$. Since, according to Eq. (\ref{eq:q:p LJ in terms of q0m p0m}) $m=2$, the $4:6$ Lissajous figure completes its oscillation along the whole curve twice as fast as the $2:3$ figure, making the frequency twice as large.} 
    \label{fig:CLJ-p2q3-p4q6-comparison}
\end{figure}

Another essential characteristic is when $q$ and $p$ are no longer coprime, i.e. $q=q_0m$, $p=p_0m$. This is the case in which there are higher harmonic frequencies. When plotted, the ratio between the frequencies simplifies to the ratio of the corresponding coprime case, but now the frequency of the higher harmonic oscillator is $m$ times faster, accounting for the increased mechanical energy in the system. This can be seen mathematically,
\begin{align}
\label{eq:q:p LJ in terms of q0m p0m}
\begin{split}
x(t) & =A\cos{(q_0m\omega_0t)},
\\
y(t) & =B\cos{(p_0m\omega_0t-\phi)}.
\end{split}
\end{align}
Figure \ref{fig:CLJ-p2q3-p4q6-comparison} shows that the 4:6 and 2:3 Lissajous figures follow the same paths at the same $\phi$ values. In a sense, in the 4:6 Lissajous figure, there is two extrema on the $y-$axis and three extrema on the $x-$axis, but in another sense, in the same time interval, the 4:6 figure will hit every point the 2:3 figure hits, twice. The fact that evaluating the simplified fraction of $q/p$ applies to all possible frequency ratios. This is mentioned as a comparison to the striking difference in the quantum mechanical case presented later, where a non-coprime Lissajous ratio, $p:q$, is evidence of a coherent superposition of various wavefunctions. One full phase period of a $p:q$ Lissajous figure, where the phase is with $y(t)$, is completed in the range $0\leq\phi\leq2\pi/q$. This can be seen from the fact that the 1:1 Lissajous figure completes its cycle on $0\leq\phi\leq2\pi$, the 1:2 Lissajous figure on $0\leq\phi\leq\pi$, and the 2:3 Lissajous figure on $0\leq\phi\leq2\pi/3$.

\begin{figure}[H]
    \centering
    \includegraphics[width=\textwidth]{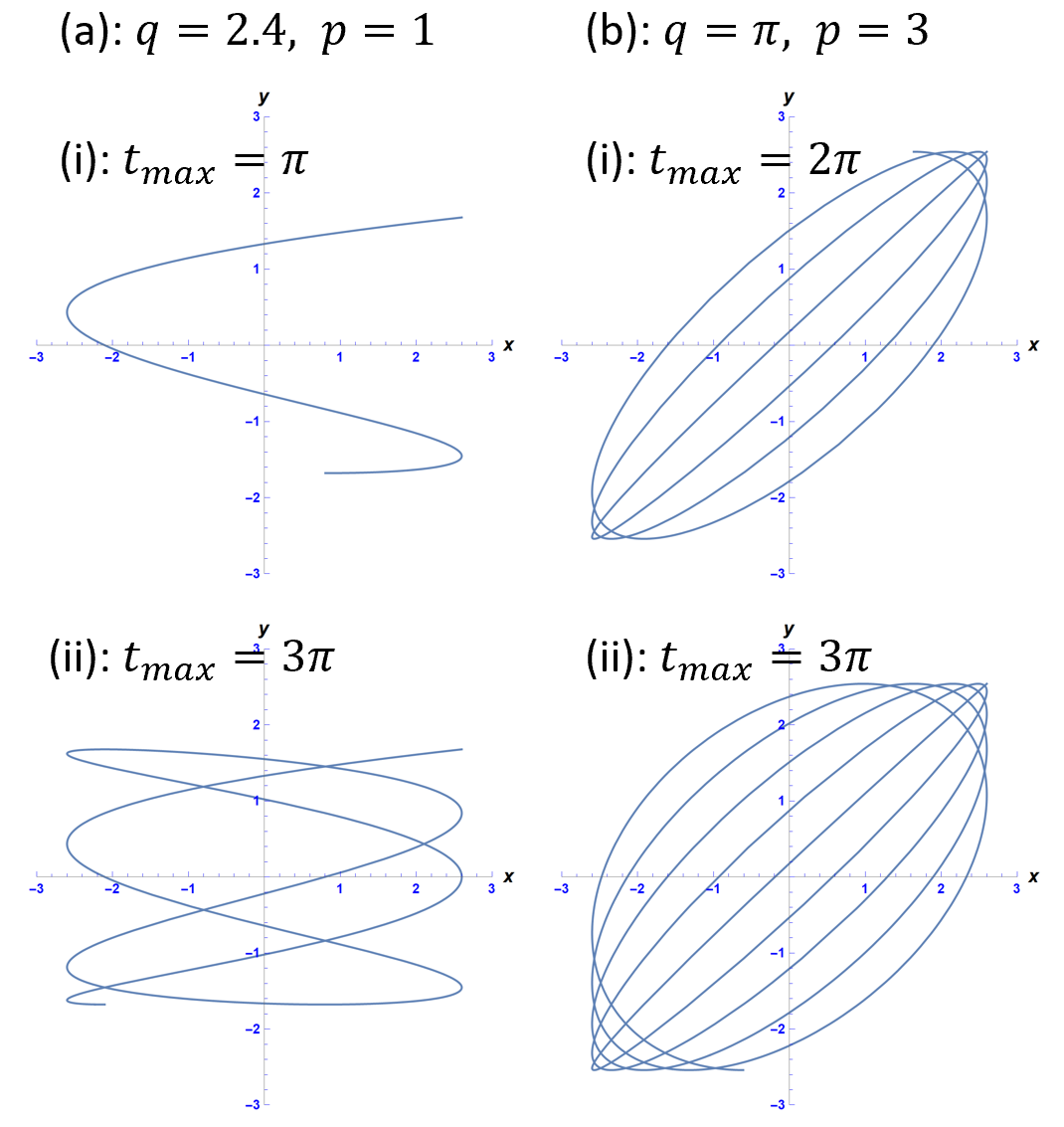}
    
    \caption{Shown are open curves with non-commensurate frequency ratios. The ratios were chosen to be almost commensurate to demonstrate how open curves exhibit similar evolution to Lissajous figures but never trace back onto themselves. (a) has a frequency ratio close to the 1:2 Lissajous figure, and it can be seen that similar curves appear as time runs on. (b) has a frequency ratio that is close to the 1:1 Lissajous figure. It is even more clear that open curves trace similar paths to Lissajous figures, as we can see the curve starts as a line and travels an elliptical path, with each revolution losing eccentricity to become closer to a circle. } 
    \label{fig:CLJ-openfigures}
\end{figure}

Thus far, all of the figures shown have been \textit{closed} curves. A Lissajous figure is always a closed curve due to the ratio $q/p$ being a rational fraction. If $q/p$ is a rational fraction, then $q$ and $p$ are said to be commensurable. If $q/p$ is not a rational fraction, $q$ and $p$ are non-commensurable, and the parametric equations yield an \textit{open} curves, and they are not Lissajous figures. These figures don't travel along the same shape periodically; instead, their paths decay in a certain way related to how close the ratio, $q/p$, is to a rational number. In other words, elements of Lissajous figures can be seen in open curves with similar ratios. Fig. (\ref{fig:CLJ-openfigures}) shows two examples of open curves; $q=2.4, p=1$ (a) and $q=\pi, p=3$ (b). Fig. (\ref{fig:CLJ-openfigures})(a)(i) has a maximum time of $\pi$ and the shape of the 1:2 Lissajous figure (with zero phase) is traced once. The figure can be seen not staying on the 1:2 trajectory. Fig. (\ref{fig:CLJ-openfigures})(a)(ii) has a maximum time of $3\pi$, and now another 1:2 figure can be seen, this time with a phase close to $3\pi/8$. This makes sense since the closest commensurate ratio to this is 1:2. Fig. (\ref{fig:CLJ-openfigures})(b)(i-ii) have maximum times of $2\pi$ and $3\pi$ respectively. We see that these "cycle" through different 1:1 Lissajous figures (ellipses), which agrees with the fact that the $\pi$:3 ratio is close to 1:1.

\section{Quantum Lissajous Figures Via Projection}
\label{chap3_QLJviaProjection}
The main goal of this section is to work our way into the principal results of this research, Quantum Lissajous figures by way of projecting a degenerate subspace of the 2DHO onto a two-mode coherent state. This is not just a jump from classical Lissajous figures to quantum Lissajous figures, but rather, a journey through the history of quantum mechanics that connects to classical concepts in more than one way. A logical first step is to assume that a 2-D quantum harmonic oscillator with commensurate frequencies will exhibit properties of Lissajous figures, as the 2-D classical harmonic oscillator does, but this is not all that is needed. We start by quantizing the Hamiltonian function from section \ref{chap2_classicalLJ} by way of the Dirac prescription. Next, we introduce key concepts of quantum mechanics, like number (Fock) states and ordinary coherent (Glauber) states. Then, the description of a semi-classical Lissajous state is given. Finally, we wrap this section up by describing our derivation of fully quantum Lissajous states, via projection.
\subsection{Quantum Mechanical Formulation of the 2DHO}
\subsubsection{Dirac Prescription}
The first step to bringing this formalism to Quantum Mechanics is to generalize the Poisson Bracket. The generalization of the Poisson Bracket allows for the quantization of variables, such as the position, momentum, and the Hamiltonian. Specifically, we are working towards the raising and lowering operators, $\hat{a}^\dagger$ and $\hat{a}$. Dirac's prescription states that the Poisson bracket is simply the quantum commutator divided by $i\hbar$\cite{dirac_generalized_1950}. We define a quantum operator $\hat{A}$ that corresponds to the classical $\alpha$

\begin{align}
\label{eq:dirac prescription}
\begin{split}
\{\alpha,\alpha^*\} \quad\Rightarrow\quad \frac{[\hat{A},\hat{A}^\dagger]}{i\hbar},
\end{split}
\end{align}
where this substitution applies for all variables that would be inside the bracket. The commutator $[\hat{A},\hat{A}^\dagger]$ goes as

\begin{align}
\label{eq:commutator}
\begin{split}
[\hat{A},\hat{A}^\dagger]=\hat{A}\hat{A}^\dagger-\hat{A}^\dagger\hat{A},
\end{split}
\end{align}
Eq. (\ref{eq:dirac prescription}) allows for the retrieval of the Heisenberg equation by way of Hamilton's equation in terms of the Poisson bracket, Eq. (\ref{eqn:hamiltons eqn in poisson bracket formulation}). Applying the Dirac prescription to Eq. (\ref{eqn:hamiltons eqn in poisson bracket formulation}), and assuming the function $u$ is an observable, it becomes an operator: 

\begin{align}
\label{eqn:heisenberg eqn}
\begin{split}
\frac{d\hat{u}}{dt}&=\frac{[\hat{u},\hat{H}]}{i\hbar}+\frac{\partial\hat{u}}{\partial t}
\\
&=\frac{i}{\hbar}[\hat{H},\hat{u}]+\frac{\partial\hat{u}}{\partial t}.
\end{split}
\end{align}
The Heisenberg equation determines the time dependence of an observable\cite{mandel_optical_1995}. This is appropriately named due to the fact that time dependent operators is characteristic of the Heisenberg Picture of Quantum Mechanics\cite{sakurai_modern_2017}. We know that $\{\alpha,\alpha^*\}=-i$, so plugging this into the Dirac prescription gives

\begin{align}
\label{eq:dirac prescription 2}
\begin{split}
\{\alpha,\alpha^*\}=-i \quad\Rightarrow\quad \frac{[\hat{A},\hat{A}^\dagger]}{i\hbar}=-i,
\end{split}
\end{align}
which shows that the commutator of the operator $\hat{A}$ is 

\begin{align}
\label{eq:commutator of A}
\begin{split}
[\hat{A},\hat{A}^\dagger]=\hbar.
\end{split}
\end{align}
To get to that Bosonic ladder operators, $\hat{a}$, a rescaling of $\hat{A}$ is required. Letting $\hat{A}=\sqrt{\hbar}\hat{a}_x$,

\begin{align}
\label{eq:commutator of ax}
\begin{split}
[\hat{A},\hat{A}^\dagger]=\hbar[\hat{a}_x,\hat{a}_x^\dagger]&=\hbar,
\\
[\hat{a}_x,\hat{a}_x^\dagger]&=1.
\end{split}
\end{align}
For $\beta$, we define a quantum operator that is equal to $\sqrt{\hbar}\hat{a}_y$, and come to

\begin{align}
\label{eq:commutator of ay}
\begin{split}
[\hat{a}_y,\hat{a}_y^\dagger]&=1.
\end{split}
\end{align}
We can apply the Dirac prescription on the canonical coordinates as well

\begin{align}
\label{eq:dirac prescription x and px}
\begin{split}
\{q_i,p_i\}=1 \quad\Rightarrow\quad \frac{[\hat{q}_i,\hat{p}_i]}{i\hbar}=1,
\end{split}
\end{align}
which yields the canonical commutation relation

\begin{align}
\label{eq:canonical commutation relation x}
\begin{split}
[\hat{q}_i,\hat{p}_i]=i\hbar.
\end{split}
\end{align}
Note that the $i$ on the right hand side of Eqs. (\ref{eq:dirac prescription x and px}) and (\ref{eq:canonical commutation relation x}) is the imaginary number, the subscript $i$ is an index where $i=x,y$. For the $x$ and $y$ coordinates,

\begin{align}
\label{eq:canonical commutation relation y}
\begin{split}
[\hat{x},\hat{p}_x]=[\hat{y},\hat{p}_y]=i\hbar.
\end{split}
\end{align}
The Dirac prescription is more straight forward for the canonical and normal coordinates, they all become quantum operators. From Eq. (\ref{eq:normal mode coordinates 2}) we get

\begin{align}
\label{eq:normal coords to operators}
\begin{split}
\alpha & =\sqrt{\frac{m_x\omega_x}{2}}x+\frac{i}{\sqrt{2m_x\omega_x}}p_x\quad\Rightarrow\quad \hat{A}=\sqrt{\frac{m_x\omega_x}{2}}\left(\hat{x}+\frac{i}{m_x\omega_x}\hat{p}_x\right),
\\
\beta & =\sqrt{\frac{m_y\omega_y}{2}}y+\frac{i}{\sqrt{2m_y\omega_y}}p_y\quad\Rightarrow\quad \hat{B}=\sqrt{\frac{m_y\omega_y}{2}}\left(\hat{y}+\frac{i}{m_y\omega_y}\hat{p}_y\right),
\end{split}
\end{align}
and again, rescaling using $\hat{A}=\sqrt{\hbar}\hat{a}_x$ and $\hat{B}=\sqrt{\hbar}\hat{a}_y$

\begin{align}
\label{eq:operators to quantum operators}
\begin{split}
\hat{A}=\sqrt{\frac{m_x\omega_x}{2}}\left(\hat{x}+\frac{i}{m_x\omega_x}\hat{p}_x\right)\quad\Rightarrow\quad \hat{a}_x=\sqrt{\frac{m_x\omega_x}{2\hbar}}\left(\hat{x}+\frac{i}{m_x\omega_x}\hat{p}_x\right),
\\
\hat{B}=\sqrt{\frac{m_y\omega_y}{2}}\left(\hat{y}+\frac{i}{m_y\omega_y}\hat{p}_y\right)\quad\Rightarrow\quad \hat{a}_y=\sqrt{\frac{m_y\omega_y}{2\hbar}}\left(\hat{y}+\frac{i}{m_y\omega_y}\hat{p}_y\right).
\end{split}
\end{align}
Now with the inclusion of $\hbar$, these operators are inherently quantum mechanical. The presence of $\hbar$ is a sign of a quantum mechanical system, due to the fact that $\hbar$ is a proportionality constant between a photon's energy and angular frequency, $E=\hbar\omega$, and between a photon's momentum and angular wave number, $p=\hbar k$\cite{noauthor_planck_nodate}. The position and momentum operators can be found from Eq. (\ref{eq:operators to quantum operators})

\begin{align}
\label{eqn:p,x operators}
\begin{split}
\hat{p}_x=i\sqrt{\frac{\hbar m_x\omega_x}{2}}\left(\hat{a}_x^\dagger-\hat{a}_x\right),
\\
\hat{x}=\sqrt{\frac{\hbar}{2 m_x\omega_x}}\left(\hat{a}_x+\hat{a}_x^\dagger\right),
\end{split}
\end{align}

\begin{align}
\label{eqn:p,y operators}
\begin{split}
\hat{p}_y=i\sqrt{\frac{\hbar m_y\omega_y}{2}}\left(\hat{a}_y^\dagger-\hat{a}_y\right),
\\
\hat{y}=\sqrt{\frac{\hbar}{2 m_y\omega_y}}\left(\hat{a}_y+\hat{a}_y^\dagger\right).
\end{split}
\end{align}
The Hamiltonian function, Eq. (\ref{eqn:hamiltonian 2dho}), in operator form is

\begin{align}
\label{eqn:quantum hamiltonian 2dho}
\begin{split}
\hat{H}=\frac{\hat{p}_x^2}{2m_x}+\frac{1}{2}m_x\omega_x^2\hat{x}^2+\frac{\hat{p}_y^2}{2m_y}+\frac{1}{2}m_y\omega_y^2\hat{y}^2.
\end{split}
\end{align}
One might assume that the quantum Hamiltonian in terms of Ladder operators will be of the same form as Eq. (\ref{eqn:hamiltonian 2dho normal coords}), but this is not the case. The operators $\hat{x}$ and $\hat{p}_x$ do not commute, and so the cross-terms of $\hat{a}_x^\dagger\hat{a}_x$ do not add to zero: 

\begin{align}
\label{eqn:a^dagger a}
\begin{split}
\hat{a}_x^\dagger\hat{a}_x&=\frac{m_x\omega_x}{2\hbar}\left( \hat{x}^2+\frac{i}{m_x\omega_x}\hat{x}\hat{p}_x-\frac{i}{m_x\omega_x}\hat{p}_x\hat{x}+\frac{1}{m_x^2\omega_x^2}\hat{p}_x^2\right)
\\
&=\frac{m_x\omega_x}{2\hbar}\left( \hat{x}^2+\frac{i}{m_x\omega_x}[\hat{x},\hat{p}_x]+\frac{1}{m_x^2\omega_x^2}\hat{p}_x^2\right)
\\
&=\frac{m_x\omega_x}{2\hbar}\left( \hat{x}^2-\frac{\hbar}{m_x\omega_x}+\frac{1}{m_x^2\omega_x^2}\hat{p}_x^2\right).
\end{split}
\end{align}
To get rid of the cross-terms, the Hermitian conjugate of $\hat{a}_x^\dagger\hat{a}_x$, $\hat{a}_x\hat{a}_x^\dagger$, should be looked at:

\begin{align}
\label{eqn:a a^dagger}
\begin{split}
\hat{a}_x\hat{a}_x^\dagger&=\frac{m_x\omega_x}{2\hbar}\left( \hat{x}^2-\frac{i}{m_x\omega_x}\hat{x}\hat{p}_x+\frac{i}{m_x\omega_x}\hat{p}_x\hat{x}+\frac{1}{m_x^2\omega_x^2}\hat{p}_x^2\right)
\\
&=\frac{m_x\omega_x}{2\hbar}\left( \hat{x}^2-\frac{i}{m_x\omega_x}[\hat{x},\hat{p}_x]+\frac{1}{m_x^2\omega_x^2}\hat{p}_x^2\right)
\\
&=\frac{m_x\omega_x}{2\hbar}\left( \hat{x}^2+\frac{\hbar}{m_x\omega_x}+\frac{1}{m_x^2\omega_x^2}\hat{p}_x^2\right).
\end{split}
\end{align}
Adding Eqs. (\ref{eqn:a^dagger a}) and (\ref{eqn:a a^dagger}) yields

\begin{align}
\label{eqn:a^dagger a + a a^dagger}
\begin{split}
\hat{a}_x^\dagger\hat{a}_x+\hat{a}_x\hat{a}_x^\dagger&=\frac{m_x\omega_x}{2\hbar}\left( \hat{x}^2-\frac{\hbar}{m_x\omega_x}+\frac{1}{m_x^2\omega_x^2}\hat{p}_x^2+\hat{x}^2+\frac{\hbar}{m_x\omega_x}+\frac{1}{m_x^2\omega_x^2}\hat{p}_x^2\right)
\\
&=\frac{m_x\omega_x}{\hbar}\left( \hat{x}^2+\frac{1}{m_x^2\omega_x^2}\hat{p}_x^2\right).
\end{split}
\end{align}
Multiplying Eq. (\ref{eqn:a^dagger a + a a^dagger}) by $\hbar\omega_x/2$ gives

\begin{align}
\label{eqn:hbar omega/2(a^dagger a + a a^dagger)}
\begin{split}
\hat{H}_x=\frac{\hbar\omega_x}{2}(\hat{a}_x^\dagger\hat{a}_x+\hat{a}_x\hat{a}_x^\dagger)=\frac{1}{2}m_x\omega^2_x\hat{x}^2+\frac{\hat{p}_x^2}{2m_x}.
\end{split}
\end{align}
This is the $x$ component of the Hamiltonian operator, Eq. (\ref{eqn:quantum hamiltonian 2dho}). Recell Eq. (\ref{eq:commutator of ax}), expanding the commutator bracket

$$[\hat{a}_x,\hat{a}_x^\dagger]=\hat{a}_x\hat{a}_x^\dagger-\hat{a}_x^\dagger\hat{a}_x=1,$$
and isolating $\hat{a}_x\hat{a}_x^\dagger$ gives

$$\hat{a}_x\hat{a}_x^\dagger=\hat{a}_x^\dagger\hat{a}_x+1.$$
Plugging this into Eq. (\ref{eqn:hbar omega/2(a^dagger a + a a^dagger)})

\begin{align}
\label{eqn:Hx}
\begin{split}
\hat{H}_x&=\frac{\hbar\omega_x}{2}(2\hat{a}_x^\dagger\hat{a}_x+1)
\\
&=\hbar\omega_x\left(\hat{a}_x^\dagger\hat{a}_x+\frac{1}{2}\right).
\end{split}
\end{align}
The $y$ component of the Hamiltonian is found by analogy

\begin{align}
\label{eqn:Hy}
\begin{split}
\hat{H}_y&=\frac{\hbar\omega_y}{2}(\hat{a}_y^\dagger\hat{a}_y+\hat{a}_y\hat{a}_y^\dagger)=\frac{1}{2}m_y\omega^2_y\hat{y}^2+\frac{\hat{p}_y^2}{2m_y}
\\
&=\hbar\omega_y\left(\hat{a}_y^\dagger\hat{a}_y+\frac{1}{2}\right).
\end{split}
\end{align}
The total Hamiltonian, in terms of $\hat{a}_x$ and $\hat{a}_y$ is then

\begin{align}
\label{eqn:quantum hamiltonian 2dho in terms of a and a^dagger}
\begin{split}
\hat{H}&=\hbar\omega_x\left(\hat{a}_x^\dagger\hat{a}_x+\frac{1}{2}\right)+\hbar\omega_y\left(\hat{a}_y^\dagger\hat{a}_y+\frac{1}{2}\right).
\end{split}
\end{align}
Now if we assume, as we did for Eqs. (\ref{eq:q:p LJ in terms of qp}) and (\ref{eqn:hamiltonian 2dho normal coords with q and p}), that $\omega_x=q\omega_0$ and $\omega_y=p\omega_0$, Eq. (\ref{eqn:quantum hamiltonian 2dho in terms of a and a^dagger}) becomes

\begin{align}
\label{eqn:quantum hamiltonian 2dho in terms of a and a^dagger and p and q}
\begin{split}
\hat{H}&=\hbar\omega_0\left(q\hat{a}_x^\dagger\hat{a}_x+p\hat{a}_y^\dagger\hat{a}_y+\frac{q+p}{2}\right).
\end{split}
\end{align}

\subsubsection{Number States (Fock States)}
The raising and lowering operators, $\hat{a}_i^\dagger$ and $\hat{a}_i$, where $i=x,y$, are named as such due to the result of when they're applied to the Number states, $\ket{n}$,

\begin{align}
\label{eqn:a on number states}
\begin{split}
\hat{a}_x\ket{n}=\sqrt{n}\ket{n-1},
\end{split}
\end{align}

\begin{align}
\label{eqn:adagger on number states}
\begin{split}
\hat{a}_x^\dagger\ket{n}=\sqrt{n+1}\ket{n+1}.
\end{split}
\end{align}
If we apply $\hat{a}_x^\dagger$ to Eq. (\ref{eqn:a on number states}), the result is

\begin{align}
\label{eqn:adagger on a on number states}
\begin{split}
\hat{a}_x^\dagger(\hat{a}_x\ket{n})&=\sqrt{n}\hat{a}_x^\dagger\ket{n-1}\\
&=\sqrt{n}\sqrt{n-1+1}\ket{n}\\
&=n\ket{n}.
\end{split}
\end{align}
This combination of operators, $\hat{a}_x^\dagger\hat{a}_x$, is the same form as what appears in the quantum harmonic oscillator's Hamiltonian. It is called the Number operator

\begin{align}
\label{eqn:number operator}
\begin{split}
\hat{n}=\hat{a}_x^\dagger\hat{a}_x.
\end{split}
\end{align}
Eq. (\ref{eqn:adagger on a on number states}) shows that the number state is an eigenstate of the number operator. Explicitly, the eigenvalue equation is

\begin{align}
\label{eqn:number state eigenvalue eqn}
\begin{split}
\hat{n}\ket{n}=n\ket{n},
\end{split}
\end{align}
where $n$ is a nonnegative integer, i.e. $n=0,1,2,...$. Number states $n\geq 1$ can be constructed from the ground state, $\ket{0}$

\begin{align}
\label{eqn:number state in terms of grd state}
\begin{split}
\ket{n}=\frac{(\hat{a}_x^\dagger)^n}{\sqrt{n!}}\ket{0}.
\end{split}
\end{align}
It is useful to know the commutation relation between $\hat{n}$ and $\hat{a}_x$

\begin{align}
\label{eqn:N and a commutator}
\begin{split}
[\hat{n},\hat{a}_x]&=\hat{n}\hat{a}_x-\hat{a}_x\hat{n}
\\
&=\hat{a}_x^\dagger\hat{a}_x\hat{a}_x-\hat{a}_x\hat{a}_x^\dagger\hat{a}_x
\\
&=(\hat{a}_x^\dagger\hat{a}_x-\hat{a}_x\hat{a}_x^\dagger)\hat{a}_x
\\
&=[\hat{a}_x^\dagger,\hat{a}_x]\hat{a}_x
\\
&=-\hat{a}_x,
\end{split}
\end{align}
and between $\hat{n}$ and $\hat{a}_x^\dagger$

\begin{align}
\label{eqn:N and adagger commutator}
\begin{split}
[\hat{n},\hat{a}_x^\dagger]&=\hat{a}_x^\dagger.
\end{split}
\end{align}
Consider the 1-dimensional Hamiltonian given by Eq. (\ref{eqn:Hx}). It can be rewritten in terms of the number operator

\beq
\hat{H}_x=\hbar\omega_x\left(\hat{n}+\frac{1}{2}\right).
\label{eqn:Hx in terms of number operator}
\eeq
By inspection, we see that $\ket{n}$ is also an eigenstate of Eq. (\ref{eqn:Hx in terms of number operator}) 

\begin{align}
\label{eqn:1DHO eigenvalue eqn}
\begin{split}
\hat{H}_x\ket{n}&=\hbar\omega_x\left(\hat{n}+\frac{1}{2}\right)\ket{n}\\
&=\hbar\omega_x\left(n+\frac{1}{2}\right)\ket{n},
\end{split}
\end{align}

with an eigenvalue
\begin{align}
\label{eqn:1DHO eigenvalue}
\begin{split}
E_n=\hbar\omega_x\left(n+\frac{1}{2}\right).
\end{split}
\end{align}
The wavefunction of the Fock state is found by projection of the configuration space state $\ket{x}$

\beq
\psi_n(\nu)=\braket{x}{n}=\frac{1}{\sqrt{n!2^n}}\left(\frac{m_x\omega_x}{\pi\hbar}\right)^{1/4}H_n(\nu)e^{-\nu^2/2},
\label{eqn:1dho x wavefcn}
\eeq
where

\beq
\nu=\sqrt{\frac{m_x \omega_x}{\hbar}}x,
\label{eqn:1dho x wavefcn variable}
\eeq
and $H_n(\nu)$ is the Hermite polynomial. Back to the 2-dimensional case, we choose the convention where the $x$ operators are applied to the $\ket{n}$ number state, and the $y$ operators are applied to the $\ket{m}$ number state. Eq. (\ref{eqn:quantum hamiltonian 2dho in terms of a and a^dagger and p and q}) in terms of number operators becomes

\begin{align}
\label{eqn:quantum hamiltonian 2dho in terms of a and a^dagger and p and q and number operators}
\begin{split}
\hat{H}&=\hbar\omega_0\left(q\hat{n}+p\hat{m}+\frac{q+p}{2}\right).
\end{split}
\end{align}

\subsubsection{Coherent States of the Free Field (Glauber States)}
As per the theme of this study, we wish to find classical-like behavior in the quantum regime. The Glauber states are minimum uncertainty states, which already makes them more classical than other states, but this fact comes to front explicitly in Heisenberg's uncertainty principle, in which the canonical position and momentum operators are directly involved\cite{arecchi_atomic_1972}. The coherent state is an eigenstate of the lowering operator, such that,  

\begin{align}
\label{eqn:coherent state eigenvalue eqn}
\begin{split}
\hat{a}_x\ket{\alpha}=\alpha\ket{\alpha},
\end{split}
\end{align}
where $\alpha$ is a complex number. The coherent state in the number state basis is given as\cite{gerry_introductory_2005}

\begin{align}
\label{eqn:coherent state in number basis}
\begin{split}
\ket{\alpha}=e^{-\abs{\alpha}^2/2}\sum^\infty_{n=0}\frac{\alpha^n}{\sqrt{n!}}\ket{n}.
\end{split}
\end{align}

The general uncertainty principle applies to three observables that satisfy the commutation relation

$$[\hat{A},\hat{B}]=i\hat{C},$$
then the uncertainty for some quantum state is

\begin{align}
\label{eqn:general uncertainty principle}
\begin{split}
\expval{(\hat{\Delta A})^2}\expval{(\hat{\Delta B})^2}&\geq\abs{\frac{1}{2i}\expval{[\hat{A},\hat{B}]}}^2\\
&\geq\frac{1}{4}\abs{\expval{\hat{C}}}^2,
\end{split}
\end{align}
where $\hat{(\Delta A)}^2$ is the variance of an operator $\hat{A}$

$$\hat{(\Delta A)}^2=\left(\hat{A}-\expval{\hat{A}}\right)^2,$$
and the expectation value of the variances works out as such:

\begin{align}
\label{eqn:variance of exp val}
\begin{split}
\expval{(\hat{\Delta A})^2}&=\expval{\left(\hat{A}-\expval{\hat{A}}\right)^2}\\
&=\expval{\hat{A}^2-2\hat{A}\expval{\hat{A}}+\expval{\hat{A}}^2}\\
&=\expval{\hat{A}^2}-2\expval{\hat{A}}^2+\expval{\hat{A}}^2\\
&=\expval{\hat{A}^2}-\expval{\hat{A}}^2.
\end{split}
\end{align}
To retrieve Heisenberg's uncertainty principle, we use the canonical commutation relation given by Eq. (\ref{eq:canonical commutation relation x}), where $\hat{A}=\hat{q}_i$, $\hat{B}=\hat{p}_i$, and $\hat{C}=\hbar\hat{I}$

\begin{align}
\label{eqn:heisenberg uncertainty principle}
\begin{split}
\expval{(\hat{\Delta q_i})^2}\expval{(\hat{\Delta p_i})^2}&\geq\frac{1}{4}\abs{\expval{\hbar\hat{I}}}^2\\
&\geq\frac{\hbar^2}{4}.
\end{split}
\end{align}
From here, we consider the $x$ direction for continuity, and it is easy to show that for the coherent state $\ket{\alpha}$, the Heisenberg uncertainty principle is minimized,

\begin{align}
\label{eqn:heisenberg uncertainty principle coherent state}
\begin{split}
\expval{(\hat{\Delta x})^2}{\alpha}\expval{(\hat{\Delta p_x})^2}{\alpha}&=\frac{\hbar}{2m_x\omega_x}\frac{\hbar m_x\omega_x}{2}\\
&=\frac{\hbar^2}{4},
\end{split}
\end{align}
verifying that the coherent state is a minimum uncertainty state. We can build coherent states from the number state basis, with the introduction of the displacement operator,

\begin{align}
\label{eqn:displacement operator}
\begin{split}
\hat{D}(\alpha)=e^{\alpha\hat{a}_x^\dagger-\alpha^*\hat{a}_x}.
\end{split}
\end{align}
The displacement operator is a unitary operator, i.e. the multiplication of itself and its Hermitian conjugate yields the identity operator,

\begin{align}
\label{eqn:displacement operator unitarity}
\begin{split}
\hat{D}(\alpha)\hat{D}^\dagger(\alpha)=\hat{I}.
\end{split}
\end{align}
It is also worth noting that 

\begin{align}
\label{eqn:displacement operator property 1}
\begin{split}
\hat{D}^\dagger(\alpha)=\hat{D}^{-1}(\alpha)=\hat{D}(-\alpha),
\end{split}
\end{align}
and also

\begin{align}
\label{eqn:displacement operator property 2}
\begin{split}
\hat{D}^\dagger(\alpha)\hat{a}_x\hat{D}(\alpha)&=\hat{a}_x+\alpha,
\\
\hat{D}^\dagger(\alpha)\hat{a}_x^\dagger\hat{D}(\alpha)&=\hat{a}_x^\dagger+\alpha^*,
\\
\hat{D}(\alpha+\beta)&=\hat{D}(\alpha)\hat{D}(\beta)e^{-i\operatorname{Im}(\alpha\beta^*)}.
\end{split}
\end{align}
The properties in Eqs. (\ref{eqn:displacement operator property 1}) and (\ref{eqn:displacement operator property 2}) are found with the use of the Baker-Campbell-Hausdorff lemma\cite{gerry_introductory_2005}; for two arbitrary operators $\hat{A}$ and $\hat{B}$,

\begin{align}
\label{eqn:BCH lemma}
\begin{split}
e^{i\lambda\hat{A}}\hat{B}e^{-i\lambda\hat{A}}=\hat{B}+i\lambda[\hat{A},\hat{B}]+\frac{(i\lambda)^2}{2!}[\hat{A},[\hat{A},\hat{B}]]+...
\end{split}
\end{align}
This is not to be confused with the Baker-Campbell-Hausdorff Theorem\cite{gerry_introductory_2005}; for two operators $\hat{O}$ and $\hat{P}$ that do not commute, i.e. $[\hat{O},\hat{P}]\neq 0$, if 

$$[\hat{O},[\hat{O},\hat{P}]]=[\hat{P},[\hat{O},\hat{P}]]=0,$$
then 

\begin{align}
\label{eqn:BCH theorem}
\begin{split}
e^{\hat{O}+\hat{P}}&=e^{-[\hat{O},\hat{P}]/2}e^{\hat{O}}e^{\hat{P}}\\
&=e^{[\hat{O},\hat{P}]/2}e^{\hat{P}}e^{\hat{O}}.
\end{split}
\end{align}
The Baker-Campbell-Hausdorff Theorem allows for Eq. (\ref{eqn:displacement operator}) to be expressed in its normally ordered form\cite{arecchi_atomic_1972},

\begin{align}
\label{eqn:displacement operator normally ordered}
\begin{split}
\hat{D}(\alpha)=e^{-\abs{\alpha}^2/2}e^{\alpha\hat{a}_x^\dagger}e^{\alpha\hat{a}_x}.
\end{split}
\end{align}
The normally ordered form of the displacement operator is a tool used to reach Eq. (\ref{eqn:coherent state in number basis}). 
If the normally ordered displacement operator is applied to the ground state $\ket{n=0}$, we retrieve the coherent state

\begin{align}
\label{eqn:coherent state from ground state}
\begin{split}
e^{-\abs{\alpha}^2/2}e^{\alpha\hat{a}_x^\dagger}e^{\alpha\hat{a}_x}\ket{0}&=e^{-\abs{\alpha}^2/2}e^{\alpha\hat{a}_x^\dagger}(1+\alpha\hat{a}_x+\frac{(\alpha\hat{a}_x)^2}{2}+...)\ket{0}\\
&=e^{-\abs{\alpha}^2/2}\sum^\infty_{n=0}\frac{(\alpha\hat{a}_x^\dagger)^n}{n!}\ket{0}\\
&=e^{-\abs{\alpha}^2/2}\sum^\infty_{n=0}\frac{\alpha^n}{\sqrt{n!}}\ket{n}\\
&=\ket{\alpha}.
\end{split}
\end{align}

\subsection{Semi-classical Dynamics}
We have discussed the properties of the single-mode ordinary coherent states. To argue for the possibility of quantum Lissajous figures, it is only natural to begin with a two-mode ordinary coherent state since ordinary coherent states are minimum uncertainty states and classical Lissajous figures arise from a 2-dimensional system of orthogonal harmonic oscillators. 

\begin{align}
\label{eqn:2d coherent state}
\begin{split}
\ket{\alpha,\beta}&=e^{-\abs{\alpha}^2/2}e^{-\abs{\beta}^2/2}\sum^\infty_{n=0}\sum^\infty_{m=0}\frac{\alpha^n\beta^m}{\sqrt{n!m!}}\ket{n,m}.\\
\end{split}
\end{align}
Eq. (\ref{eqn:2d coherent state}) as is, produces a probability density function that is a Gaussian displaced from the origin of configuration space by $|\alpha|$ in the $x$-direction and $|\beta|$ in the $y$-direction. Therefore, it does not immediately give indication of being related to Lissajous figures.

\begin{figure}[H]
    \centering
    \includegraphics[width=\textwidth]{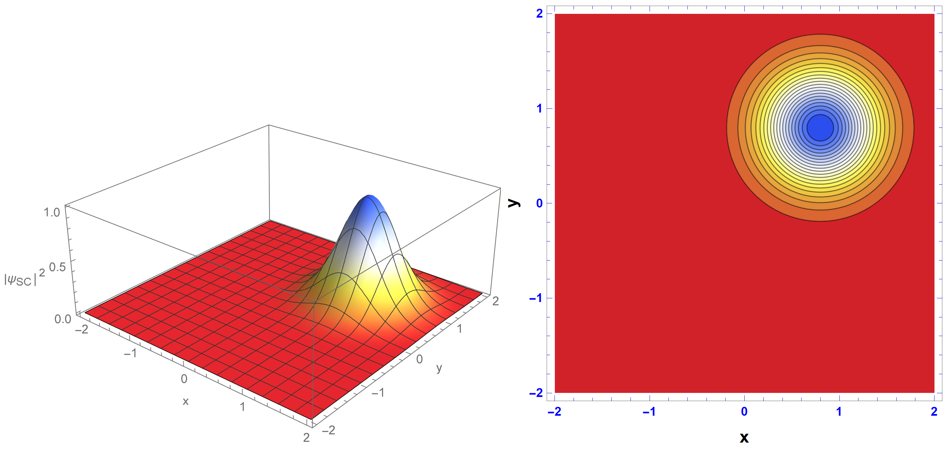}
    
    \caption{The contour and 3D plots of the probability density of a two-mode ordinary coherent state. The centroid of the wave packet is a Gaussian displaced from the origin of configuration space by $|\alpha|$ in the $x$-direction and $|\beta|$ in the $y$-direction.} 
    \label{fig:LJ-figure-2mode-coherentstate}
\end{figure}
Where it becomes related, is when the time evolution operator is applied. Working in 1-dimension for simplicity, the time evolution operator is given by

\begin{align}
\label{eqn:time evolution operator}
\begin{split}
\hat{U}_x(t)=e^{-i\hat{H}_xt/\hbar},
\end{split}
\end{align}
and like Eq. (\ref{eqn:displacement operator}), it is unitary. $\hat{H}_x$ is given by Eq. (\ref{eqn:Hx}). Applying the time evolution operator to the ordinary coherent state gives

\begin{align}
\label{eqn:time evolution operator on the coherent state}
\begin{split}
\hat{U}_x(t)\ket{\alpha}&=e^{-i\hat{H}_xt/\hbar}\ket{\alpha}\\
&=e^{-\abs{\alpha}^2/2}\sum^\infty_{n=0}\frac{\alpha^n}{\sqrt{n!}}e^{-i\hat{H}_xt/\hbar}\ket{n}\\
&=e^{-\abs{\alpha}^2/2}\sum^\infty_{n=0}\frac{\alpha^n}{\sqrt{n!}}e^{-i\omega_xt(n+\frac{1}{2})}\ket{n}\\
&=e^{-i\omega_xt/2}e^{-\abs{\alpha}^2/2}\sum^\infty_{n=0}\frac{(\alpha e^{-i\omega_xt})^n}{\sqrt{n!}}\ket{n}\\
&=e^{-i\omega_xt/2}\ket{\alpha e^{-i\omega_xt}},
\end{split}
\end{align}
where the complex phase out front can be neglected as it does not contribute to normalization. It is clear that this state is not stationary, i.e. the probability density is time-dependent. Moving into two dimensions, the 2-dimensional time evolution operator is given as

\begin{align}
\label{eqn:2d time evolution operator}
\begin{split}
\hat{U}(t)=e^{-i\hat{H}t/\hbar}=e^{-i(\hat{H}_x+\hat{H}_y)t/\hbar},
\end{split}
\end{align}
and when applied to the two-mode coherent state yields

\begin{align}
\label{eqn:2d time dep coherent state}
\begin{split}
\ket{\alpha e^{-i\omega_xt},\beta e^{-i\omega_yt}}&=e^{-\abs{\alpha}^2/2}e^{-\abs{\beta}^2/2}\sum^\infty_{n=0}\sum^\infty_{m=0}\frac{(\alpha e^{-i\omega_xt})^n(\beta e^{-i\omega_yt})^m}{\sqrt{n!m!}}\ket{n,m},\\
\end{split}
\end{align}
which is simply the multiplication of two single-mode coherent states. The configuration space wavefunction, $\braket{x,y}{\alpha e^{-i\omega_xt},\beta e^{-i\omega_yt}}$, is given by

\begin{align}
\label{eqn:2d time dep coherent state wavefcn}
\begin{split}
\Psi_{SC}(\nu,\eta,\alpha,\beta;t)&=e^{-(\abs{\alpha}^2+\abs{\beta}^2)/2}\sum^\infty_{n=0}\sum^\infty_{m=0}\frac{(\alpha e^{-i\omega_xt})^n(\beta e^{-i\omega_yt})^m}{\sqrt{n!m!}}\psi_n(\nu)\psi_m(\eta),\\
\end{split}
\end{align}
where $\psi_n(\nu)$ is given by Eq. (\ref{eqn:1dho x wavefcn}), and by analogy $\psi_m(\eta)$ is 

\beq
\psi_m(\eta)=\braket{y}{m}=\frac{1}{\sqrt{m!2^m}}\left(\frac{m_y\omega_y}{\pi\hbar}\right)^{1/4}H_m(\eta)e^{-\eta^2/2},
\label{eqn:1dho y wavefcn}
\eeq
with

\beq
\eta=\sqrt{\frac{m_y \omega_y}{\hbar}}y.
\label{eqn:1dho y wavefcn variable}
\eeq
For the rest of this study we set $m=1$ and $\hbar=1$ such that $\nu=\sqrt{\omega_x}x$ and $\eta=\sqrt{\omega_y}y$. An alternative form of Eq. (\ref{eqn:2d time dep coherent state wavefcn}) can be found by employing the use of the generating function of Hermite polynomials[arfken ref]. That is,

\beq
e^{z^2-(z-w)^2}=\sum^\infty_{n=0}H_n(z)\frac{w^n}{n!}.
\label{eqn:generating function hermite poly}
\eeq
The time-dependent two-mode coherent state is then

\begin{align}
\label{eqn:2d time dep coherent state wavefcn hermite poly}
\begin{split}
\Psi_{SC}^{(\omega_y,\omega_x)}(x,y,\alpha,\beta;t)&=\frac{\sqrt{\omega_x\omega_y}}{\pi}e^{-(\abs{\alpha}^2+\abs{\beta}^2)/2}e^{-(\omega_xx^2+\omega_yy^2)/2}\\
&\times e^{(\sqrt{2\omega_x}x\alpha e^{-i\omega_xt}+\sqrt{2\omega_y}y\beta e^{-i\omega_yt})}e^{-(\alpha^2e^{-i2\omega_xt}+\beta^2e^{-i2\omega_yt})}.\\
\end{split}
\end{align}
To begin, we consider the \textit{isotropic} case where $\omega_x=\omega_y=\omega_0$. Eq. (\ref{eqn:2d time dep coherent state wavefcn}) becomes

\begin{align}
\label{eqn:isotropic Psi SC}
\begin{split}
\Psi_{SC}^{(\omega_0)}(x,y,\alpha,\beta;t)&=e^{-(\abs{\alpha}^2+\abs{\beta}^2)/2}\sum^\infty_{n=0}\sum^\infty_{m=0}\frac{(\alpha e^{-i\omega_0t})^n(\beta e^{-i\omega_0t})^m}{\sqrt{n!m!}}\psi_n^{(\omega_0)}(x)\psi_m^{(\omega_0)}(y)\\
&=\frac{\omega_0}{\pi}e^{-(\abs{\alpha}^2+\abs{\beta}^2)/2}e^{-\omega_0(x^2+y^2)/2}e^{\sqrt{2\omega_0}e^{-i\omega_0t}(x\alpha +y\beta)}\\
&\times e^{-e^{-i2\omega_0t}(\alpha^2+\beta^2)}.\\
\end{split}
\end{align}
where the single mode wavefunctions are

\beq
\psi_n^{(\omega_0)}(x)=\braket{x}{n}=\frac{1}{\sqrt{n!2^n}}\left(\frac{\omega_0}{\pi}\right)^{1/4}H_n(\sqrt{\omega_0}x)e^{-\omega_0x^2/2},
\label{eqn:1DHO x wavefcn}
\eeq
and 
\beq
\psi_m^{(\omega_0)}(y)=\braket{y}{m}=\frac{1}{\sqrt{m!2^m}}\left(\frac{\omega_0}{\pi}\right)^{1/4}H_m(\sqrt{\omega_0}y)e^{-\omega_0y^2/2}.
\label{eqn:1DHO y wavefcn}
\eeq
Plotting the probability density
\begin{align}
\label{eqn:isotropic Psi SC prob dens}
\begin{split}
\rho_{SC}^{(\omega_0)}(x,y,\alpha,\beta;t)&=\frac{\omega_0^2}{\pi^2}e^{-(\abs{\alpha}^2+\abs{\beta}^2)}e^{-\omega_0(x^2+y^2)}e^{2\sqrt{2\omega_0}(x\alpha \cos{(\omega_0 t)} +y\abs{\beta}\cos{(\omega_0 t-\phi)})}\\
&\times e^{-2(\alpha^2\cos{(2\omega_0 t)}+\abs{\beta}^2\cos{(2\omega_0 t-2\phi)})},\\
\end{split}
\end{align}
we expect the average value of the Gaussian wave packet to move like a classical particle, at the same velocity of that particle\cite{sakurai_modern_2017}. This is Ehrenfest's Theorem, which says that quantum mechanical expectation values follow the classical laws of motion\cite{mcintyre_quantum_2012}. We have already shown these states to be non-stationary, so there mustn't be a steady current density, $\Vec{\nabla}\cdot\Vec{J}\neq0$. According to the continuity equation
\beq
\Vec{\nabla}\cdot \Vec{J}+\frac{\partial\rho}{\partial t}=0,
\label{eq:continuity eqn}
\eeq
the divergence of the probability current density can only be zero if the probability density is \textbf{time-independent}, but these states are time-dependent. In order to satisfy the continuity equation, the divergence of the current density must offset the non-zero value of the partial derivative of $\rho$ with respect to time. Having a non-steady current makes perfect sense in a classical setting, a classical oscillator would have acceleration, breaking the "steady flow" of velocity. 

\begin{figure}[H]
    \centering
    \includegraphics[width=\textwidth]{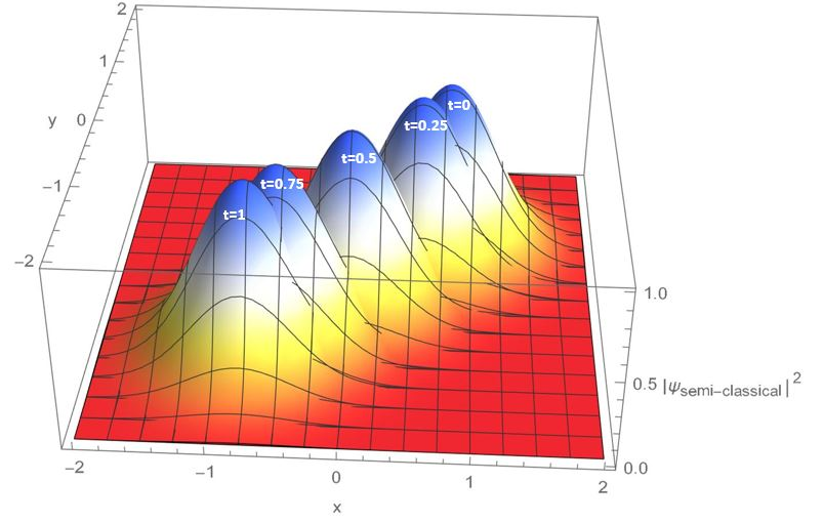}
    
    \caption{Presented is the time evolution of the isotropic two-mode coherent state for complex amplitudes $\alpha=\beta=1$. Each peak shown is a separate function, overlaid together to show the movement through time, with each being labeled with its unique time value. This case corresponds to that of Fig. \ref{fig:CLJ-p1q1-phiVARIED}(a). It is shown that the centroid of the Gaussian travels the exact path of the simplest Lissajous curve and directly verifies the application of Ehrenfest's Theorem with a linear potential.} 
    \label{fig:11SCLJ-beta=1}
\end{figure}
The verification of Ehrenfest's Theorem does not stop at Fig. \ref{fig:11SCLJ-beta=1}. All possible phase differences between $\alpha$ and $\beta$ will exhibit this behavior, as well as other coprime ratios of frequency. These are what are known as semi-classical Lissajous states, coming from the fact that we are dealing with the most classical of quantum states and that probability density mimics the shape of Lissajous figures through time. The quantum Lissajous figures via projection, proposed below, encode the Lissajous figures in an even more fundamental way while being stationary and having a steady current flow. Before we reach that section of this section, we will continue with more examples of these semi-classical states. Next is to examine some semi-classical examples where the phase difference between the complex amplitudes is non-zero.

\begin{figure}[H]
    \centering
    \includegraphics[width=\textwidth]{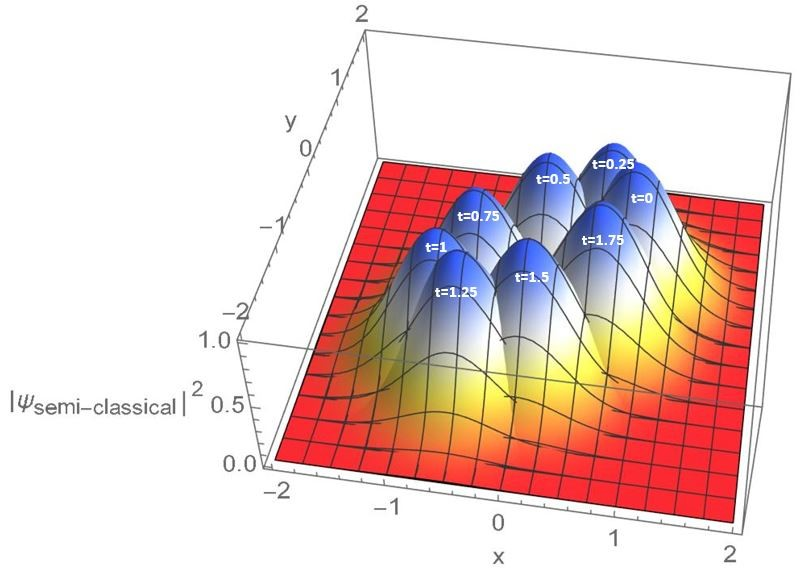}
    
    \caption{Presented is the time evolution of the isotropic two-mode coherent state for complex amplitudes $\alpha=1$ and $\beta=e^{i\pi/4}$. Each peak shown is a separate function, overlaid together to show the movement through time, with each being labeled with its unique time value. This case corresponds to that of Fig. \ref{fig:CLJ-p1q1-phiVARIED}(c). It is shown that the centroid of the Gaussian travels the exact path of the elliptical Lissajous curve and directly verifies the application of Ehrenfest's Theorem with a linear potential.} 
    \label{fig:11SCLJ-beta=eipiover4}
\end{figure}

\begin{figure}[H]
    \centering
    \includegraphics[width=\textwidth]{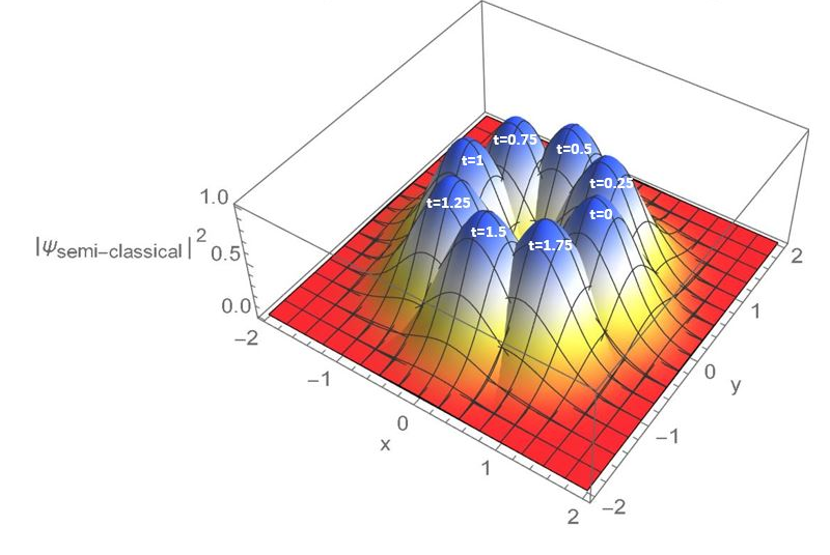}
    
    \caption{Presented is the time evolution of the isotropic two-mode coherent state for complex amplitudes $\alpha=1$ and $\beta=e^{i\pi/2}$. Each peak shown is a separate function, overlaid together to show the movement through time, with each being labeled with its unique time value. This case corresponds to that of Fig. \ref{fig:CLJ-p1q1-phiVARIED}(e). It is shown that the centroid of the Gaussian travels the exact path of the circular Lissajous curve and directly verifies the application of Ehrenfest's Theorem with a linear potential.} 
    \label{fig:11SCLJ-beta=eipiover2}
\end{figure}
Now we consider $\omega_x=q\omega_0$ and $\omega_y=p\omega_0$, where $q$ and $p$ are integers that form a commensurate frequency ratio between $\omega_x$ and $\omega_y$. Eq. (\ref{eqn:2d time dep coherent state wavefcn}) becomes

\begin{align}
\label{eqn:aniso Psi SC}
\begin{split}
\Psi_{SC}^{(p,q)(\omega_0)}(x,y,\alpha,\beta;t)&=e^{-(\abs{\alpha}^2+\abs{\beta}^2)/2}\sum^\infty_{n=0}\sum^\infty_{m=0}\frac{(\alpha e^{-iq\omega_0t})^n(\beta e^{-ip\omega_0t})^m}{\sqrt{n!m!}}\psi_n^{(q)(\omega_0)}(x)\psi_m^{(p)(\omega_0)}(y)\\
&=\frac{\omega_0\sqrt{qp}}{\pi}e^{-(\abs{\alpha}^2+\abs{\beta}^2)/2}e^{-(q\omega_0x^2+p\omega_0y^2)/2}\\
&\times e^{(\sqrt{2q\omega_0}x\alpha e^{-iq\omega_0t}+\sqrt{2p\omega_0}y\beta e^{-ip\omega_0t})}e^{-(\alpha^2e^{-i2q\omega_0t}+\beta^2e^{-i2p\omega_0t})},\\
\end{split}
\end{align}
where the single-mode harmonic oscillator wavefunctions are

\beq
\psi_n^{(q)(\omega_0)}(x)=\braket{x}{n}=\frac{1}{\sqrt{n!2^n}}\left(\frac{q\omega_0}{\pi}\right)^{1/4}H_n(\sqrt{q\omega_0}x)e^{-q\omega_0x^2/2},
\label{eqn: aniso 1DHO x wavefcn}
\eeq
and 
\beq
\psi_m^{(p)(\omega_0)}(y)=\braket{y}{m}=\frac{1}{\sqrt{m!2^m}}\left(\frac{p\omega_0}{\pi}\right)^{1/4}H_m(\sqrt{p\omega_0}y)e^{-p\omega_0y^2/2}.
\label{eqn:aniso 1DHO y wavefcn}
\eeq
Plotting the probability density,
\begin{align}
\label{eqn:2d time dep coherent state prob density}
\begin{split}
\rho_{SC}^{(p,q)(\omega_0)}(x,y,\alpha,\beta;t)&=\frac{\omega_0\sqrt{qp}}{\pi}e^{-(\abs{\alpha}^2+\abs{\beta}^2)}e^{-(q\omega_0x^2+p\omega_0y^2)}\\
&\times e^{2\sqrt{2\omega_0}(\sqrt{q}x\alpha\cos{(q\omega_0t)}+\sqrt{p}y\abs{\beta}\cos{(p\omega_0t-\phi)})}\\
&\times e^{-2(\alpha^2\cos{(2q\omega_0t)}+\abs{\beta}^2\cos{(2p\omega_0t-2\phi)})},\\
\end{split}
\end{align}
we expect this to act exactly like the isotropic case, in that the centroid of the Gaussian travels along the path of corresponding classical Lissajous figures, but now for ratios that are not 1:1. Therefore, Ehrenfest's Theorem is doubly justified in the cases of semi-classical Lissajous states.

\begin{figure}[H]
    \centering
    \includegraphics[width=\textwidth]{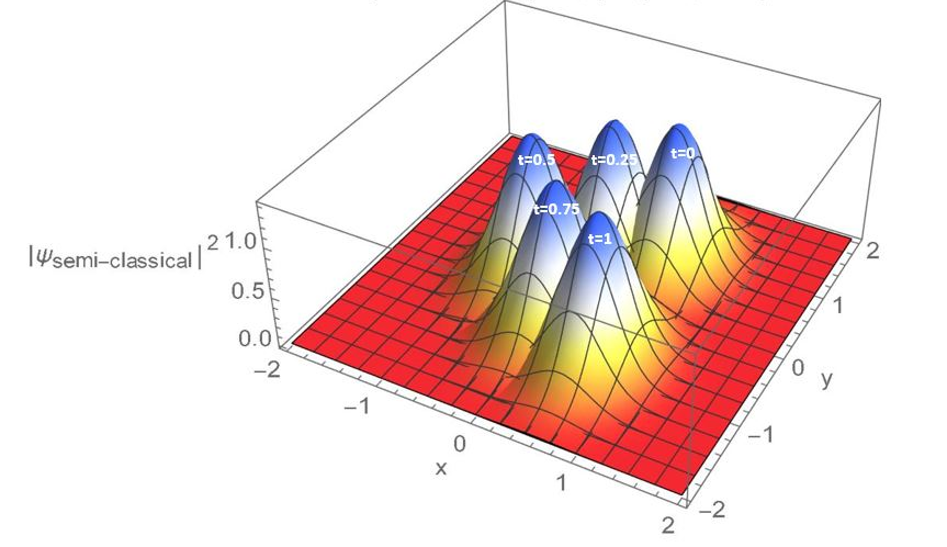}
    
    \caption{Presented is the time evolution of the anisotropic two-mode coherent state for complex amplitudes $\alpha=1$ and $\beta=1$. Each peak shown is a separate function, overlaid together to show the movement through time, with each being labeled with its unique time value. This case corresponds to that of Fig. \ref{fig:CLJ-p1q2-phiVARIED}(a). It is shown that the centroid of the Gaussian travels the exact path of the circular Lissajous curve and directly verifies the application of Ehrenfest's Theorem with a linear potential.} 
    \label{fig:21SCLJ-beta=1}
\end{figure}

\begin{figure}[H]
    \centering
    \includegraphics[width=\textwidth]{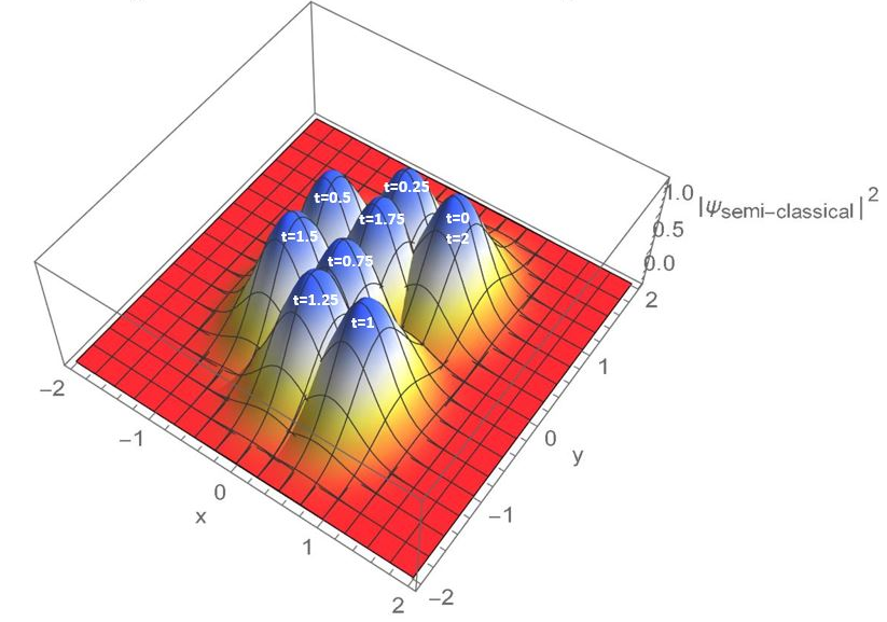}
    
    \caption{Presented is the time evolution of the anisotropic two-mode coherent state for complex amplitudes $\alpha=1$ and $\beta=e^{i\pi/8}$. Each peak shown is a separate function, overlaid together to show the movement through time, with each being labeled with its unique time value. This case corresponds to that of Fig. \ref{fig:CLJ-p1q2-phiVARIED}(b). It is shown that the centroid of the Gaussian travels the exact path of the circular Lissajous curve and directly verifies the application of Ehrenfest's Theorem with a linear potential.} 
    \label{fig:21SCLJ-beta=eipiover8}
\end{figure}

\begin{figure}[H]
    \centering
    \includegraphics[width=\textwidth]{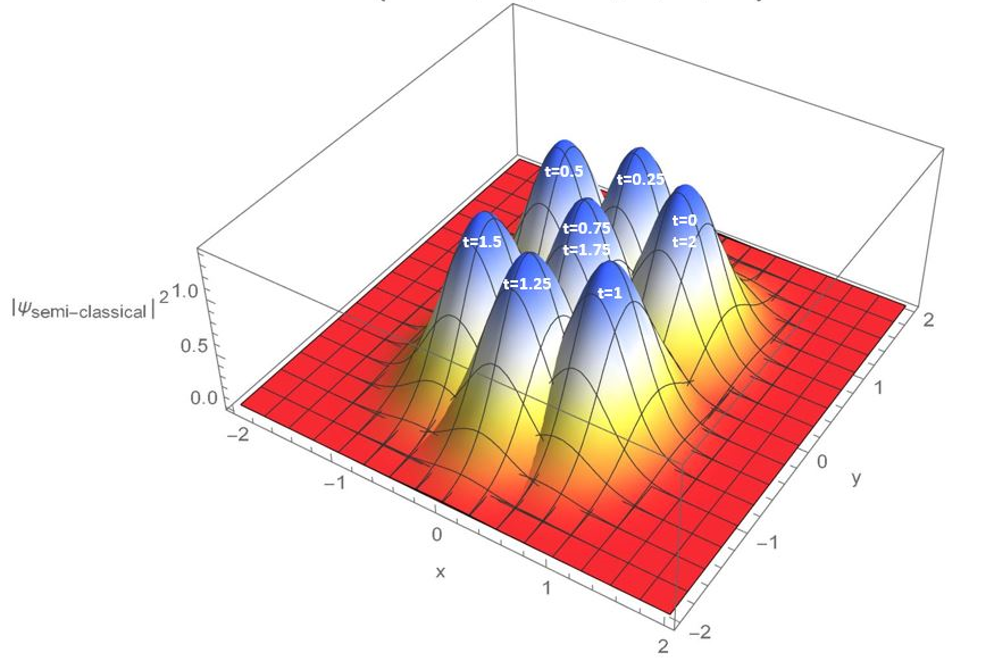}
    
    \caption{Presented is the time evolution of the anisotropic two-mode coherent state for complex amplitudes $\alpha=1$ and $\beta=e^{i\pi/4}$. Each peak shown is a separate function, overlaid together to show the movement through time, with each being labeled with its unique time value. This case corresponds to that of Fig. \ref{fig:CLJ-p1q2-phiVARIED}(c). It is shown that the centroid of the Gaussian travels the exact path of the circular Lissajous curve and directly verifies the application of Ehrenfest's Theorem with a linear potential.} 
    \label{fig:21SCLJ-beta=eipiover4}
\end{figure}

Once again, it is of much importance to note that according to Ehrenfest's Theorem, the centroid of the probability density holds its Gaussian shape while evolving along the trajectory of classical Lissajous curves, as seen in Figs. \ref{fig:11SCLJ-beta=1}-\ref{fig:21SCLJ-beta=eipiover4}. There is an even more striking form of Lissajous states in quantum mechanics: stationary states with steady current densities. These states will encode the Lissajous trajectories in a more fundamental and quantum mechanical fashion.

\subsection{Quantum Lissajous Figures-Fundamental Cases}
This section contains the analysis of states that we have shown to be inherently quantum mechanical analogs to classical Lissajous figures. Unlike the previously mentioned semi-classical Lissajous states, these states encode whole classical Lissajous figures in a corresponding probability density. Time evolution is trivial when applying to a degenerate subspace, so the quantum Lissajous states are stationary with steady current density. Two separate classes of quantum Lissajous states will be analyzed: "fundamental" and "higher harmonic" quantum Lissajous states. The fundamental case's constraint is that the frequencies must be coprime, whereas the higher harmonic case is categorized by having non-coprime, but still commensurate frequencies. Some examples of fundamental cases are $q=1:p=1$, $q=2:p=1$, and $q=3:p=2$; and higher harmonic cases include, but are not limited to, $q=2:p=2$, $q=3:p=3$, and $q=4:p=2$. Recall, when reviewing the classical Lissajous figures, cases with higher harmonic-like ratios simplified to the coprime cases and only rescaled the size of the figures based on increased mechanical energy. For quantum Lissajous states, this is strikingly different. The \textit{higher harmonic} quantum Lissajous figures do not simplify to the corresponding fundamental quantum Lissajous figure, they become linear superpositions of fundamental quantum Lissajous figures with certain complex phases. 

Starting back up with with Eq. (\ref{eqn:quantum hamiltonian 2dho in terms of a and a^dagger and p and q and number operators}), this Hamiltonian acts on a composite Fock state, $\ket{n,m}$.

\begin{align}
\label{eqn:quantum hamiltonian 2dho eigenvalue eqn}
\begin{split}
\hat{H}\ket{n,m}&=\omega_0\left(q\hat{n}+p\hat{m}+\frac{q+p}{2}\right)\ket{n,m}\\
&=\omega_0\left(qn+pm+\frac{q+p}{2}\right)\ket{n,m}.
\end{split}
\end{align}
The Fock state remains unchanged after the operation of the Hamiltonian, and we retrieve the energy eigenvalue,

\begin{align}
\label{eqn:quantum hamiltonian 2dho eigenvalue}
\begin{split}
E_{n,m}&=\omega_0\left(qn+pm+\frac{q+p}{2}\right).
\end{split}
\end{align}

\subsubsection{Fundamental Isotropic Case}
The fundamental isotropic quantum Lissajous figure is categorized by its one-to-one frequency ratio, $p=q=1$. Applying this to Eq. (\ref{eqn:quantum hamiltonian 2dho in terms of a and a^dagger and p and q}) yields the simplest form of the 2DHO Hamiltonian,

\beq
\hat{H}=\omega_0\left(\hat{a}_x^\dagger\hat{a}_x+\hat{a}_y^\dagger\hat{a}_y+1\right).
\label{eqn:2D iso ham}
\eeq
The fundamental isotropic Hamiltonian has SU(2) as its dynamical symmetry (degeneracy) group. Through the introduction of the Schwinger Realization\cite{schwinger_angular_2015},

\beq
\hat{J}_1=\frac{1}{2}(\hat{a}_x^\dagger\hat{a}_y+\hat{a}_x\hat{a}_y^\dagger),\quad \hat{J}_2=\frac{1}{2i}(\hat{a}_x^\dagger\hat{a}_y-\hat{a}_x\hat{a}_y^\dagger), \quad \hat{J}_3=\frac{1}{2}(\hat{a}_x^\dagger\hat{a}_x-\hat{a}_y^\dagger\hat{a}_y),
\label{eq:Schwinger realization}
\eeq
and

\beq
\hat{J}_+=\hat{J}_1+i\hat{J}_2=\hat{a}_x^\dagger\hat{a}_y,\quad \hat{J}_-=\hat{J}_1-i\hat{J}_2=\hat{a}_x\hat{a}_y^\dagger,
\label{eq:Schwinger realization2}
\eeq
we can express Eq. (\ref{eqn:2D iso ham}) in terms of angular momentum operators. Eqs. (\ref{eq:Schwinger realization}) and (\ref{eq:Schwinger realization2}) obey the commutation relation

\beq
\left[\hat{J}_i,\hat{J}_j\right]=i\varepsilon_{ijk}\hat{J}_k, \quad i,j=1,2,3.
\label{eq:J_123 commutators}
\eeq
We also introduce the operator

\beq
\hat{J}_0=\frac{1}{2}(\hat{a}_x^\dagger\hat{a}_x+\hat{a}_y^\dagger\hat{a}_y),
\label{eq:J0}
\eeq
which commutes with all of the operators in Eqs. (\ref{eq:Schwinger realization}) and (\ref{eq:Schwinger realization2})

\beq
\left[\hat{J}_0,\hat{J}_i\right]=0, \quad i=1,2,3.
\label{eq:J0 to J_123 commutator}
\eeq
An important object in Lie algebra, and specifically in the realm of angular momentum operators is the Casimir operator\cite{gilmore_lie_2005}. The Casimir operator for this algebra is

\beq
\hat{J}^2=\hat{J}_1^2+\hat{J}_2^2+\hat{J}_3^2=\hat{J}_0(\hat{J}_0+1),
\label{eq:casimir operator}
\eeq
which means that $\hat{J}_0$ can be considered the Casimir operator. We can rewrite the fundamental isotropic Hamiltonian, Eq. (\ref{eqn:2D iso ham}) as

\beq
\hat{H}=\omega_0(2\hat{J}_0+1).
\label{eq:quantum hamiltonian 2dho in terms of J_0}
\eeq
Equation (\ref{eq:quantum hamiltonian 2dho in terms of J_0}) shows that the Hamiltonian is a function of the Casimir operator, which is a characteristic feature of degeneracy groups. The ability for the Hamiltonian to be expressed in terms of $\hat{J}_0$ shows the rotational symmetry of the isotropic oscillator. For a system of $N$ quanta, the corresponding degenerate eigenstates are $\ket{K}_x\ket{N-K}_y$ for $K=0,1,...,N$, $N=1,2,...,\infty$, and where

\beq
\ket{K,N-K}=\frac{\left(\hat{a}_x^\dagger\right)^K\left(\hat{a}_y^\dagger\right)^{N-K}}{\sqrt{K!(N-K)!}}\ket{0,0}.
\label{eq:degenerate fock states}
\eeq
The degenerate states defined in equation (\ref{eq:degenerate fock states}) satisfy the eigenvalue equation

\beq
2\hat{J}_0\ket{K,N-K}=(\hat{a}_x^\dagger\hat{a}_x+\hat{a}_y^\dagger\hat{a}_y)\ket{K,N-K}=N\ket{K,N-K},
\label{eq:2D eigenvalue equation}
\eeq
with a degeneracy of $N+1$, owing to the fact that there are $N+1$ states of the form $\ket{K,N-K}$ for $0\leq K\leq N$. 

It is well known that the SU(2) coherent states\cite{arecchi_atomic_1972}, by analogy with ordinary coherent states, are formed by applying a rotation operator on the ground state in the angular momentum basis, where the rotation operator takes the place of the displacement operator in the ordinary coherent state case. The rotation operator

\beq
\hat{R}_{\theta,\phi}=e^{z\hat{J}_+-z^*\hat{J}_-}=e^{\zeta\hat{J}_+}e^{\ln{(1+|\zeta|^2)}\hat{J}_z}e^{-\zeta^*\hat{J}_-},
\label{eq:rotation operator}
\eeq
where $z=\frac{\theta}{2}e^{-i\phi}$ and $\zeta=e^{-i\phi}\tan(\theta/2), (0\leq\theta\leq\pi,0\leq\phi\leq 2\pi)$, is the one appropriate for rotating the spin-down state into an SU(2) coherent state localized on the Bloch sphere at azimuthal angle, $\phi$, and latitude angle, $\theta$, as measured relative to the south pole.

\begin{figure}[H]
    \centering
    \includegraphics[width=\textwidth]{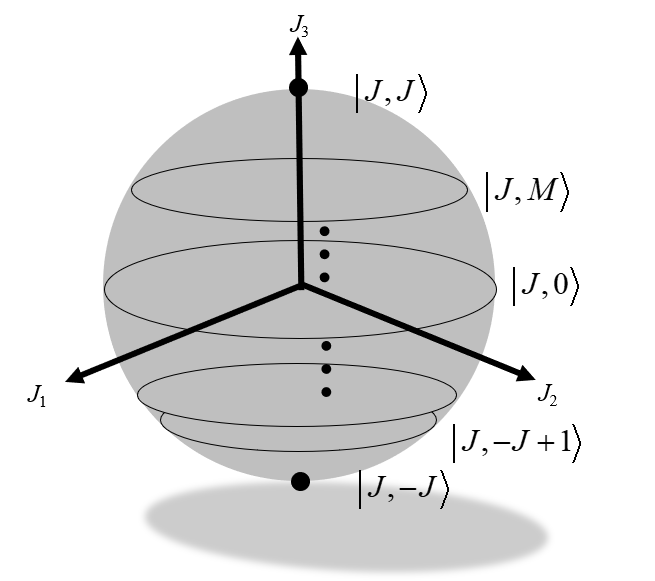}
    
    \caption{Bloch Sphere representation of $\hat{J}_3$ basis states for spin-$J$ angular momentum. In this figure, $J$ is taken to be an integer. The ground state is represented by a point on the Bloch sphere located at the south pole.} 
    \label{fig:BlochSphere}
\end{figure}

\begin{figure}[H]
    \centering
    \includegraphics[width=\textwidth]{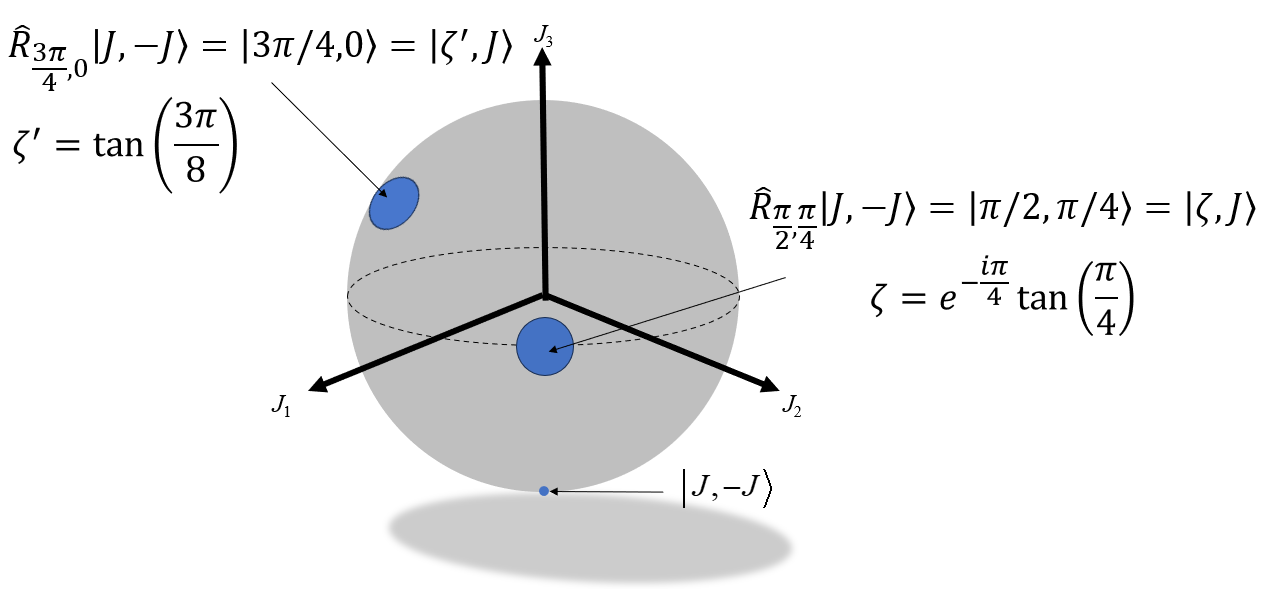}
    
    \caption{Bloch Sphere representations for a couple of spin-$J$ SU(2) coherent states resulting from the indicated rotations on the ground state. The circular patches indicate the support of a distribution characterizing the linear superposition of basis states. The size and shape of such a patch should characterize the uncertainties in the Cartesian components of angular momentum. This figure is merely conceptual, and no care has been taken to scale these patches accurately. The rotations have been done in the angular momentum basis. See Appendix \ref{appendix a} for the relationship between the angular momentum and Fock state bases.} 
    \label{fig:BlochSphere2}
\end{figure}

The Bloch sphere is a 2-dimensional surface embedded in a 3-dimensional angular momentum space that gives a representation of angular momentum states. In particular, regions on the Bloch sphere represent states in a spin-$J$ system. For example, in Fig. \ref{fig:BlochSphere}, we show several of the eigenstates of $\hat{J}_3$ in the basis in which the operator is diagonal. In general, states on the Bloch sphere correspond to extended regions owing to the uncertainty principle for angular momentum operators, which is why states of well-defined $\hat{J}_3$ are represented by latitude lines on the Bloch sphere. On such a line, the eigenvalue of $\hat{J}_3$ is well-defined, but the values of $\hat{J}_1$ and $\hat{J}_2$ are distributed randomly around the latitude line. The spin-up and spin-down states are exceptional in that they correspond to points on the Bloch sphere, which, strictly speaking, seem to violate the uncertainty principle. However, these points should be thought of as limiting cases of latitude lines with uncertain $\hat{J}_1$ and $\hat{J}_2$.
In Fig. \ref{fig:BlochSphere2}, we represent the action of the rotation operator, Eq. (\ref{eq:rotation operator}), on the ground state, $\ket{J,-J}$, to form an SU(2) coherent state localized around the desired location on the Bloch sphere.

We now use the idea of rotation on the Bloch sphere and the connection between the $\hat{J}_3$ angular momentum basis states and the two-mode Fock states within a degenerate subspace of the 2DHO to illustrate the identification of the isotropic quantum Lissajous states with the SU(2) coherent states. The second equality in Eq. (\ref{eq:rotation operator}) is the normally ordered form of the rotation operator and is found using the disentangling theorem for angular momentum operators\cite{demkov_yn_definition_nodate}. The Schwinger Realization\cite{schwinger_angular_2015} allows us to treat this angular momentum system as two uncoupled quantum harmonic oscillators\cite{sakurai_modern_2017} where the number of quanta is equal to double the maximum value of angular momentum, $N=2J$. Since $\hat{J}_+=\hat{a}_x^\dagger\hat{a}_y$, the $y$ oscillator initially holds all $N$ quanta. Applying the rotation operator on the state $\ket{J,-J}=\ket{0,N}$,

\beq
\begin{split}
\ket{\zeta,N}=\hat{R}_{\theta,\phi}\ket{0,N}&=\left(\frac{1}{1+|\zeta|^2}\right)^{N/2}e^{\zeta\hat{a}_x^\dagger\hat{a}_y}\ket{0,N}
\\
&=\left( 1+\abs{\zeta}^2 \right)^{-N/2}\sum^N_{K=0}\binom{N}{K}^{1/2}\zeta^K\ket{K,N-K},
\end{split}
\label{eq:su2 coherent state oscillator}
\eeq
which shows that the SU(2) coherent states are formed over the degenerate states of the two-mode harmonic oscillator.

We now introduce our formulation of the quantum Lissajous states for the isotropic 2DHO by considering the projection of the ordinary two-mode coherent state onto a degenerate subspace of the system, thereby, guaranteeing the result is a stationary state. The projection operator onto the degenerate subspace has the form, for a given $N$ of

\beq
\Pi_N=\sum^N_{K=0}\ket{K}_x\ket{N-K}_y\bra{K}_x\bra{N-K}_y.
\label{eq:degenerate state projection operator}
\eeq
Applying Eq. (\ref{eq:degenerate state projection operator}) to Eq. (\ref{eqn:2d coherent state}), and disregarding the global factors,

\beq
\Pi_N\ket{\alpha,\beta}=\sum^N_{K=0}\frac{(\alpha/\beta)^K}{\sqrt{K!(N-K)!}}\ket{K,N-K},
\label{eq:degenerates state proj onto coherent states}
\eeq
which when normalized becomes,
\beq
\begin{split}
\ket{\zeta,N}&=\left( 1+\abs{\zeta}^2 \right)^{-N/2}\sum^N_{K=0}\binom{N}{K}^{1/2}\zeta^K\ket{K,N-K},\\
\end{split}
\label{eq:NORMALIZED degenerates state proj onto coherent states}
\eeq
where $\zeta=e^{-i\phi}\abs{\alpha}/\abs{\beta}$. Comparing Eq. (\ref{eq:NORMALIZED degenerates state proj onto coherent states}) with Eq. (\ref{eq:su2 coherent state oscillator}), we see that the projected state is indeed an SU(2) coherent state. If we consider this state on the Bloch sphere, we can draw the correspondences that $\phi$, the relative phase between $\alpha$ and $\beta$ in the projected state is identified with the azimuthal angle appropriate to the Bloch sphere representation of this state. Further, the ratio, $\abs{\alpha}/\abs{\beta}$ is identified with $\tan{(\theta/2)}$. It is important to keep in mind that these analogies, while compelling, are just analogies. The Hilbert space represented by the Bloch sphere is different than the Hilbert space of the 2DHO. We employ this analogy as a means of understanding the SU(2) structure of the projected states of the isotropic 2DHO, but, the eventual connection with Lissajous figures occurs the configuration space of the 2DHO, not on the Bloch sphere.

Applying the time evolution operator to Eq. (\ref{eq:NORMALIZED degenerates state proj onto coherent states}), it is clear that the time evolution behaves like a global phase

\beq
e^{-i\hat{H}t}\ket{\zeta,N}=e^{-i\omega_0(N+1)t}\ket{\zeta,N},
\label{eq:time evolution onto isotropic qlj state}
\eeq
thus, the probability density is time-independent. Eq. (\ref{eq:NORMALIZED degenerates state proj onto coherent states}) is in agreement with Eq. (\ref{eq:su2 coherent state oscillator}), successfully reaching the SU(2) coherent state via projection.

To see the connection with Lissajous figures, we must study the configuration space probability density for projected states. To do this we must form the configuration space wavefunction. The wavefunction associated with equation (\ref{eq:NORMALIZED degenerates state proj onto coherent states}) is found by projection onto configuration space, $\ket{x,y}$:

\beq
\begin{split}
\Psi_N^{(\omega_0)}(x,y,\zeta)&=\left( 1+\abs{\zeta}^2 \right)^{-N/2}\sum^N_{K=0}\binom{N}{K}^{1/2}\zeta^K\psi_K^{(\omega_0)}\left(x\right)\psi_{N-K}^{(\omega_0)}\left(y\right),\\
\end{split}
\label{eq:2D wavefunction from coherent states}
\eeq
where the 1-dimensional wavefunctions are separable and given by,

\beq
\psi_K^{(\omega_0)}\left(x\right)=\frac{1}{\sqrt{2^K K!}}\left(\frac{\omega_0}{\pi}\right)^{1/2}e^{-\omega_0 x^2/2}H_K\left(x\sqrt{\omega_0}\right),
\label{eq:x wavefunction}
\eeq
and

\beq
\psi_{N-K}^{(\omega_0)}\left(y\right)=\frac{1}{\sqrt{2^{N-K} (N-K)!}}\left(\frac{\omega_0}{\pi}\right)^{1/2}e^{-\omega_0 y^2/2}H_{N-K}\left(y\sqrt{\omega_0}\right).
\label{eq:y wavefunction}
\eeq
These are the usual wavefunctions of a 1DHO, which involve Gaussian factors and Hermite polynomials, $H_K\left(x\sqrt{\omega_0}\right)$ and $H_{N-K}\left(y\sqrt{\omega_0}\right)$. The probability density associated with equation (\ref{eq:2D wavefunction from coherent states}) is

\begin{align}
\begin{split}
\rho_N^{(\omega_0)}(x,y,\zeta)&=\abs{\Psi_N^{(\omega_0)}(x,y,\zeta)}^2\\
&=\abs{\left( 1+\abs{\zeta}^2 \right)^{-N/2}\sum^N_{K=0}\binom{N}{K}^{1/2}\zeta^K\psi_K^{(\omega_0)}\left(x\right)\psi_{N-K}^{(\omega_0)}\left(y\right)}^2.
\label{eq:probability density}
\end{split}
\end{align}

As we will demonstrate in section \ref{chap4_PrincipleResults}, where we present our major results, the support of the distribution given in Eq. (\ref{eq:probability density}) lies along the classical Lissajous figure for the isotropic 2DHO. It is for this reason that we describe our projected states as "quantum Lissajous states."

Throughout this literature, we assume $\alpha$ to be real and take $\beta=\abs{\beta}e^{i\phi}$ so that $\zeta=\abs{\zeta}e^{-i\phi}$ with $\abs{\zeta}=\abs{\alpha}/\abs{\beta}$. This assumption does not restrict the cases which can be studied since the phase is transferable to either complex amplitude. The phase, $\phi$, and amplitude, $\abs{\zeta}$, affect the eccentricity and orientation of the Lissajous figure along which a quantum Lissajous state is localized. The quantum Lissajous state is a stationary state, which is completely due to the projection onto the degenerate states, so it will no longer move in time. Instead, the whole probability density will localize along the classical Lissajous ellipse. Graphical evidence of these characteristics will be shown in the next section. 

\subsubsection{Fundamental Anisotropic Case}

The role of projection onto degenerate states becomes much more significant for the fundamental anisotropic oscillator. The anisotropic Hamiltonian is given by Eq. (\ref{eqn:quantum hamiltonian 2dho in terms of a and a^dagger and p and q})

\begin{align}
\label{eqn:quantum hamiltonian 2dho in terms of a and a^dagger and p and q again}
\begin{split}
\hat{H}&=\omega_0\left(q\hat{a}_x^\dagger\hat{a}_x+p\hat{a}_y^\dagger\hat{a}_y+\frac{q+p}{2}\right).
\end{split}
\end{align}
With the introduction of the factors of $p$ and $q$, we can still use the Schwinger realization\cite{schwinger_angular_2015} to represent the Hamiltonian in terms of angular momentum operators, but we cannot represent the Hamiltonian in terms of only the Casimir operator,

\beq
\label{eqn:aniso hamiltonian in terms of ang mom operators}
\hat{H}=\omega_0\left((q+p)\hat{J}_0+(q-p)\hat{J}_3+\frac{q+p}{2}\right).
\eeq
As a result, the anisotropic Hamiltonian no longer commutes with the rotation operator that generates an SU(2) coherent state on the Bloch sphere, breaking the rotational symmetry of the system. Therefore, appealing to the correspondence we developed above between SU(2) coherent states on the Bloch sphere and the projected states of the 2DHO, we expect that the quantum Lissajous states for the anisotropic oscillator \textbf{are not SU(2) coherent states.} In other words, because the anisotropic Hamiltonian doesnt commute with the rotation operator, Eq. (\ref{eq:rotation operator}), there is no way to start with the ground state and rotate into an anisotropic quantum Lissajous state while remaining in the same degenerate subspace. The derivations that follow will mirror those for the fundamental isotropic 2DHO. 

The Hamiltonian for the anisotropic oscillator is given by Eq. (\ref{eqn:quantum hamiltonian 2dho in terms of a and a^dagger and p and q again}), where $p$ and $q$ are coprime numbers, and $\omega_x=q\omega_0$ and $\omega_y=p\omega_0$ are commensurate frequencies. The degenerate eigenstates of the Hamiltonian, Eq. (\ref{eqn:quantum hamiltonian 2dho in terms of a and a^dagger and p and q again}), are given by $\ket{pK,q(N-K)}$, for $K=0,1,2,...,N$ and $N=1,2,...,\infty$, and where

\beq
\ket{pK,q(N-K)}=\frac{\left(\hat{a}_x^\dagger\right)^{pK}\left(\hat{a}_y^\dagger\right)^{q(N-K)}}{\sqrt{(pK)!(q(N-K))!}}\ket{0,0}.
\label{eq:aniso degenerate fock states}
\eeq
The fundamental anisotropic Hamiltonian satisfies its own eigenvalue equation,

\beq
\label{eqn:aniso 2DHO eigenvalue eqn}
\left[q\hat{a}_x^\dagger\hat{a}_x+p\hat{a}_y^\dagger\hat{a}_y\right]\ket{pK,q(N-K)}=Npq\ket{pK,q(N-K)},
\eeq
with an eigenvalue of $Npq$, with a degeneracy of $N+1$. This is a result of accidental degeneracy since there is no rotational symmetry. The projection operator for the fundamental anisotropic oscillator is,

\beq
\label{eqn:aniso projection operator}
\Pi_{Npq}=\sum_{K=0}^N\ket{pK}_x\ket{q(N-K)}_y\bra{pK}_x\bra{q(N-K)}_y.
\eeq
Performing the projection on $\ket{\alpha,\beta}$ and disregarding global factors,

\beq
\label{eqn:aniso 2DHO unnormalized coherent state}
\Pi_{Npq}\ket{\alpha,\beta}=\sum_{K=0}^N\frac{(\alpha^p/\beta^q)^K}{\sqrt{(pK)!(q(N-K))!}}\ket{pK,q(N-K)},
\eeq
and when normalized, becomes

\beq
\label{eqn:aniso 2DHO coherent state with xi}
\ket{\xi_{pq},N,p,q}=N_{N,p,q}\sum_{K=0}^N\left(\frac{(qN)!}{(pK)!(q(N-K))!}\right)^{1/2}\xi_{pq}^K\ket{pK,q(N-K)},
\eeq
where $\xi_{pq}=\alpha^p/\beta^q$ and the normalization constant is

\beq
\label{eqn:aniso 2DHO coherent state norm const with xi}
N_{N,p,q}=\left[\sum_{L=0}^N\frac{(qN)!}{(pL)!(q(N-L))!}\abs{\xi_{pq}}^{2L}\right]^{-1/2}.
\eeq
Eq. (\ref{eqn:aniso 2DHO coherent state with xi}) is no longer an SU(2) coherent state, as expected. Although the degeneracy is still $N+1$, characteristic of SU(2), now the rotational symmetry of the system has been broken. This is evident from the fact that the coefficient inside the summation cannot be simplified to the binomial coefficient, as would have to be the case for an SU(2) coherent state. Consequently, $p=q=1$ in the normalization constant yields the summation recognized as the binomial theorem and leads to the isotropic oscillator's normalization constant, $( 1+\abs{\zeta}^2 )^{-N/2}$. Some authors have attempted to formulate "generalized SU(2) coherent states", but we find this to be inaccurate\cite{gorska_correspondence_2006,moran_coherent_2019,kumar_commensurate_2008,chen_vortex_2003}. It is true that setting $p=q=1$ in Eq. (\ref{eqn:aniso 2DHO coherent state with xi}) gives back the SU(2) coherent state, but, Eq. (\ref{eqn:aniso 2DHO coherent state with xi}) is \textbf{not} a "generalized SU(2) coherent state" due to the lack of rotational symmetry; $p=q=1$ is the characteristic feature that preserves rotational symmetry to the isotropic oscillator.

Applying the time evolution operator to Eq. (\ref{eqn:aniso 2DHO coherent state with xi}), the time evolution behaves like a global phase

\beq
\label{eqn:time evol anisotropic}
e^{-i\hat{H}t}\ket{\xi_{pq},N,p,q}=e^{-i\omega_0(Npq+\frac{q+p}{2})t}\ket{\xi_{pq},N,p,q}
\eeq
thus, these are stationary states with a steady probability current density. The wavefunction associated with Eq. (\ref{eqn:aniso 2DHO coherent state with xi}) is again found by projection onto configuration space, $\ket{x,y}$:

\beq
\label{eqn:aniso 2DHO wavefcn}
\Psi_N^{(p,q)(\omega_0)}(x,y,\xi_{pq})=N_{N,p,q}\sum_{K=0}^N\left(\frac{(qN)!}{(pK)!(q(N-K))!}\right)^{1/2}\xi_{pq}^K\psi^{(q)(\omega_0)}_{pK}(x)\psi^{(p)(\omega_0)}_{q(N-K)}(y),
\eeq
where the 1-dimensional wavefunctions are still separable, but more complicated:

\beq
\label{eqn:aniso 2DHO x wavefcn}
\psi^{(q)(\omega_0)}_{pK}(x)=\frac{1}{\sqrt{2^{pK}(pK)!}}\left(\frac{q\omega_0}{\pi}\right)^{1/4}e^{-q\omega_0x^2/2}H_{pK}(x\sqrt{q\omega_0}),
\eeq
and,

\beq
\label{eqn:aniso 2DHO y wavefcn}
\psi^{(p)(\omega_0)}_{q(N-K)}(y)=\frac{1}{\sqrt{2^{q(N-K)}(q(N-K))!}}\left(\frac{p\omega_0}{\pi}\right)^{1/4}e^{-p\omega_0y^2/2}H_{q(N-K)}(y\sqrt{p\omega_0}).
\eeq
It can be seen that the anisotropic 1-D wavefunctions hold the same form as their isotropic counterparts, but now with factors of $p$ and $q$ that change the proportionality of the equations. The current density for Eq. (\ref{eqn:aniso 2DHO wavefcn}) is given by 

\beq
\begin{split}
\rho^{(p,q)(\omega_0)}_N(x,y,\xi_{pq}) & =\abs{\Psi^{(p,q)(\omega_0)}_N(x,y,\xi_{pq})}^2
\\
&=\abs{N_{N,p,q}\sum_{K=0}^N\left(\frac{(qN)!}{(pK)!(q(N-K))!}\right)^{1/2}\xi_{pq}^K\psi^{(q)(\omega_0)}_{pK}(x)\psi^{(p)(\omega_0)}_{q(N-K)}(y)}^2.
\end{split}
\label{eq:aniso 2DHO probability density}
\eeq
We keep the convention of holding $\alpha$ to be real and taking $\beta$ to be complex, $\beta^q=\abs{\beta}^qe^{iq\phi}$. The anisotropic complex amplitude is then $\xi=e^{-iq\phi}\abs{\alpha}^p/\abs{\beta}^q=\abs{\xi}e^{-iq\phi}$. Eq. (\ref{eq:aniso 2DHO probability density}) produces patterns localized along $p:q$ classical Lissajous figures of commensurate frequency ratios. The phase, $\phi$, now changes the shape of the Lissajous curve, i.e. the path that is traced. Just like classical Lissajous curves, the phase cycles through different shapes that Lissajous figures evolve in and out of. The quantity $\abs{\xi_{pq}}$ changes the orientation of the figures in configuration space. Graphical examples will be shown in section \ref{chap4_PrincipleResults} to show the roles of phase and orientation.

\subsection{Quantum Lissajous Figures-Higher Harmonic Cases}

The theory in section 3.3 refer to states having frequency ratios where $p$ and $q$ are coprime numbers. We refer to these states as \textbf{fundamental} quantum Lissajous states of the isotropic and anisotropic 2DHO. We refer to frequency ratios that are not coprime as \textbf{higher harmonic} quantum Lissajous states. Classically, there is no unique higher harmonic Lissajous curve, as the frequency ratio reduces to its lowest terms, or in other words, reduces to the fundamental case. This can be seen in Fig. (\ref{fig:CLJ-p2q3-p4q6-comparison}). There is no visual change to the higher harmonic classical figures, but there is an increase in the total mechanical energy of the system. The quantum mechanical higher harmonic cases do not reduce as their classical counterparts do, but they can be explained in terms of the fundamental quantum Lissajous states.

\subsubsection{Isotropic Higher Harmonic Quantum Lissajous State}

Recall that we've shown the fundamental isotropic 2HDO states to be an SU(2) coherent state\cite{arecchi_atomic_1972}. Using Eqns. (\ref{eqn:aniso 2DHO coherent state with xi}) and (\ref{eqn:aniso 2DHO coherent state norm const with xi}) with $p=q=m>1$,

\beq
\label{eqn:aniso 2DHO coherent state with xi and m}
\ket{\xi_{mm},N,m,m}=N_{N,m,m}\sum_{K=0}^N\binom{mN}{mK}^{1/2}\xi_{mm}^K\ket{mK,m(N-K)},
\eeq
where $\xi_{mm}=\zeta^m=\left(\alpha/\beta\right)^m$ and the normalization constant is

\beq
\label{eqn:aniso 2DHO coherent state norm const with xi and m}
N_{N,m,m}=\left[\sum_{L=0}^N\binom{mN}{mL}\abs{\xi_{mm}}^{2L}\right]^{-1/2},
\eeq
it is easy to show that the m\textsuperscript{th} harmonic isotropic quantum Lissajous state can be written as a linear superposition of fundamental isotropic quantum Lissajous states having; $mN$ quanta, frequencies that are $m$ times the frequency of higher harmonic state, $m\omega_0$, and complex amplitudes that are weighted by factors that are the distinct the m\textsuperscript{th} roots of unity,

\beq
\ket{\abs{\zeta}^me^{-im\phi},N,m}=N_{Nm}\sum^{m-1}_{n=0}\ket{\abs{\zeta} e^{-i(2\pi n+m\phi)/m},mN}.
\label{eq:iso higher harmonic quantum lissajous state}
\eeq
The wavefunction corresponding to Eq. (\ref{eq:iso higher harmonic quantum lissajous state}) accounts  for the $mN$ quanta, frequency characteristics and the roots of unity of this superposition

\beq
\Psi^{(m,m)(\omega_0)}_N(x,y,\abs{\zeta}^me^{-im\phi})=N_{Nm}\sum^{m-1}_{n=0}\Psi_{mN}^{(m\omega_0)}(x,y,\abs{\zeta} e^{-i(2\pi n+m\phi)/m}).
\label{eq:iso higher harmonic quantum lissajous wavefunction}
\eeq
The normalization constant of the isotropic higher harmonic 2DHO turns out to be the ratio of two normalization constants, both of which can be found using Eq. (\ref{eqn:aniso 2DHO coherent state norm const with xi}), and multiplied by $1/m$,

\beq
N_{Nm}=\frac{N_{N,m,m}}{mN_{mN,1,1}}=\frac{1}{m}\frac{\left(\sum^N_{L=0}\binom{mN}{mL}\abs{\zeta}^{2mL}\right)^{-1/2}}{(1+\abs{\zeta}^2)^{-mN/2}}.
\label{eq:iso higher harmonic norm const}
\eeq
The constant in the numerator is Eq. (\ref{eqn:aniso 2DHO coherent state norm const with xi}) with $p=q=m$ and $N$ quanta, while the one in the denominator is Eq. (\ref{eqn:aniso 2DHO coherent state norm const with xi}) with $p=q=1$ and $mN$ quanta. It is obvious that the normalization constant on the left hand side of Eq. (\ref{eq:iso higher harmonic quantum lissajous state}) is $N_{N,m,m}$, thus we see it included on the right hand side. $N_{mN,1,1}$ is included in the denominator to eliminate the normalization constants of the states in the superposition. The states given by Eq. (\ref{eq:iso higher harmonic quantum lissajous state}) are generalizations of the SU(2) cat states, which turn out to be the isotropic higher harmonic quantum Lissajous states for $m=2$\cite{he_fast_2023}, and also generalizations of the SU(2) compass states, which are the isotropic higher harmonic quantum Lissajous states for $m=4$\cite{akhtar_sub-planck_2021}. Once again, the isotropic higher harmonic quantum Lissajous states are superpositions of SU(2) coherent states, which in and of itself is non-classical. The concept of superposition gives rise to quantum interference, which will be shown graphically in section \ref{chap4_PrincipleResults}. The interference terms can be seen by writing out the probability density

\beq
\begin{split}
\abs{\Psi^{(m,m)(\omega_0)}_N}^2&=\abs{N_{Nm}}^2\left[ \sum_{j=0}^{m-1}\Psi_{mN}^{(m\omega_0)(2\pi j/m)*}\right] \left[ \sum_{k=0}^{m-1}\Psi_{mN}^{(m\omega_0)(2\pi k/m)}\right]
\\
&=\abs{N_{Nm}}^2\left[\sum_{j=k}\abs{\Psi_{mN}^{(m\omega_0)(2\pi j/m)}}^2+\sum_{j\neq k}\Psi_{mN}^{(m\omega_0)(2\pi j/m)*}\Psi_{mN}^{(m\omega_0)(2\pi k/m)}\right].
\end{split}
\label{eq:hh iso prob dens}
\eeq
The summation with $j=k$ are the individual probability densities in the superposition and the summation with $j\neq k$ gives rise to the quantum interference between the wavefunctions in the superposition. Each wavefunction interferes with all the others separately.
\subsubsection{Anisotropic Higher Harmonic Quantum Lissajous State}

Now consider the anisotropic higher harmonic quantum Lissajous states , for which $r=q/p\neq1$. Interpolating on the isotropic case, one can figure that the superposition now involves fundamental anisotropic quantum Lissajous states, Eq. (\ref{eqn:aniso 2DHO coherent state with xi}), instead of the fundamental isotropic states. Let $q_0$ and $p_0$ be coprime numbers where $q=mq_0$ and $p=mp_0$, which leads to $r=q/p=q_0/p_0\neq1$. Using Eq. (\ref{eqn:aniso 2DHO coherent state with xi}), it is easy to show that the m\textsuperscript{th} harmonic anisotropic quantum Lissajous state can be written as a linear superposition of fundamental anisotropic quantum Lissajous states having; $mN$ quanta, frequencies that are $m$ times the frequency of higher harmonic state, $m\omega_0$, and complex amplitudes that are weighted by factors that are the distinct the m\textsuperscript{th} roots of unity,

\beq
\ket{\abs{\xi_{pq}}e^{-iq\phi},N,p,q}=N_{Nm,p_0,q_0}^{(r)}\sum^{m-1}_{n=0}\ket{\abs{\xi_{p_0q_0}} e^{-i(2\pi n+q_0m\phi)/m},mN,p_0,q_0},
\label{eq:aniso higher harmonic quantum lissajous state}
\eeq
with the wavefunction

\beq
\Psi_N^{(p,q)(\omega_0)}(x,y,\abs{\xi_{pq}}e^{-iq\phi})=N_{Nm,p_0,q_0}^{(r)}\sum^{m-1}_{n=0}\Psi_{mN}^{(p_0,q_0)(m\omega_0)}(x,y,\abs{\xi_{p_0q_0}}e^{-i(2\pi n+q_0m\phi)/m}),
\label{eq:aniso higher harmonic quantum lissajous state wavefcn}
\eeq
where, similar to the isotropic higher harmonic 2DHO, the normalization constant is the ratio of two normalization constants of Eq. (\ref{eqn:aniso 2DHO coherent state norm const with xi}),

\beq
N_{Nm,p_0,q_0}^{(r)}=\frac{1}{m}\frac{N_{N,p,q}}{N_{mN,p_0,q_0}}=\frac{1}{m}\frac{\left(\sum^{N}_{L=0}\frac{(qN)!}{(pL)!(qN-qL)!}\abs{\xi_{pq}}^{2L}\right)^{-1/2}}{\left(\sum^{mN}_{L=0}\frac{(q_0mN)!}{(p_0L)!(q_0mN-q_0L)!}\abs{\xi_{p_0q_0}}^{2L}\right)^{-1/2}}.
\label{eq:aniso higher harmonic norm const}
\eeq
The relationship between $\xi_{pq}$ and $\xi_{p_0q_0}$ is simply $\xi_{pq}=\frac{\alpha^p}{\beta^q}=\frac{\alpha^{mp_0}}{\beta^{mq_0}}=\left(\frac{\alpha^{p_0}}{\beta^{q_0}}\right)^m=\xi_{p_0q_0}^m$. Eq. (\ref{eq:aniso higher harmonic quantum lissajous state}) is similar to Eq. (\ref{eq:iso higher harmonic quantum lissajous state}) in the sense that the higher harmonic state can be decomposed as linear superpositions of fundamental quantum Lissajous states with phases corresponding to $m$\textsuperscript{th} roots-of-unity, but the fundamental states on the right-hand-side of Eq. (\ref{eq:aniso higher harmonic quantum lissajous state}), $\ket{\xi_{p_0q_0} e^{2\pi ni/m},mN,p_0,q_0}$, are no longer SU(2) coherent states. It is true that for a state of $Npq$ quanta, there is an $N+1$-fold degeneracy which is characteristic of SU(2), but in the anistropic case, the rotational symmetry is broken in contrast with the case of the isotropic 2DHO. For $p=q=m$ Eq. (\ref{eq:aniso higher harmonic quantum lissajous state}) becomes Eq. (\ref{eq:iso higher harmonic quantum lissajous state}). The probability density of the anisotropic higher harmonic quantum Lissajous state can also be written in such a way to separate interference terms from the terms that encompass the individual probability densities, refer to Eq. (\ref{eq:hh iso prob dens}).

\subsection{Probability Current Density for the Quantum Lissajous States}

The probability current density is a vector quantity that describes the flow of probability. It can be thought of in the context of hydrodynamics (electromagnetism) as the flow of fluid (charge)\cite{sakurai_modern_2017}. The probability current density along with the probability density, due to local conservation of probability, must obey the continuity equation

\beq
\Vec{\nabla}\cdot \Vec{J}+\frac{\partial\rho}{\partial t}=0.
\label{eq:iso continuity eqn }
\eeq
The already discussed probability density is given by
\beq
\rho=\abs{\Psi}^2,
\label{eq:prob dens}
\eeq
and the probability current density in general is
\beq
\Vec{J}=\operatorname{Im}\left[ \Psi^*\Vec{\nabla}\Psi\right].
\label{eq:prob current density}
\eeq
To understand more about how the probability current density is characterized, consider a wavefunction of the form

$$\Psi=\Xi(\Vec{x},t) e^{iS(\Vec{x},t)},$$
where $\Xi$ and $S$ are real numbers that are functions of space and time generally. The probability density becomes
$$\rho(\Vec{x},t)=\Xi^2(\Vec{x},t),$$
and by using Eq. (\ref{eq:prob current density}), the probability current density is

\beq
\Vec{J}(\Vec{x},t)=\Xi^2(\Vec{x},t)\Vec{\nabla}S(\Vec{x},t)=\rho(\Vec{x},t)\Vec{\nabla}S(\Vec{x},t).
\label{eq:spatial variation of phase}
\eeq
Eq. (\ref{eq:spatial variation of phase}) tells us that the spatial variation of the phase defines the behavior of the probability current density\cite{sakurai_modern_2017}. Eqs. (\ref{eq:time evolution onto isotropic qlj state}) and (\ref{eqn:time evol anisotropic}) show that the quantum Lissajous states formed via projection on their respective degenerate subspaces are stationary states; thus the probability density is time-independent, and the continuity equation for these states becomes

\beq
\Vec{\nabla}\cdot \Vec{J}=0,
\label{eq:steady current}
\eeq
which says the current density is steady. Two distinct behaviors emerge due to Eq. (\ref{eq:steady current}); the current density can vanish ($\Vec{J}=0$), or the current density can be non-zero in a way consistent with laminar flow of probability ($\Vec{J}\neq0$, but $\Vec{\nabla}\cdot \Vec{J}=0$). In the case of $\Vec{J}=0$, quantum interference is maximal, with fringes appearing in the probability density $\rho$, perpendicular to the motion of the current density, called \textbf{static states}. In the case of $\Vec{J}\neq0$, there are varying amounts of quantum interference, and representative examples will be shown of the emergence of quantum interference in section \ref{chap4_PrincipleResults}. A non-zero current density will require the divergence of the current to vanish to satisfy the continuity equation. These are \textbf{vortex states}. In section \ref{chap4_PrincipleResults}, graphical results will be shown, along with conceptual descriptions, to develop a connection between the nature of non-zero, steady probability current density and the emergence of quantum interference in the quantum Lissajous states for both isotropic and anisotropic oscillators and for both fundamental and higher harmonic cases.

\section{Representative Examples of Principal Results}
\label{chap4_PrincipleResults}
Section \ref{chap4_PrincipleResults} is intended as an abridged archive of several quantum Lissajous figures illustrating the important non-classical features (quantum interference, natural emergence of roots-of-unity superpositions) and allowing for the comprehensive comparison of the fundamental cases with the corresponding classical Lissajous figures. We will present graphical results based upon numerical renderings of our analytical results from section \ref{chap3_QLJviaProjection} to demonstrate the important mathematical features of the quantum Lissajous figures formed via projection onto a degenerate subspace. The comparison between the quantum and classical cases (where possible) is exquisite in its accuracy, justifying the nomenclature we have chosen for the states we have studied here.

\subsection{Fundamental Isotropic Quantum Lissajous State}
\begin{figure}[H]
    \centering
    \includegraphics[width=\textwidth]{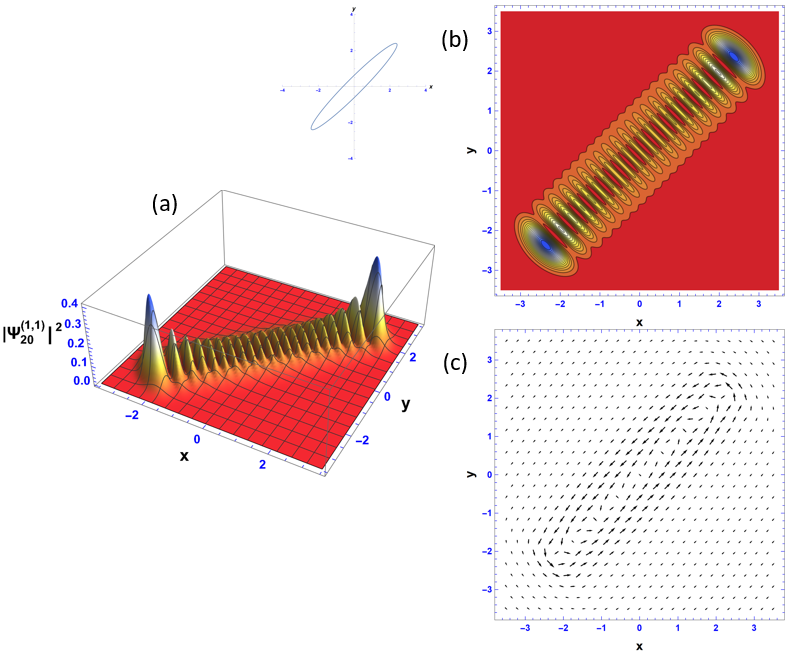}
    
    \caption{The surface (a) and contour (b) plots of the probability density of the isotropic quantum Lissajous state with $\alpha=1$ and $\beta=e^{i\pi/12}$. The associated classical Lissajous figure is shown to the left of (b) for comparison. The corresponding probability current density (c) is steady and propagates in the CCW direction. The interference fringes are no longer fully resolved, and each fringe bleeds into the adjacent fringes. It can be seen in (c) that the innermost current is beginning to vanish due to the current on the opposite co-vertex of the ellipse, forming a vortex.} 
    \label{fig:LJ-figure-N20p1q1alpha1-betaExpiPi12}
\end{figure}

\begin{figure}[H]
    \centering
    \includegraphics[width=\textwidth]{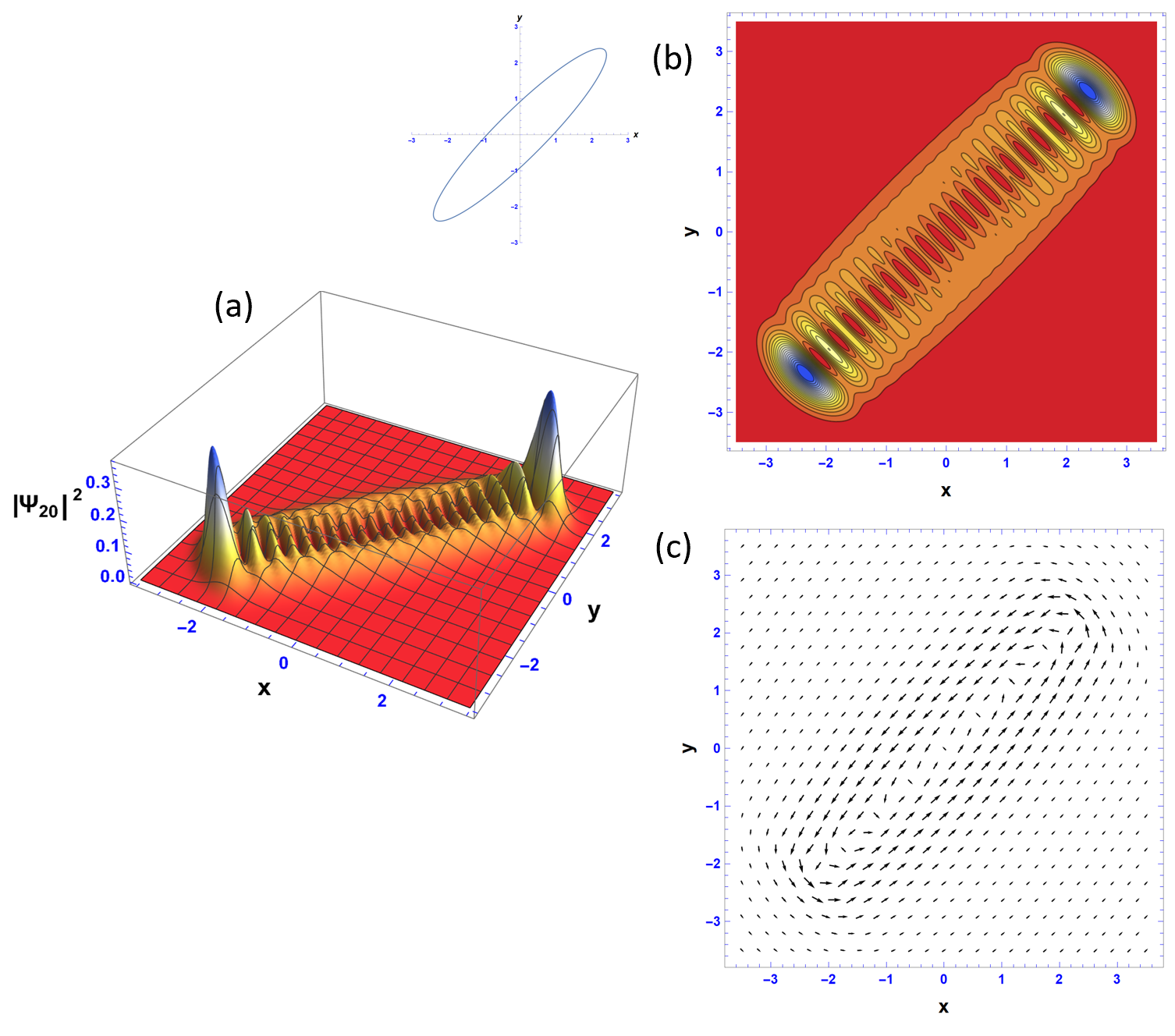}
    
    \caption{The surface (a) and contour (b) plots of the probability density of the isotropic quantum Lissajous state with $\alpha=1$ and $\beta=e^{i\pi/8}$. The associated classical Lissajous figure is shown to the left of (b) for comparison. The corresponding probability current density (c) is steady and propagates in the CCW direction, thus it forms a vortex. Interference fringes are still present, but it can be seen that the centroid of the counter-propagating current is beginning to separate, and the center fringes begin to become less resolved.} 
    \label{fig:LJ-figure-N20p1q1alpha1-betaExpiPi8}
\end{figure}

\begin{figure}[H]
    \centering
    \includegraphics[width=\textwidth]{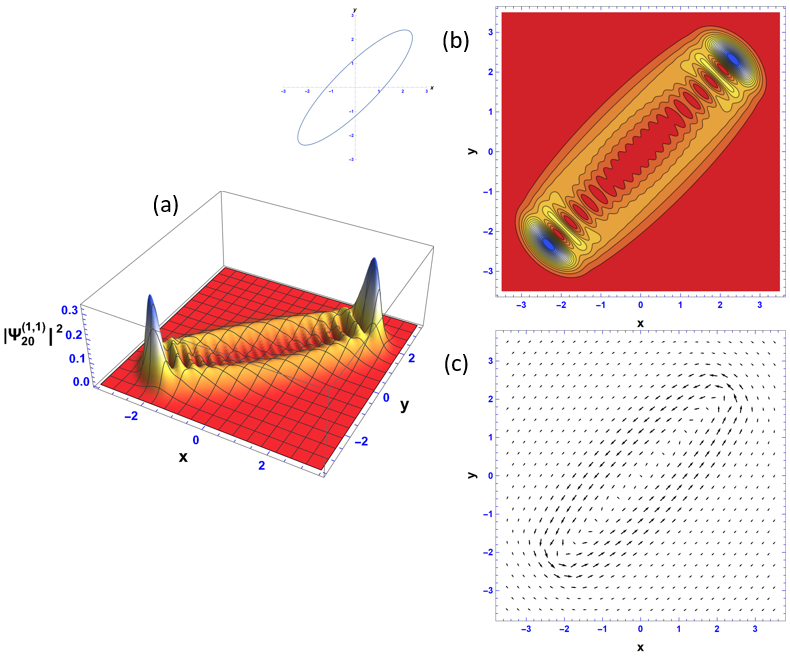}
    
    \caption{The surface (a) and contour (b) plots of the probability density of the isotropic quantum Lissajous state with $\alpha=1$ and $\beta=e^{i\pi/6}$. The associated classical Lissajous figure is shown to the left of (b) for comparison. The corresponding probability current density plot (c) is steady and propagates in the CCW direction, thus it forms a vortex. The interference fringes begin to disappear as the semi-minor axis increases and the vortex becomes more prominent. The emergence of a vortex corresponds to the vanishing of the interference fringes.} 
    \label{fig:LJ-figure-N20p1q1alpha1-betaExpiPi6}
\end{figure}

\begin{figure}[H]
    \centering
    \includegraphics[width=\textwidth]{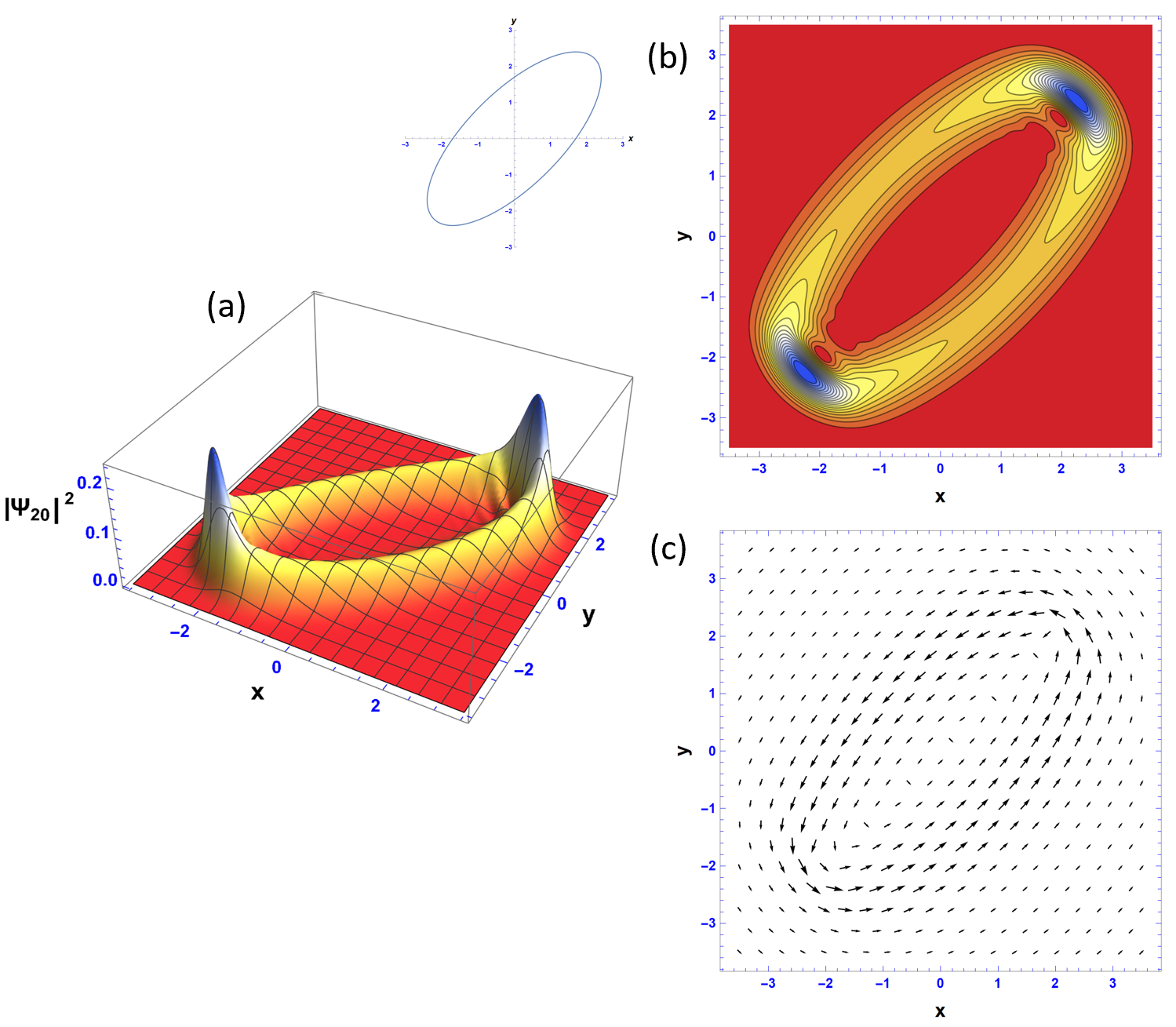}
    
    \caption{The surface (a) and contour (b) plots of the probability density of the isotropic quantum Lissajous state with $\alpha=1$ and $\beta=e^{i\pi/4}$. The associated classical Lissajous figure is shown to the left of (b) for comparison. The corresponding probability current density plot (c) is steady and propagates in the CCW direction, thus it forms a vortex. The interference fringes in the center of the figure are all gone, only remnants of fringes are present near the vertices of the ellipse, and the vortex is becoming more circular.} 
    \label{fig:LJ-figure-N20p1q1alpha1-betaExpiPi4}
\end{figure}

The results presented in this section correspond to the theory developed on the fundamental isotropic quantum Lissajous states, Eq. (\ref{eq:NORMALIZED degenerates state proj onto coherent states}), and its wavefunction, Eq. (\ref{eq:2D wavefunction from coherent states}). The probability density is calculated using Eq. (\ref{eq:probability density}) and the probability current density using Eq. (\ref{eq:prob current density}).
\begin{figure}[H]
    \centering
    \includegraphics[width=\textwidth]{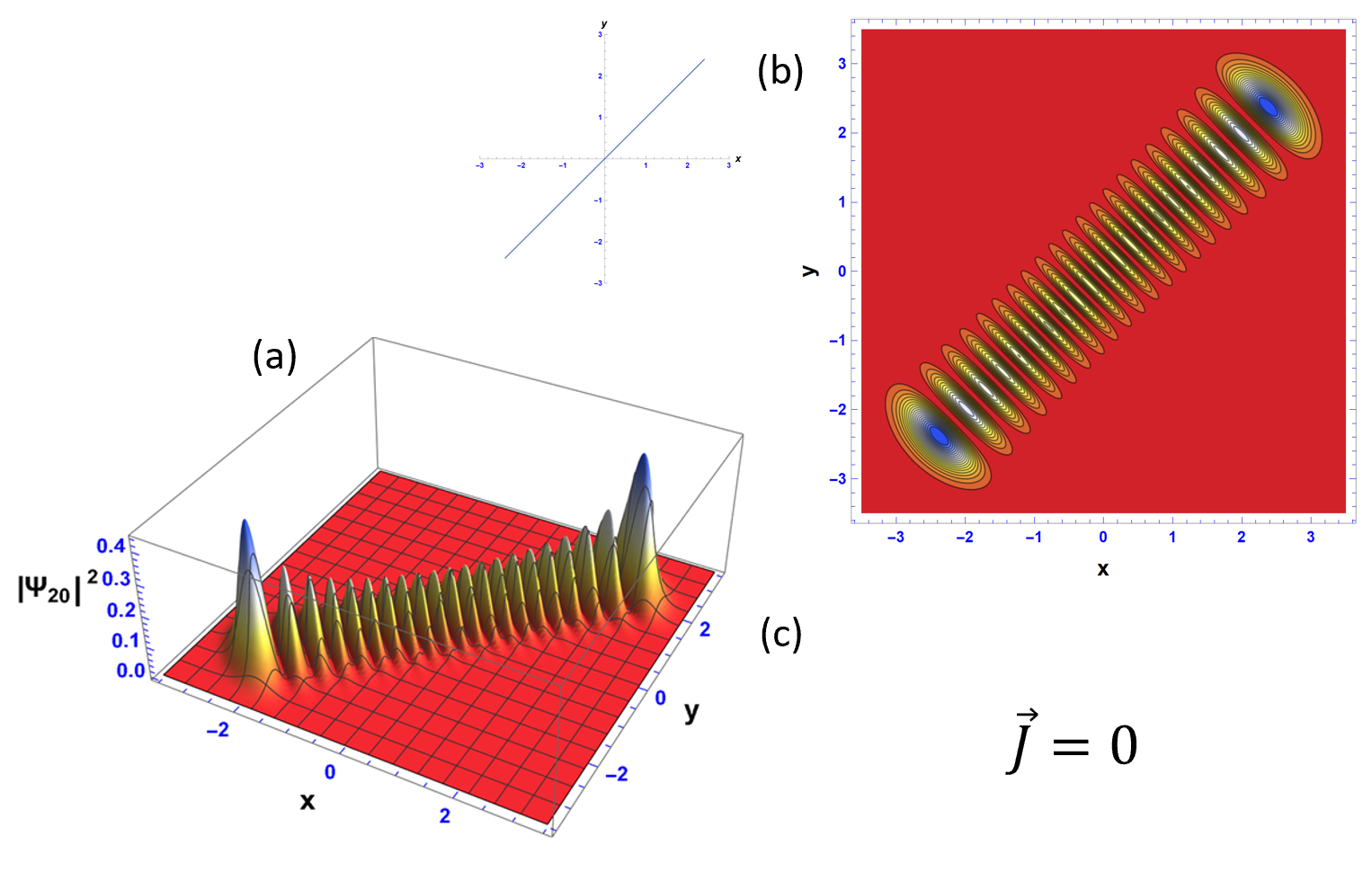}
    
    \caption{The surface (a) and contour (b) plots of the probability density of the isotropic quantum Lissajous state with $\alpha=\beta=1$. The associated classical Lissajous figure is shown to the left of (b) for comparison. The corresponding probability current density (c) vanishes due to spatial oscillation of the probability density, thus giving rise to fully resolved quantum interference fringes. This is the limiting case of a static quantum Lissajous state for the isotropic 2DHO.}
    \label{fig:LJ-figure-N20p1q1alpha1-beta1}
\end{figure}
Fig. \ref{fig:LJ-figure-N20p1q1alpha1-beta1} presents the surface (a) and contour (b) plots of the probability density, as well as the vector (c) plot of the probability current density of the fundamental isotropic quantum Lissajous state, Eq. (\ref{eq:2D wavefunction from coherent states}), for $\alpha=\beta=1$, complex amplitudes in phase. Just as in all other probability densities, the highest peaks are where the oscillator is most likely to be. The fringes that appear here are analogous to the patterns found in double-slit interference so we have referred to the probability peaks as "bright spots" and the minima as "dark spots". It can be seen that probability density localizes over the $1:1$ classical Lissajous figure for $\phi=0$ (Fig. \ref{fig:CLJ-p1q1-phiVARIED}(a)), which is pictured to the left of Fig. \ref{fig:LJ-figure-N20p1q1alpha1-beta1}(b). The probability current density completely vanishes. The classical oscillator travels along the path of a Lissajous figure; in the case of the $1:1$ classical Lissajous figure, the oscillator travels along the path of the straight line on $y=x$ and temporally oscillates on that line. For the quantum oscillator, the probability density spatially oscillates along the corresponding Lissajous figure. Because the oscillation is isolated along one line, and the probability current density vanishes, the probability current must be equal and opposite so for it to cancel out and vanish. The most striking feature in this figure is the appearance of fully resolved interference fringes. As more examples are shown, it will become increasingly clearer that quantum interference fringes appear along sections of quantum Lissajous figures where the probability current is "counter-propagating" and that quantum interference fringes are maximally resolved for vanishing probability current density. Fig. \ref{fig:LJ-figure-N20p1q1alpha1-beta1} satisfies the continuity equation, Eq. (\ref{eq:iso continuity eqn }), by having a vanishing probability current. Thus, this is a static quantum Lissajous state and, in fact, is the \textbf{\underline{static limit}} of the quantum Lissajous states for the isotropic 2DHO.
\begin{figure}[H]
    \centering
    \includegraphics[width=\textwidth]{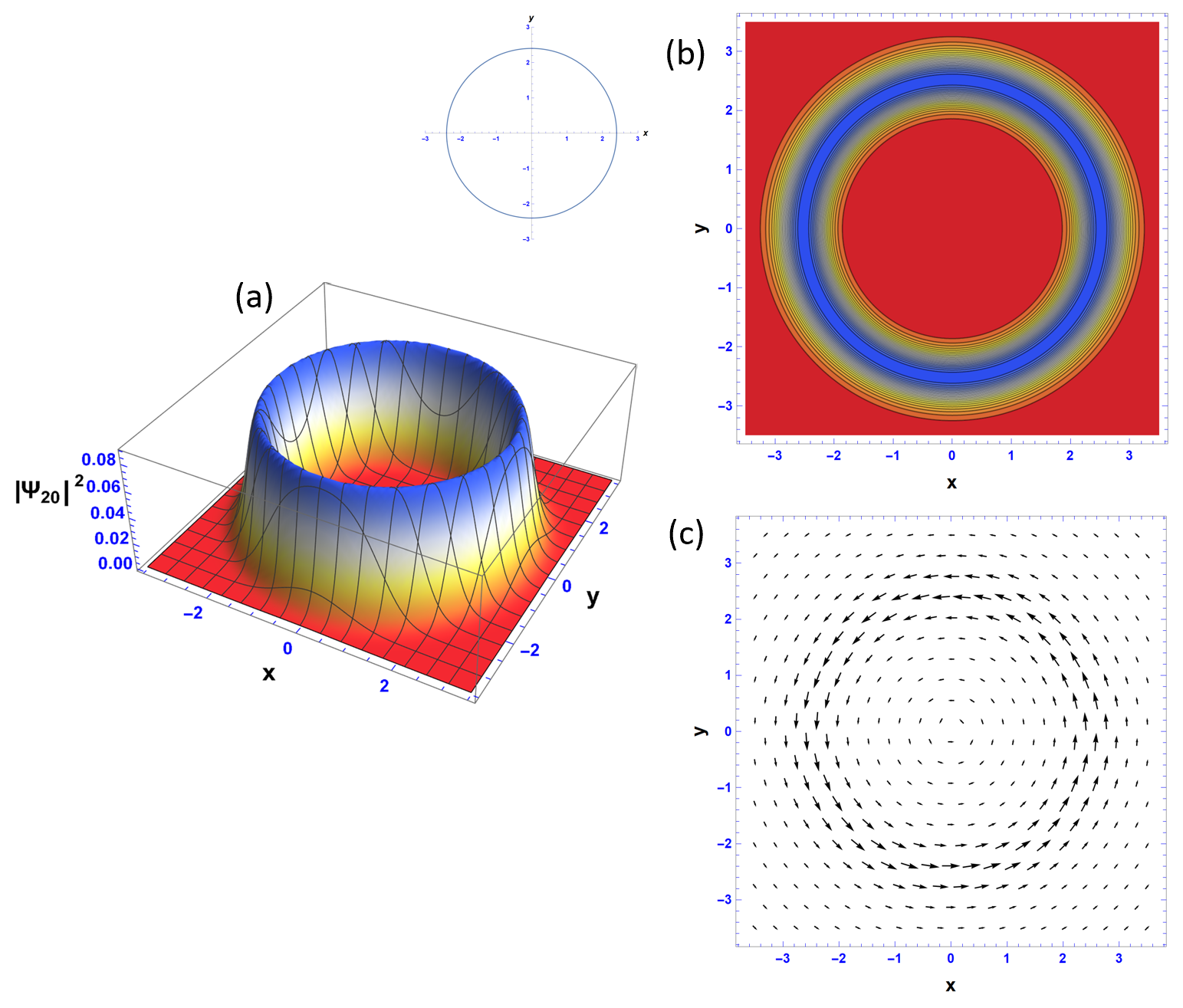}
    
    \caption{The surface (a) and contour (b) plots of the probability density of the isotropic quantum Lissajous state with $\alpha=1$ and $\beta=e^{i\pi/2}$. The associated classical Lissajous figure is shown to the left of (b) for comparison. The corresponding probability current density (c) is steady and propagates in the CCW direction, thus it forms a vortex. There is no sign of quantum interference fringes and, by looking only at the current density plot, this can be expected due to the lack of counter-propagating current. This is the limiting case of a vortex quantum Lissajous state for the isotropic 2DHO.} 
    \label{fig:LJ-figure-N20p1q1alpha1-betaExpiPi2}
\end{figure}
Fig. \ref{fig:LJ-figure-N20p1q1alpha1-betaExpiPi2} presents the surface (a) and contour (b) plots of the probability density, as well as the vector (c) plot of the probability current density of the fundamental isotropic quantum Lissajous state, Eq. (\ref{eq:2D wavefunction from coherent states}), for $\alpha=1$ and $\beta=e^{i\pi/2}$, complex amplitudes in quadrature. It can be seen that probability density localizes over the $1:1$ classical Lissajous figure for $\phi=\pi/2$, which is pictured to the left of Fig. \ref{fig:LJ-figure-N20p1q1alpha1-betaExpiPi2}(b). The probability current density is steady and flows in the counter-clockwise (CCW) direction, which matches the classical oscillator of the same phase. There are smooth and monotonic spatial variations in probability owing to the lack of quantum interference fringes. The CCW propagating current by itself is conducive to the absence of interference fringes, as there is a steady but non-zero probability current. Fig. \ref{fig:LJ-figure-N20p1q1alpha1-betaExpiPi2} satisfies the continuity equation, Eq. (\ref{eq:iso continuity eqn }), by having a divergenceless current density. This is easy to see in Fig. \ref{fig:LJ-figure-N20p1q1alpha1-betaExpiPi2}(c) as the current flows back into itself. This idea is analogous to magnetic fields by way of Gauss' law for magnetism, $\Vec{\nabla}\cdot\Vec{B}=0$, which states that magnetic monopoles do not exist because the magnetic field is divergenceless\cite{griffiths_introduction_2017}. This is the \textbf{\underline{vortex limit}} of the quantum Lissajous states for the isotropic 2DHO. Notice that from Fig. \ref{fig:LJ-figure-N20p1q1alpha1-beta1} to Fig. \ref{fig:LJ-figure-N20p1q1alpha1-betaExpiPi2}, the quantum interference fringes wash away. As the phase is increased from $\phi=0$ to $\phi=\pi/2$, we expect to see the quantum interference fringes to disappear in such a way that corresponds to the absence counter-propagation of current density. Examples of the fundamental isotropic quantum Lissajous states for phases in the range $0<\phi<\pi/2$ are shown in Figs. \ref{fig:LJ-figure-N20p1q1alpha1-betaExpiPi12}-\ref{fig:LJ-figure-N20p1q1alpha1-betaExpiPi4}, and will defend the fact that quantum Lissajous figures localize along their classical counterparts of the same phase difference, as well as depict the emergence (or disappearance) of interference fringes.

\begin{figure}[H]
    \centering
    \includegraphics[width=\textwidth]{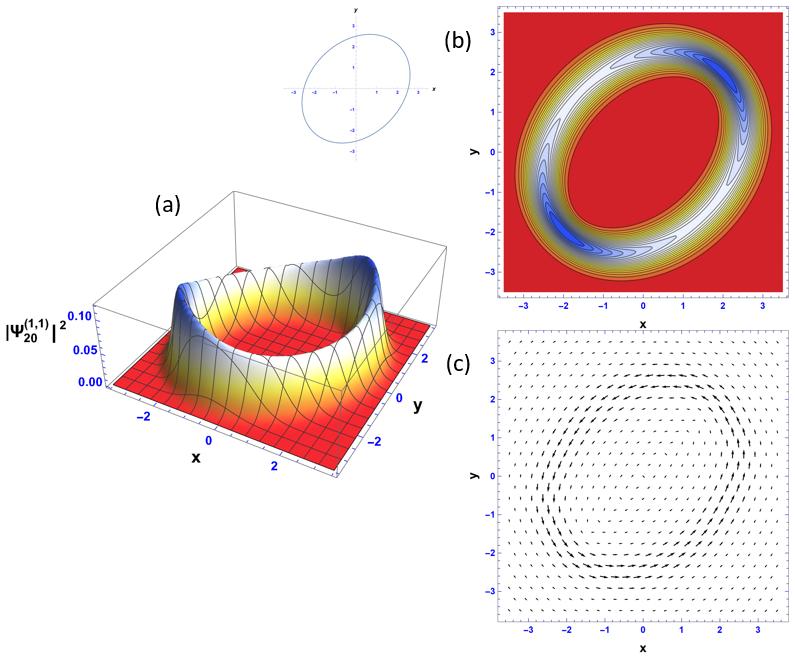}
    
    \caption{The surface (a) and contour (b) plots of the probability density of the isotropic quantum Lissajous state with $\alpha=1$ and $\beta=e^{i5\pi/12}$. The associated classical Lissajous figure is shown to the left of (b) for comparison. The corresponding probability current density plot (c) is steady and propagates in the CCW direction, thus it forms a vortex. The interference fringes in the center of the figure, as well as remnant fringes near the vertices of the ellipse, are completely gone. The peaks at the vertices of the ellipse are now less pronounced due to the uniform distribution of probability along the centroid of the elliptical shape.} 
    \label{fig:LJ-figure-N20p1q1alpha1-betaExpi5Pi12}
\end{figure}
Fig. \ref{fig:LJ-figure-N20p1q1alpha1-betaExpiPi12} presents the beginning for the transition from a line segment to an ellipse. The interference fringes are no longer fully resolved, as in Fig. \ref{fig:LJ-figure-N20p1q1alpha1-beta1}. The probability in each fringe flows out of the bright spot and into the surrounding space, meeting the adjacent fringe's probability "leak", which connects the fringes at their ends and encloses the dark spots. A vortex emerges in the shape of a highly eccentric ellipse flow CCW and the transition from a line segment to an ellipse begins. Fig. \ref{fig:LJ-figure-N20p1q1alpha1-betaExpiPi8} depicts the dissipation of the interior fringes from Fig. \ref{fig:LJ-figure-N20p1q1alpha1-beta1}, which corresponds to the formation of a vortex. The semi-minor axis is the distance between the co-vertices of the ellipse. Thus, the central interference fringes wash out first as the ellipse becomes decreasingly eccentric. Fig. \ref{fig:LJ-figure-N20p1q1alpha1-betaExpiPi6} supports the description of the behavior of Fig. \ref{fig:LJ-figure-N20p1q1alpha1-betaExpiPi8} with more pronounced features.
Continuing up to Fig. \ref{fig:LJ-figure-N20p1q1alpha1-betaExpiPi4}, the interference in the center is completely gone, with only remnants of fringes remaining near the vertices of the ellipse, and the probability current vortex has become more circular. Looking back at Fig. \ref{fig:LJ-figure-N20p1q1alpha1-betaExpiPi2}, we see that all interference fringes are gone, and the bright spots that were once at the vertices of the ellipse have now uniformly distributed themselves along the whole circular figure. With the information provided by Figs. \ref{fig:LJ-figure-N20p1q1alpha1-betaExpiPi12}-\ref{fig:LJ-figure-N20p1q1alpha1-betaExpiPi4}, it is clear that vortex quantum Lissajous states can also exhibit interference fringes when counter-propagating currents are spatially close to one another, where the vortex appearance of the interference disappearance are proportional. At the point of Fig. \ref{fig:LJ-figure-N20p1q1alpha1-betaExpi5Pi12}, the interference fringes are completely gone due to the lack of eccentricity, so there is a threshold of the measure of eccentricity that corresponds to the emergence of interference fringes. This can also be explained by the clear separation of counter-propagating probability current exhibited in Fig. \ref{fig:LJ-figure-N20p1q1alpha1-betaExpi5Pi12} and all other cases of small values of eccentricity. From the presentation of Figs. \ref{fig:LJ-figure-N20p1q1alpha1-betaExpiPi12}-\ref{fig:LJ-figure-N20p1q1alpha1-betaExpi5Pi12} the distribution of probability density is a direct result of conservation of probability and having a normalized state. If these states were not normalized, the uniform distribution of probability through changes of phase would not happen, and the net probability would be greater than one. Because the fundamental isotropic quantum Lissajous states are SU(2) coherent states, it is interesting to examine the behavior of their probability densities at different latitudes of the Bloch sphere. Recall the relationship,
$$\frac{\abs{\alpha}}{\abs{\beta}}e^{-i\phi}=\tan{(\theta/2)}e^{-i\phi},$$
where $\phi$ is the azimuthal angle and $\theta$ is the angle from the $-z$-axis.
\begin{figure}[H]
    \centering
    \includegraphics[width=\textwidth]{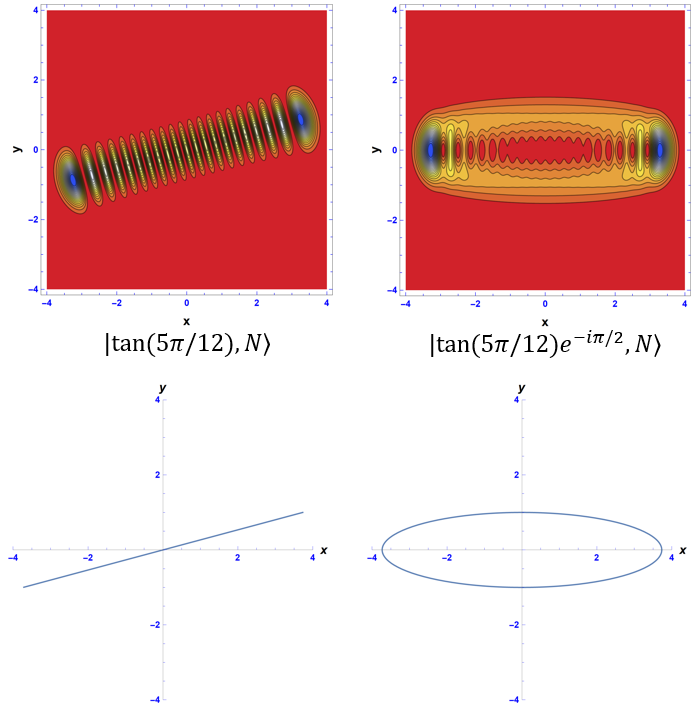}
    
    \caption{The contour plots of the probability density of the isotropic quantum Lissajous state with $\alpha=\tan{(5\pi/12)}$ and $\beta=1,e^{i\pi/2}$, along with the corresponding classical Lissajous figures with $A=\tan{(5\pi/12)}$ and $B=1,e^{i\pi/2}$. These figures are positioned $\pi/3$ radians above the equator of the Bloch sphere, $\theta=5\pi/6$. It can be seen that the same change in phase as figures on the equator of the Bloch sphere, Figs. \ref{fig:LJ-figure-N20p1q1alpha1-beta1} and \ref{fig:LJ-figure-N20p1q1alpha1-betaExpiPi2}, do not correlate to the same rate of vanishing of the quantum interference fringes. The complex amplitude, $\abs{\zeta}$, is what causes the orientation change of the figures in configuration space, but changing the phase whilst having $\abs{\zeta}\neq1$ also changes the orientation.} 
    \label{fig:LJ-figure-N20p1q1alphaTan(5Pi12)-beta(1,ExpiPi2)}
\end{figure}

Fig. \ref{fig:LJ-figure-N20p1q1alphaTan(5Pi12)-beta(1,ExpiPi2)} presents probability densities of the SU(2) coherent state with values of $\alpha=\tan{(5\pi/12)}$, $\pi/3$ radians above the equator of the Bloch sphere, and $\beta=1,e^{i\pi/2}$. Comparing these plots with Figs. \ref{fig:LJ-figure-N20p1q1alpha1-beta1} and \ref{fig:LJ-figure-N20p1q1alpha1-betaExpiPi2}, it can be seen that the case with $\beta=1$, $\ket{\tan{(5\pi/12)},N}$, rotates the figure $\pi/6$ radians CW from Fig. \ref{fig:LJ-figure-N20p1q1alpha1-beta1}. The case with a phase of $\pi/2$, $\ket{\tan{(5\pi/12)}e^{-i\pi/2},N}$ decreases the eccentricity of the figure, but not at the same rate as Fig. \ref{fig:LJ-figure-N20p1q1alpha1-betaExpiPi2}. $\ket{\tan{(5\pi/12)}e^{-i\pi/2},N}$ is also rotated $\pi/12$ radians CW such that the semi-major axis is parallel to the $x$-axis. The combination of $\tan{(\theta/2)}\neq1$ and $\phi\neq0$ rotates the Lissajous figures (both quantum and classical) an additional amount. Just as in the classical Lissajous ellipse, it can be seen that there is one extreme value on both the $x-$ and $y-$axis.
\subsection{Fundamental Anisotropic Quantum Lissajous State}
The results presented in this section correspond to the theory developed on the fundamental anisotropic quantum Lissajous states, Eq. (\ref{eqn:aniso 2DHO coherent state with xi}), and its wavefunction, Eq. (\ref{eqn:aniso 2DHO wavefcn}). The probability density is calculated using Eq. (\ref{eq:aniso 2DHO probability density}) and the probability current density using Eq. (\ref{eq:prob current density}).
\begin{figure}[H]
    \centering
    \includegraphics[width=\textwidth]{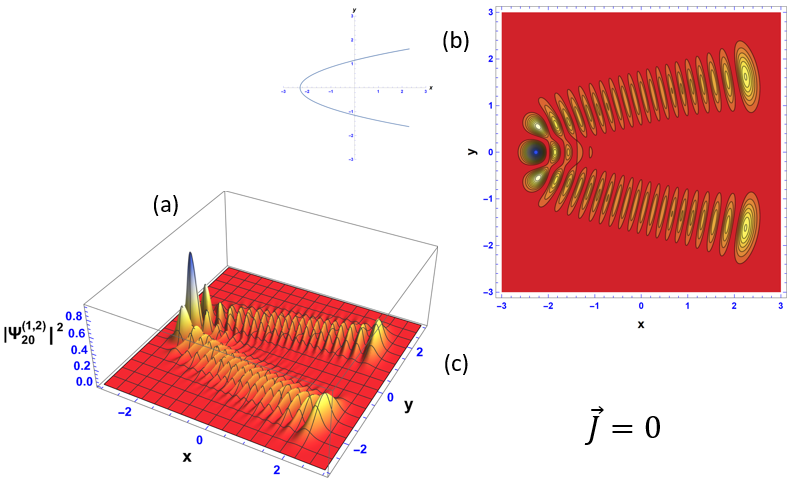}
    
    \caption{The surface (a) and contour (b) plots of the probability density of the fundamental anisotropic quantum Lissajous state with $p=1$, $q=2$, and $\alpha=\beta=1$. The associated classical Lissajous figure is shown to the left of (b) for comparison. The corresponding probability current density (c) vanishes due to the spatial oscillation of the probability current. The interference fringes are fully resolved because of this spatial oscillation, as the counter-propagating currents fully overlap and cancel out. This is the \textbf{\underline{static limit}} of the 1:2 quantum Lissajous states for the anisotropic 2DHO.} 
    \label{fig:LJ-figure-N20p1q2alpha1-beta1}
\end{figure}
Fig. \ref{fig:LJ-figure-N20p1q2alpha1-beta1} localizes over the 1:2 classical Lissajous figure for $\phi=0$, picture to the left of (b). Similarly to Fig. \ref{fig:LJ-figure-N20p1q1alpha1-beta1}, the probability current density vanishes and the interference fringes are fully resolved. The classical oscillator travels along the path of the 1:2 Lissajous figure and temporally oscillates on that curve, whereas the quantum oscillator's probability density spatially oscillates along the corresponding classical Lissajous figure. The reasoning for the fully resolved interference fringes is the same as in Fig. \ref{fig:LJ-figure-N20p1q1alpha1-beta1}, thus, this is the \textbf{\underline{static limit}} of the 1:2 quantum Lissajous states for the anisotropic 2DHO.
\begin{figure}[H]
    \centering
    \includegraphics[width=\textwidth]{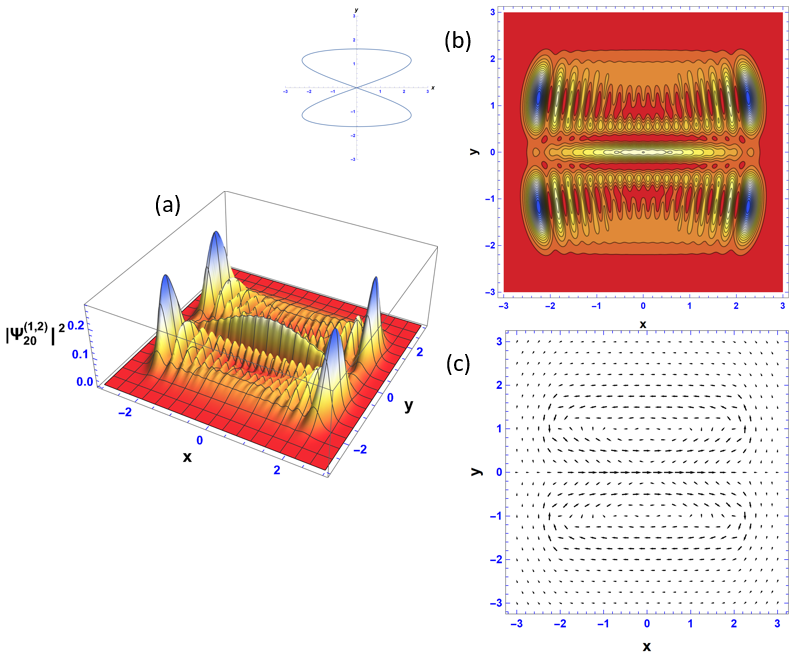}
    
    \caption{The surface (a) and contour (b) plots of the probability density of the anisotropic quantum Lissajous state with $p=1$, $q=2$, $\alpha=1$ and $\beta=e^{i\pi/4}$. The associated classical Lissajous figure is shown to the left of (b) for comparison. The corresponding probability current density (c) forms two vortices, one column of vortices with two rows. There is a noticeable fringe at the center of this figure, which, in the classical case, is an overlap in the path the classical oscillator takes. Quantum mechanically, the probability current in the $y-$direction totally cancels, leaving a stream of $x$ current between the vortices. This is the \textbf{\underline{vortex limit}} of the 1:2 quantum Lissajous states for the anisotropic 2DHO.} 
    \label{fig:LJ-figure-N20p1q2alpha1-betaExpiPi4}
\end{figure}

\begin{figure}[H]
    \centering
    \includegraphics[width=\textwidth]{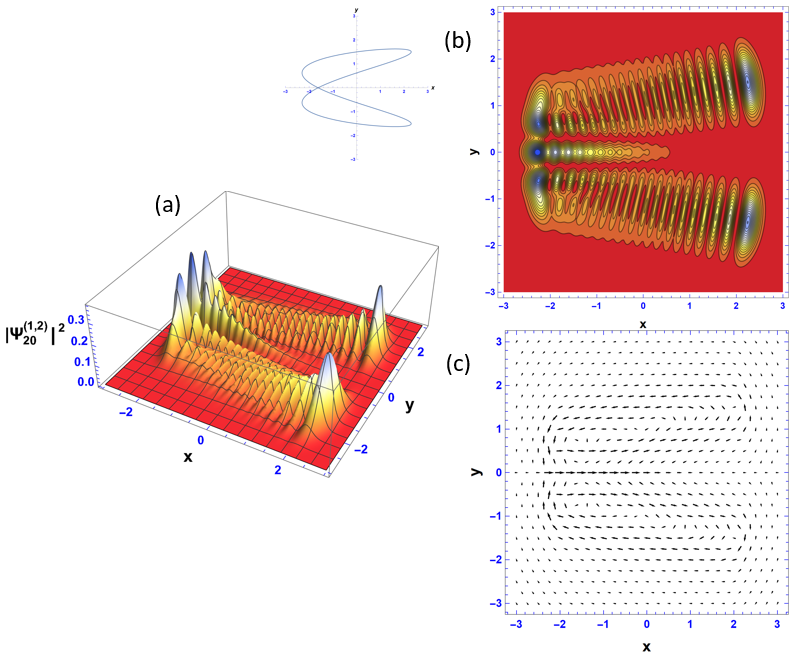}
    
    \caption{The surface (a) and contour (b) plots of the probability density of the anisotropic quantum Lissajous state with $p=1$, $q=2$, $\alpha=1$ and $\beta=e^{i\pi/8}$. The associated classical Lissajous figure is shown to the left of (b) for comparison. The corresponding probability current density (c) forms two vortices, one column of vortices with two rows. There is a noticeable fringe in the quantum Lissajous figure where there is an overlap in the classical Lissajous figure. Quantum mechanically, the probability current in the $y-$direction totally cancels, leaving a stream of $x$ current between the vortices.} 
    \label{fig:LJ-figure-N20p1q2alpha1-betaExpiPi8}
\end{figure}

\begin{figure}[H]
    \centering
    \includegraphics[width=\textwidth]{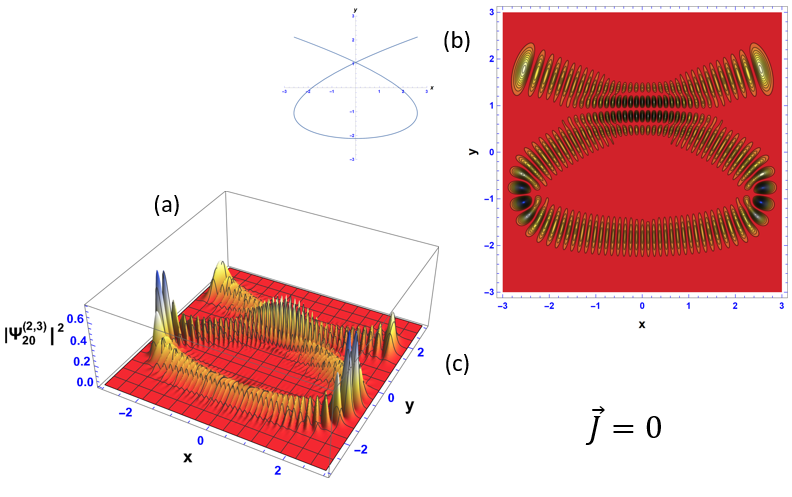}
    
    \caption{The surface (a) and contour (b) plots of the probability density of the anisotropic quantum Lissajous state with $p=2$, $q=3$ and $\alpha=\beta=1$. The associated classical Lissajous figure is shown to the left of (b) for comparison. The corresponding probability current density (c) vanishes since the counter-propagating currents flow on the same curve. Consequently, maximally resolved interference fringes are present. There is an overlap in the 2:3 classical Lissajous curve, which corresponds to the presence of interference fringes on that overlap in the quantum Lissajous figure. The fringes at the overlap of the curve are fully resolved as opposed to Figs. \ref{fig:LJ-figure-N20p1q2alpha1-betaExpiPi4} and \ref{fig:LJ-figure-N20p1q2alpha1-betaExpiPi8}. The absence of probability current at the overlap gives rise to vertical and horizontal fringes. This is the \textbf{\underline{static limit}} of the 2:3 quantum Lissajous states for the anisotropic 2DHO. } 
    \label{fig:LJ-figure-N20p2q3alpha1-beta1}
\end{figure}

\begin{figure}[H]
    \centering
    \includegraphics[width=\textwidth]{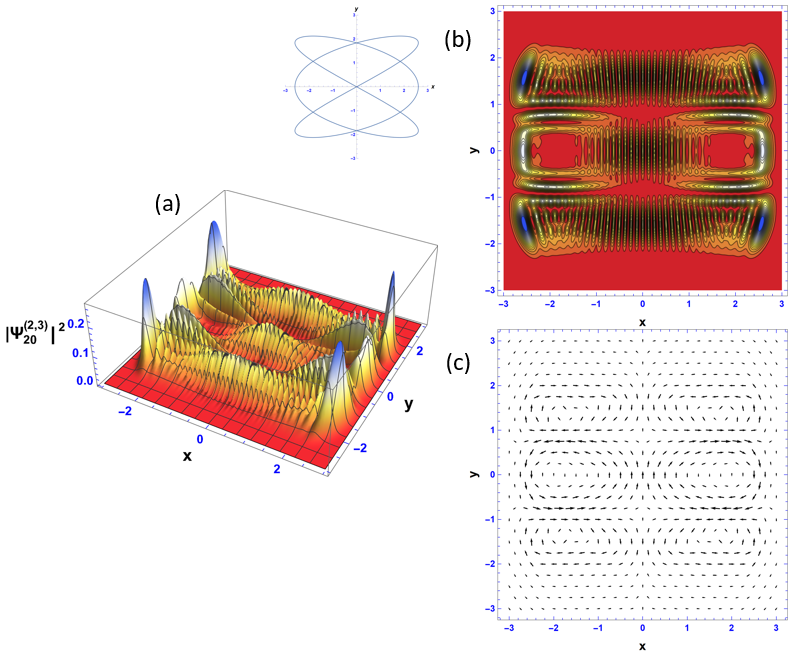}
    
    \caption{The surface (a) and contour (b) plots of the probability density of the anisotropic quantum Lissajous state with $p=2$, $q=3$, $\alpha=1$, and $\beta=e^{i\pi/6}$. The associated classical Lissajous figure is shown to the left of (b) for comparison. The corresponding probability current density (c) forms six vortices, two columns of vortices with three rows. The overlaps in the classical Lissajous figure translate to interference fringes in the quantum Lissajous figure. There are either vertical or horizontal fringes where the overlaps are, not both, depending on if the $x$ or $y$ current vanishes. This is the \textbf{\underline{vortex limit}} of the 2:3 quantum Lissajous states for the anisotropic 2DHO.} 
    \label{fig:LJ-figure-N20p2q3alpha1-betaExpiPi6}
\end{figure}

\begin{figure}[H]
    \centering
    \includegraphics[width=\textwidth]{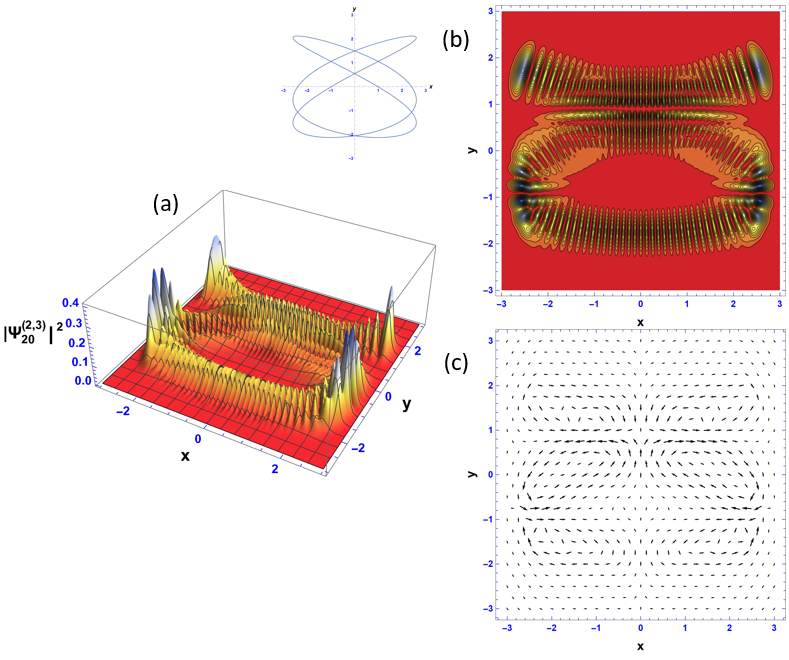}
    
    \caption{The surface (a) and contour (b) plots of the probability density of the anisotropic quantum Lissajous state with $p=2$, $q=3$, $\alpha=1$, and $\beta=e^{i\pi/12}$. The associated classical Lissajous figure is shown to the left of (b) for comparison. The corresponding probability current density (c) . } 
    \label{fig:LJ-figure-N20p2q3alpha1-betaExpiPi12}
\end{figure}
Figs. \ref{fig:LJ-figure-N20p1q2alpha1-beta1}-\ref{fig:LJ-figure-N20p2q3alpha1-betaExpiPi12} present results of the fundamental anisotropic quantum Lissajous states. The emergence of interference fringes and the connection they have with probability current density comes about in the same way as the fundamental isotropic quantum Lissajous state. A new feature that is observed in the anisotropic probability density and probability current density is the appearance of fringes at the overlaps in the classical Lissajous figure. It can be seen in Figs. \ref{fig:LJ-figure-N20p1q2alpha1-betaExpiPi4}, \ref{fig:LJ-figure-N20p1q2alpha1-betaExpiPi8}, \ref{fig:LJ-figure-N20p2q3alpha1-betaExpiPi6}, and \ref{fig:LJ-figure-N20p2q3alpha1-betaExpiPi12} that when there is an overlap in the classical Lissajous figure, this corresponds to a cancellation of probability current in the quantum Lissajous figure in one direction, leaving a horizontal (vertical) stream of current and resulting in vertical (horizontal) interference fringes. Fig. \ref{fig:LJ-figure-N20p2q3alpha1-beta1} also exhibits interference fringes at the overlap of the classical figure, but the fringes are fully resolved and there are both vertical and horizontal fringes due to the vanishing probability current density. Another new feature is the relationship between $p$ and $q$, and the number of vortices present. The number of vortices is proportional to $qp$, with $p$ columns and $q$ rows. There are two vortices in the 1:2 quantum Lissajous state and six vortices in the 2:3 quantum Lissajous state. The number of extrema still holds for quantum Lissajous figures as; $p$ extrema on the $y-$axis and $q$ extrema on the $x-$axis, so it can be deduced that $p$ extrema on the $y-$axis corresponds to $p$ columns of vortices and $q$ extrema on the $x-$axis corresponds to $q$ rows of vortices.
\subsection{Isotropic Higher Harmonic Quantum Lissajous State}
The results presented in this section correspond to the theory developed on the isotropic higher harmonic quantum Lissajous states, Eq. (\ref{eq:iso higher harmonic quantum lissajous state}), and its wavefunction, Eq. (\ref{eq:iso higher harmonic quantum lissajous wavefunction}). The probability density is calculated using Eq. (\ref{eq:hh iso prob dens}) and the probability current density using Eq. (\ref{eq:prob current density}).
\begin{figure}[H]
    \centering
    \includegraphics[width=\textwidth]{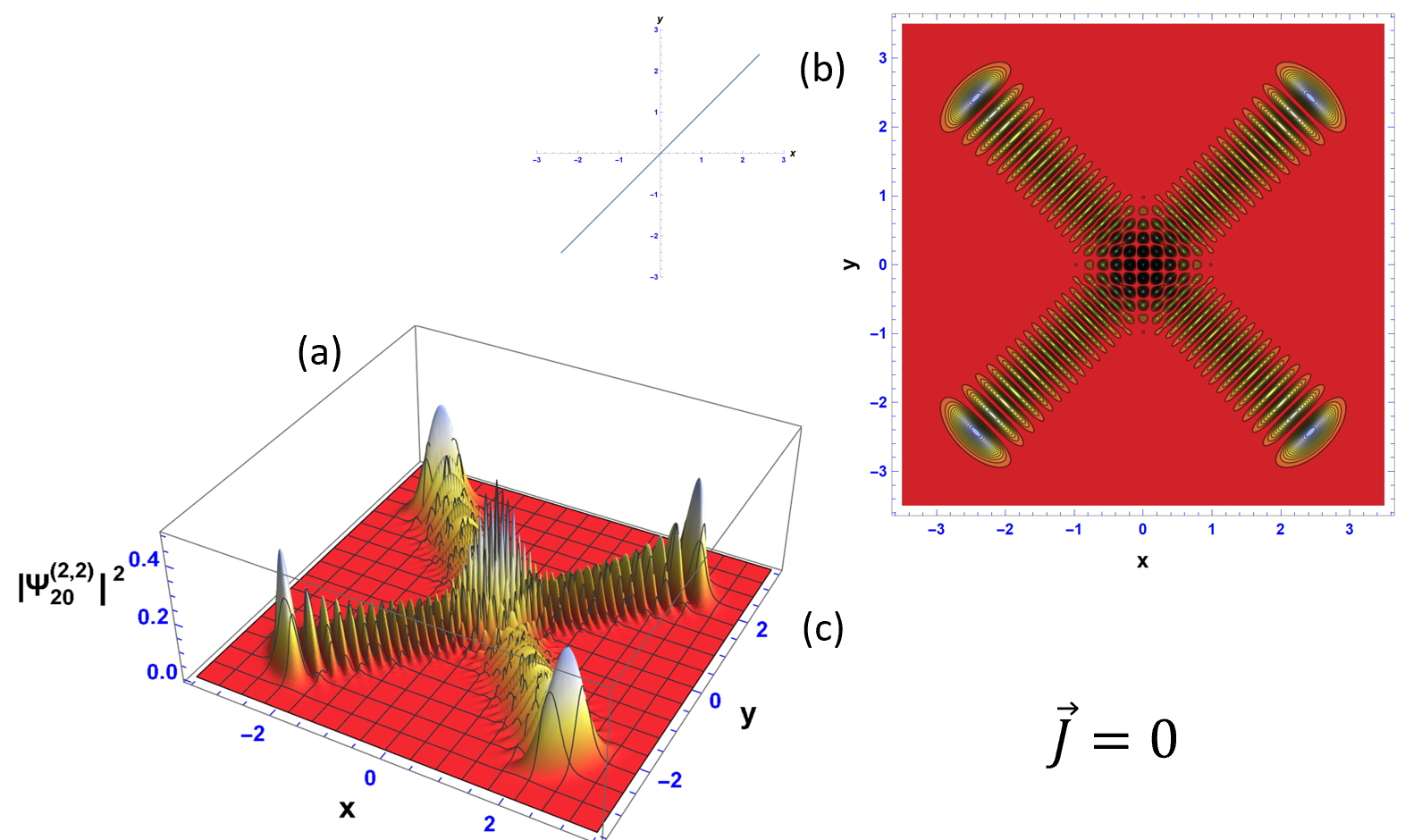}
    
    \caption{The surface (a) and contour (b) plots of the probability density of the isotropic higher harmonic quantum Lissajous state with $p=q=2$, $\alpha=\beta=1$. The associated classical Lissajous figure is shown to the left of (b) for comparison. The classical Lissajous figure is that of the 1:1 classical Lissajous figure for $\phi=0$, oscillating at double the frequency of Fig. \ref{fig:CLJ-p1q1-phiVARIED}(a). The corresponding probability current density (c) vanishes and is a \textbf{\underline{static limit}} of the 2:2 quantum Lissajous states for the isotropic higher harmonic 2DHO. The 2:2 quantum Lissajous figure is a coherent superposition of two fundamental isotropic quantum Lissajous figures having $2N$ quanta, complex amplitudes that are the square roots of unity, and oscillating at double the frequency of Fig. \ref{fig:LJ-figure-N20p1q1alpha1-beta1}. Unlike the 2:2 classical Lissajous figures, there are two distinct extrema on both the $x-$ and $y-$axis.} 
    \label{fig:LJ-figure-N20p2q2alpha1-beta1}
\end{figure}
From Fig. \ref{fig:LJ-figure-N20p2q2alpha1-beta1}, one can observe a coherent superposition of two fundamental isotropic quantum Lissajous states having complex amplitudes that are the square roots of unity. It is clear that this is a coherent superposition and not an incoherent mixture. The overlap of the two separate probability densities has developed interference fringes like the ones seen at that overlap in Fig. \ref{fig:LJ-figure-N20p2q3alpha1-beta1}. Also, based upon Eq. (\ref{eq:hh iso prob dens}), the second term is the interference term. If this were an incoherent mixture, the second term of Eq. (\ref{eq:hh iso prob dens}) would vanish, and the probability density of the 2:2 quantum Lissajous state would simply be the addition of the probability densities of the 1:1 quantum Lissajous figures having amplitudes that are the square roots of unity. As verification of Eq. (\ref{eq:hh iso prob dens}), the incoherent mixture can be plotted, 
\begin{figure}[H]
    \centering
    \includegraphics[width=\textwidth]{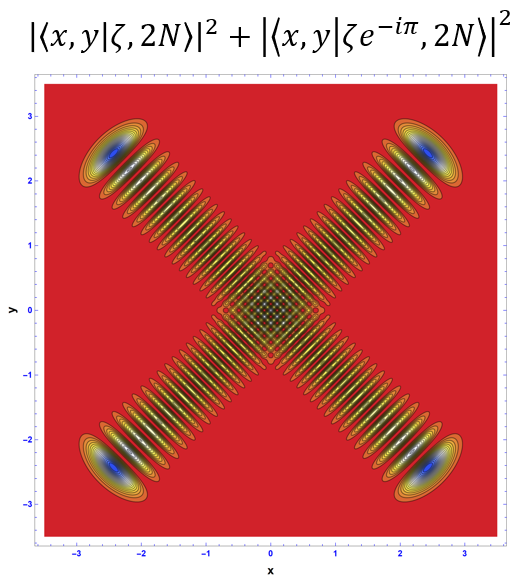}
    
    \caption{The contour plot of the incoherent mixture of the probability densities of the 1:1 quantum Lissajous states with amplitudes that are the square roots of unity, $\beta=1,e^{i\pi}$. The overlap of the figures is noticeably different than Fig. \ref{fig:LJ-figure-N20p2q2alpha1-beta1}, and does not quantum mechanically interfere. The probability value at each point of one figure is added to the probability value at the same point of the other figure. This plot does not match Fig. \ref{fig:LJ-figure-N20p2q2alpha1-beta1}(b) and is proof that the isotropic higher harmonic quantum Lissajous states are a coherent superposition and not an incoherent mixture.} 
    \label{fig:LJ-figure-N20p2q2alpha1-beta1_INCOHERENT-MIXTURE}
\end{figure}

\begin{figure}[H]
    \centering
    \includegraphics[width=\textwidth]{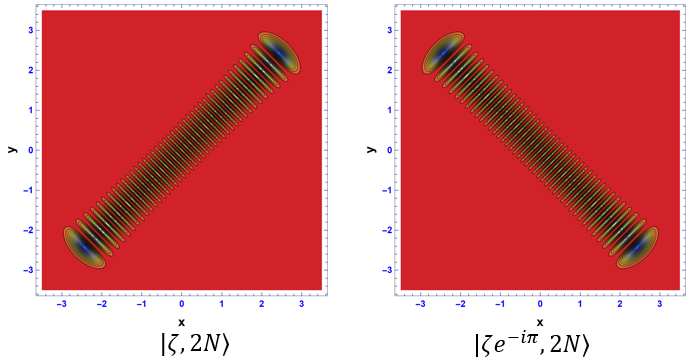}
    
    \caption{The contour plots of the probability densities of the fundamental isotropic quantum Lissajous state with $\alpha=1$ and $\beta=1,e^{i\pi}$. A coherent superposition of these probability densities with $2N$ quanta and a frequency of $2\omega_0$ makes up the 2:2 quantum Lissajous state with $\alpha=\beta=1$, Fig. \ref{fig:LJ-figure-N20p2q2alpha1-beta1}.} 
    \label{fig:LJ-figure-N40p1q1alpha1-beta(1,ExpiPi)}
\end{figure}

\begin{figure}[H]
    \centering
    \includegraphics[width=\textwidth]{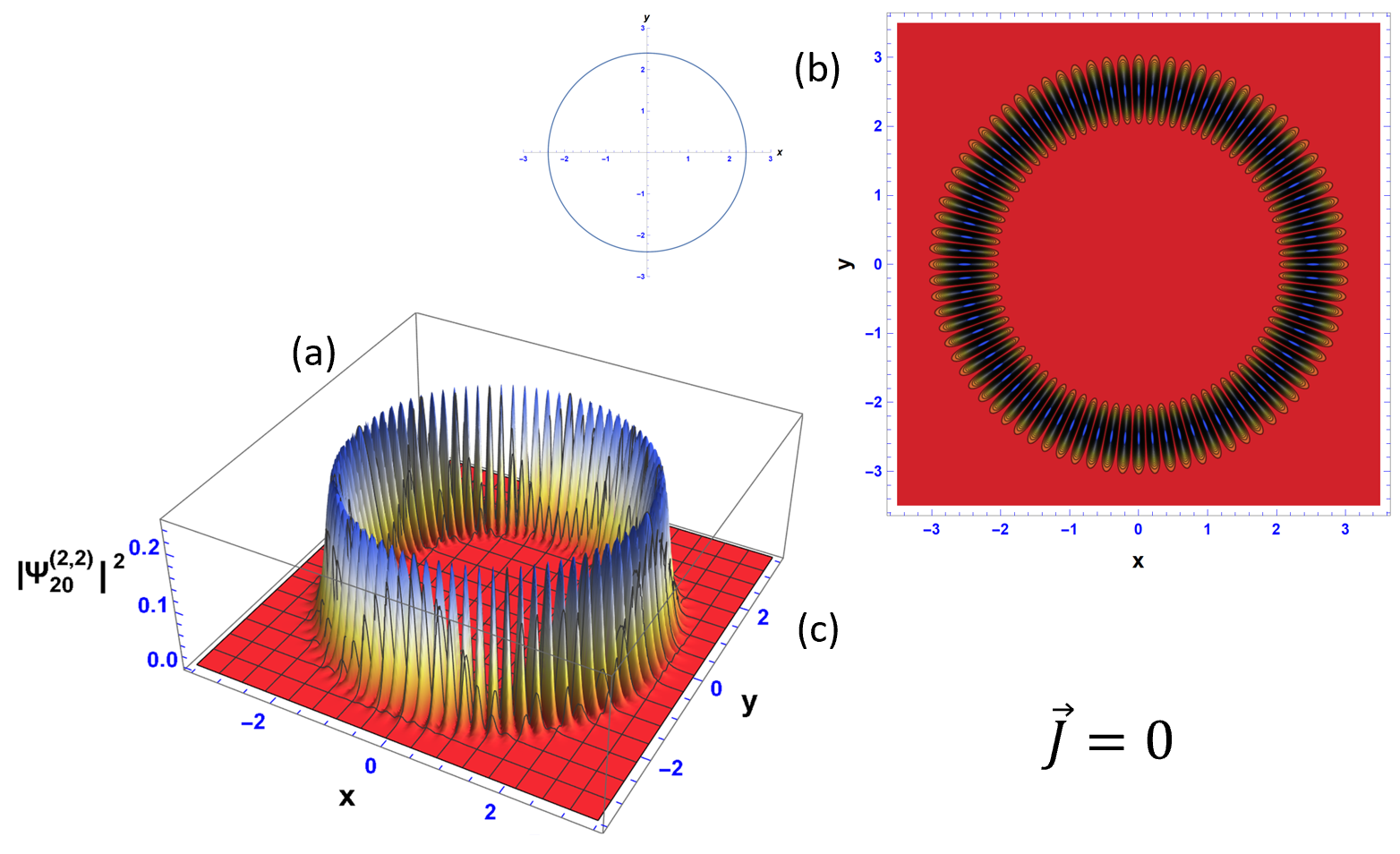}
    
    \caption{The surface (a) and contour (b) plots of the probability density of the isotropic higher harmonic quantum Lissajous state with $p=q=2$, $\alpha=1$, and $\beta=e^{i\pi/2}$. The associated classical Lissajous figure is shown to the left of (b) for comparison. The classical Lissajous figure is that of the 1:1 classical Lissajous figure for $\phi=\pi/2$, oscillating at double the frequency of the curves shown in Fig. \ref{fig:CLJ-p1q1-phiVARIED}(e). The corresponding probability current density (c) vanishes, making this a \textbf{\underline{static limit}} of the 2:2 quantum Lissajous states for the isotropic higher harmonic 2DHO. The 2:2 quantum Lissajous figure is a coherent superposition of two fundamental isotropic quantum Lissajous figures having $2N$ quanta, complex amplitudes that are the square roots of unity shifted by $\pi/2$, and oscillating at double the frequency of Figs. \ref{fig:LJ-figure-N20p1q1alpha1-beta1}-\ref{fig:LJ-figure-N20p1q1alpha1-betaExpi5Pi12}. Unlike the 2:2 classical Lissajous figures, there are two distinct extrema on both the $x-$ and $y-$axis.} 
    \label{fig:LJ-figure-N20p2q2alpha1-betaExpiPi2}
\end{figure}

\begin{figure}[H]
    \centering
    \includegraphics[width=\textwidth]{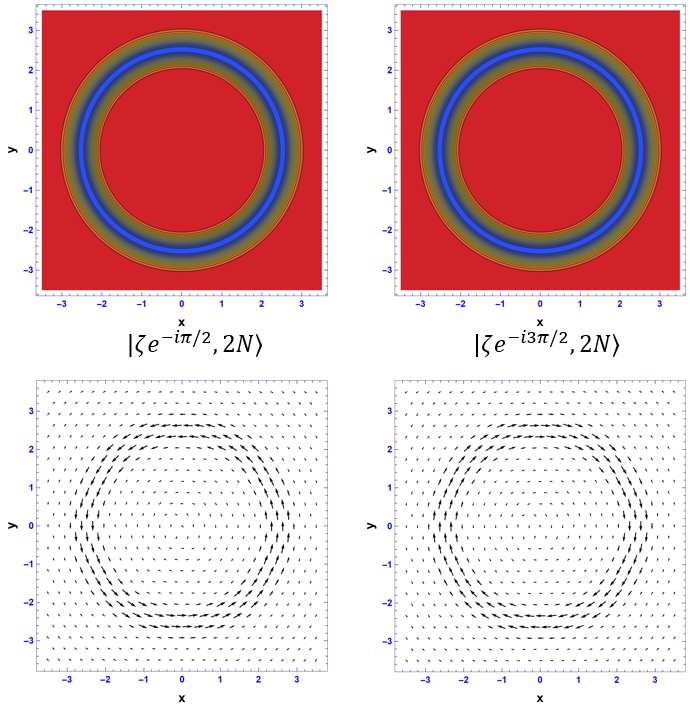}
    
    \caption{The contour plots of the probability densities and the current density plots of the fundamental isotropic quantum Lissajous state with $\alpha=1$ and $\beta=e^{i\pi/2},e^{i3\pi/2}$. A coherent superposition of these probability densities with $2N$ quanta and a frequency of $2\omega_0$ makes up the 2:2 quantum Lissajous state with $\alpha=1$ and $\beta=e^{i\pi/2}$, Fig. \ref{fig:LJ-figure-N20p2q2alpha1-betaExpiPi2}. Both probability densities localize over the same classical Lissajous figure, but the current densities are counter-propagating, leading to fully resolved interference fringes along the shape 1:1 quantum Lissajous figure when superimposing the present figures.} 
    \label{fig:LJ-figure-N40p1q1alpha1-beta(ExpiPi2,Expi3Pi2)}
\end{figure}

\begin{figure}[H]
    \centering
    \includegraphics[width=\textwidth]{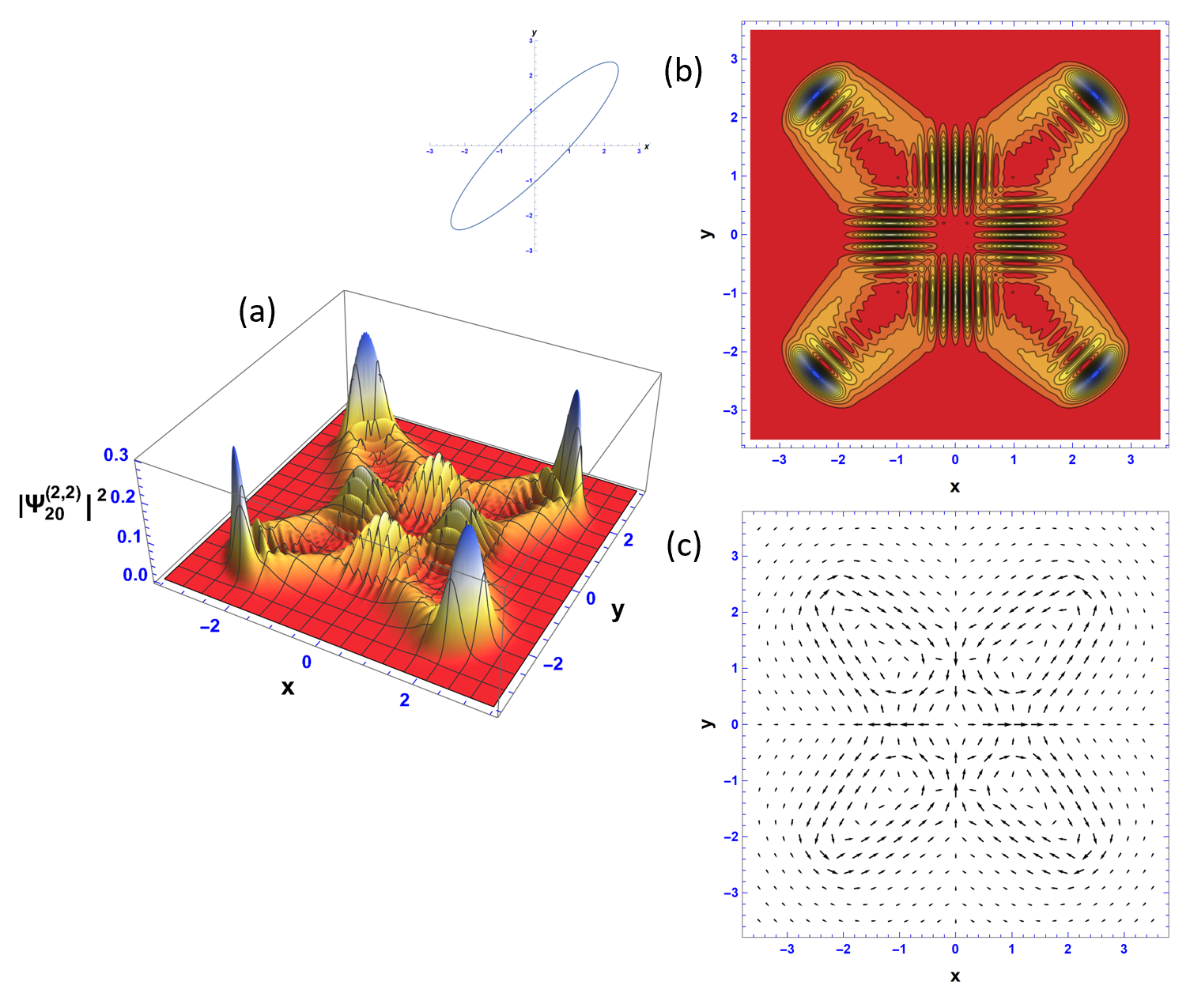}
    
    \caption{The surface (a) and contour (b) plots of the probability density of the isotropic higher harmonic quantum Lissajous state with $p=q=2$, $\alpha=1$, and $\beta=e^{i\pi/7}$. The associated classical Lissajous figure is shown to the left of (b) for comparison. The classical Lissajous figure is that of the 1:1 classical Lissajous figure for $\phi=\pi/7$, oscillating at double the frequency of the curves shown in Fig. \ref{fig:CLJ-p1q1-phiVARIED}. The corresponding probability current density (c) is non-zero and steady, making this a vortex state of the 2:2 quantum Lissajous states for the isotropic higher harmonic 2DHO. Interference fringes at the overlaps of the curves in superposition exhibit the same sort of interference as that of the fundamental anisotropic cases with non-zero probability current density. The 2:2 quantum Lissajous figure is a coherent superposition of two fundamental isotropic quantum Lissajous figures having $2N$ quanta, complex amplitudes that are the square roots of unity shifted by $\pi/7$, and oscillating at double the frequency of Figs. \ref{fig:LJ-figure-N20p1q1alpha1-beta1}-\ref{fig:LJ-figure-N20p1q1alpha1-betaExpi5Pi12}. Unlike the 2:2 classical Lissajous figures, there are two distinct extrema on both the $x-$ and $y-$axis. There are four vortices present, two columns and two rows.} 
    \label{fig:LJ-figure-N20p2q2alpha1-betaExpiPi7}
\end{figure}

\begin{figure}[H]
    \centering
    \includegraphics[width=\textwidth]{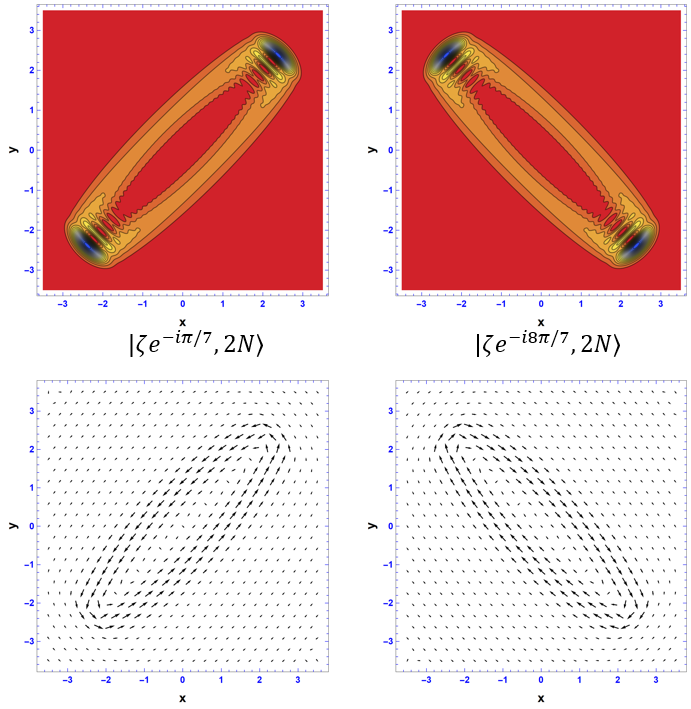}
    
    \caption{The contour plots of the probability densities and the current density plots of the fundamental isotropic quantum Lissajous state with $\alpha=1$ and $\beta=e^{i\pi/7},e^{i8\pi/7}$. A coherent superposition of these probability densities with $2N$ quanta and a frequency of $2\omega_0$ makes up the 2:2 quantum Lissajous state with $\alpha=1$ and $\beta=e^{i\pi/7}$, Fig. \ref{fig:LJ-figure-N20p2q2alpha1-betaExpiPi7}.} 
    \label{fig:LJ-figure-N40p1q1alpha1-beta(ExpiPi7,Expi8Pi7)}
\end{figure}

\begin{figure}[H]
    \centering
    \includegraphics[width=\textwidth]{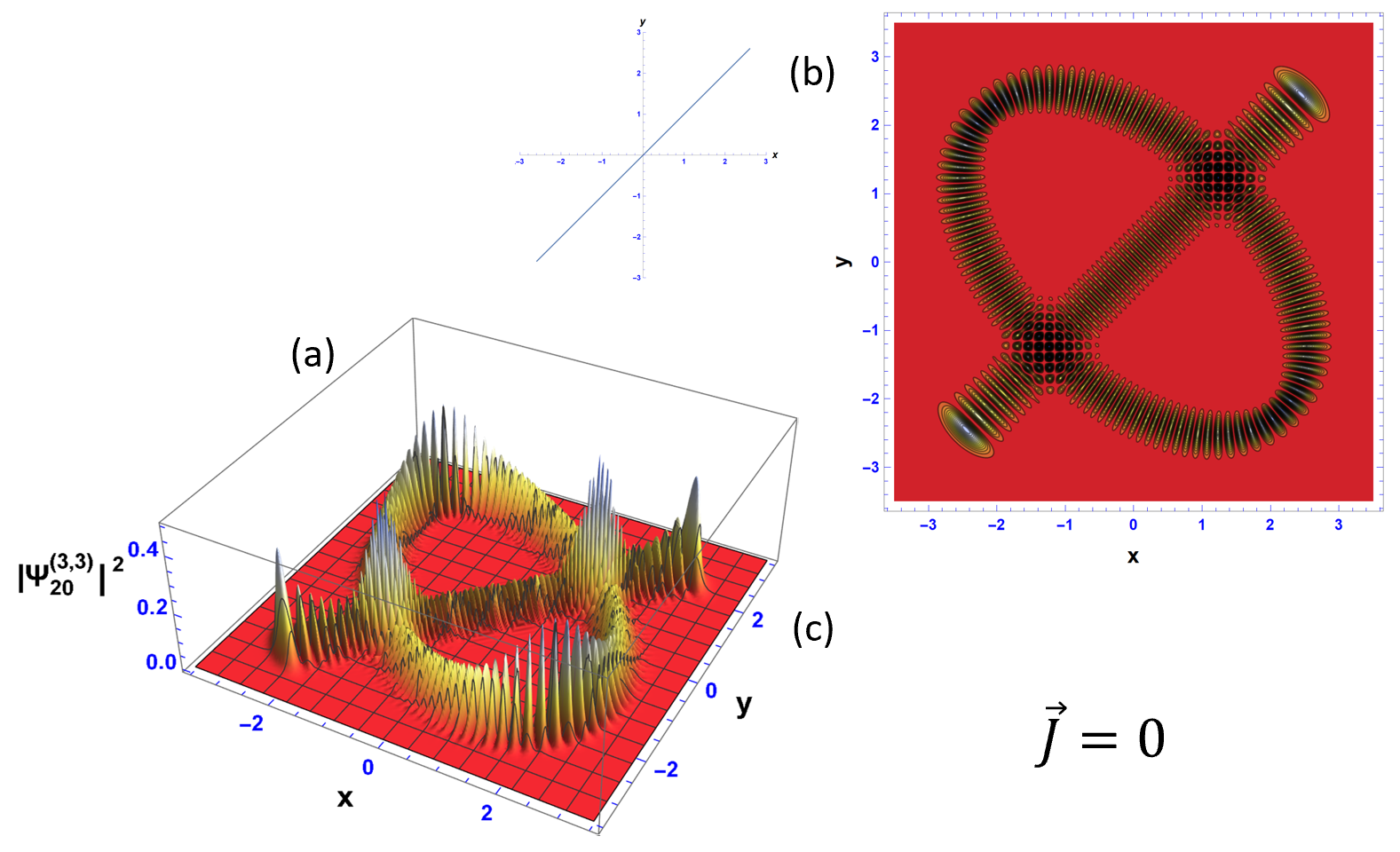}
    
    \caption{The surface (a) and contour (b) plots of the probability density of the isotropic higher harmonic quantum Lissajous state with $p=q=3$, $\alpha=1$, and $\beta=1$. The associated classical Lissajous figure is shown to the left of (b) for comparison. The classical Lissajous figure is that of the 1:1 classical Lissajous figure for $\phi=0$, oscillating at triple the frequency of the curves shown in Fig. \ref{fig:CLJ-p1q1-phiVARIED}. The corresponding probability current density (c) vanishes, making this a \textbf{\underline{static limit}} of the 3:3 quantum Lissajous states for the isotropic higher harmonic 2DHO. Interference fringes at the overlaps of the curves in superposition exhibit the same sort of interference as that of the isotropic cases of 2:2 with vanishing current density, Figs. \ref{fig:LJ-figure-N20p2q2alpha1-beta1} and \ref{fig:LJ-figure-N20p2q2alpha1-betaExpiPi2}. The 3:3 quantum Lissajous figure is a coherent superposition of three fundamental isotropic quantum Lissajous figures having $3N$ quanta, complex amplitudes that are the cube roots of unity, and oscillating at triple the frequency of Figs. \ref{fig:LJ-figure-N20p1q1alpha1-beta1}-\ref{fig:LJ-figure-N20p1q1alpha1-betaExpi5Pi12}. Unlike the 3:3 classical Lissajous figures, there are three extrema on both the $x-$ and $y-$axis.} 
    \label{fig:LJ-figure-N20p3q3alpha1-beta1}
\end{figure}

\begin{figure}[H]
    \centering
    \includegraphics[width=\textwidth]{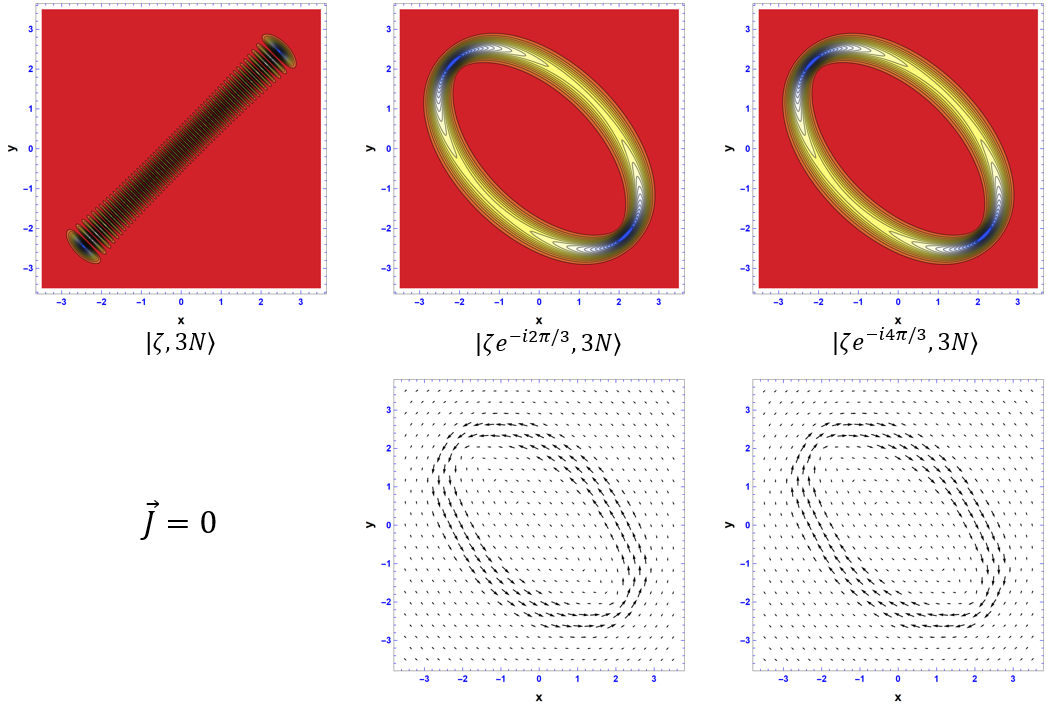}
    
    \caption{The contour plots of the probability densities and the current density plots of the fundamental isotropic quantum Lissajous state with $\alpha=1$ and $\beta=1,e^{i2\pi/3},e^{i4\pi/3}$. A coherent superposition of these probability densities with $3N$ quanta and a frequency of $3\omega_0$ makes up the 3:3 quantum Lissajous state with $\alpha=\beta=1$, Fig. \ref{fig:LJ-figure-N20p3q3alpha1-beta1}. The probability densities with $\beta=e^{i2\pi/3},e^{i4\pi/3}$ localize over the same classical Lissajous figure, but the current densities are counter-propagating, leading to fully resolved interference fringes along the shape 1:1 quantum Lissajous figure when superimposing them.} 
    \label{fig:LJ-figure-N60p1q1alpha1-beta(1,Expi2Pi3,Expi4Pi3)}
\end{figure}

\begin{figure}[H]
    \centering
    \includegraphics[width=\textwidth]{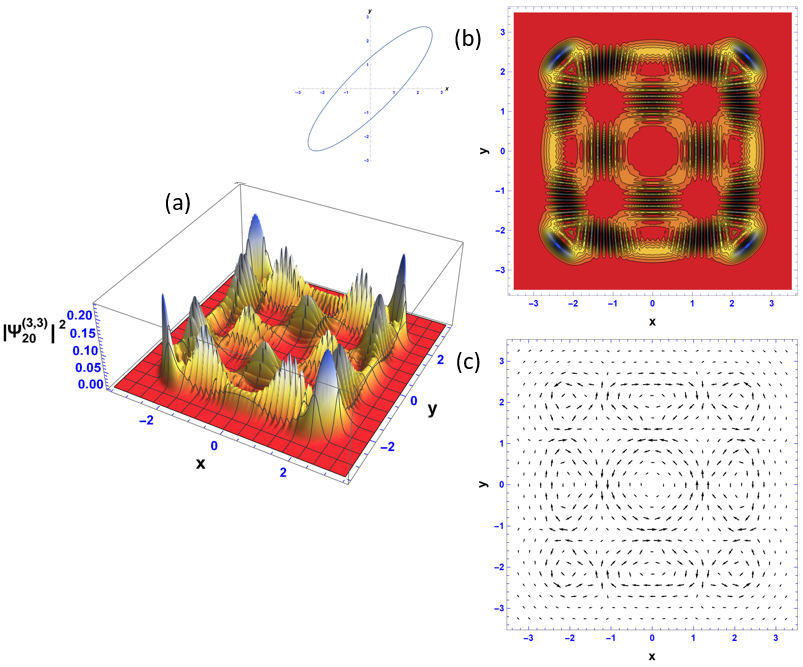}
    
    \caption{The surface (a) and contour (b) plots of the probability density of the isotropic higher harmonic quantum Lissajous state with $p=q=3$, $\alpha=1$, and $\beta=e^{i\pi/6}$. The associated classical Lissajous figure is shown to the left of (b) for comparison. The classical Lissajous figure is that of the 1:1 classical Lissajous figure for $\phi=\pi/6$, oscillating at triple the frequency of the curves shown in Fig. \ref{fig:CLJ-p1q1-phiVARIED}. The corresponding probability current density (c) is non-zero and steady, making this a vortex state, and, in fact, this is the \textbf{\underline{vortex limit}} of the 3:3 quantum Lissajous states for the isotropic higher harmonic 2DHO. Interference fringes at the overlaps of the curves in superposition exhibit the same sort of interference as that of the isotropic cases of 2:2 with non-zero current density, Fig. \ref{fig:LJ-figure-N20p2q2alpha1-betaExpiPi7}. This 3:3 quantum Lissajous figure is a coherent superposition of three fundamental isotropic quantum Lissajous figures having $3N$ quanta, complex amplitudes that are the cube roots of unity shifted by $\pi/6$, and oscillating at triple the frequency of Figs. \ref{fig:LJ-figure-N20p1q1alpha1-beta1}-\ref{fig:LJ-figure-N20p1q1alpha1-betaExpi5Pi12}. Unlike the 3:3 classical Lissajous figures, there are three extrema on both the $x-$ and $y-$axis. There are nine vortices present, three columns and three rows.} 
    \label{fig:LJ-figure-N20p3q3alpha1-betaExpiPi6}
\end{figure}

\begin{figure}[H]
    \centering
    \includegraphics[width=\textwidth]{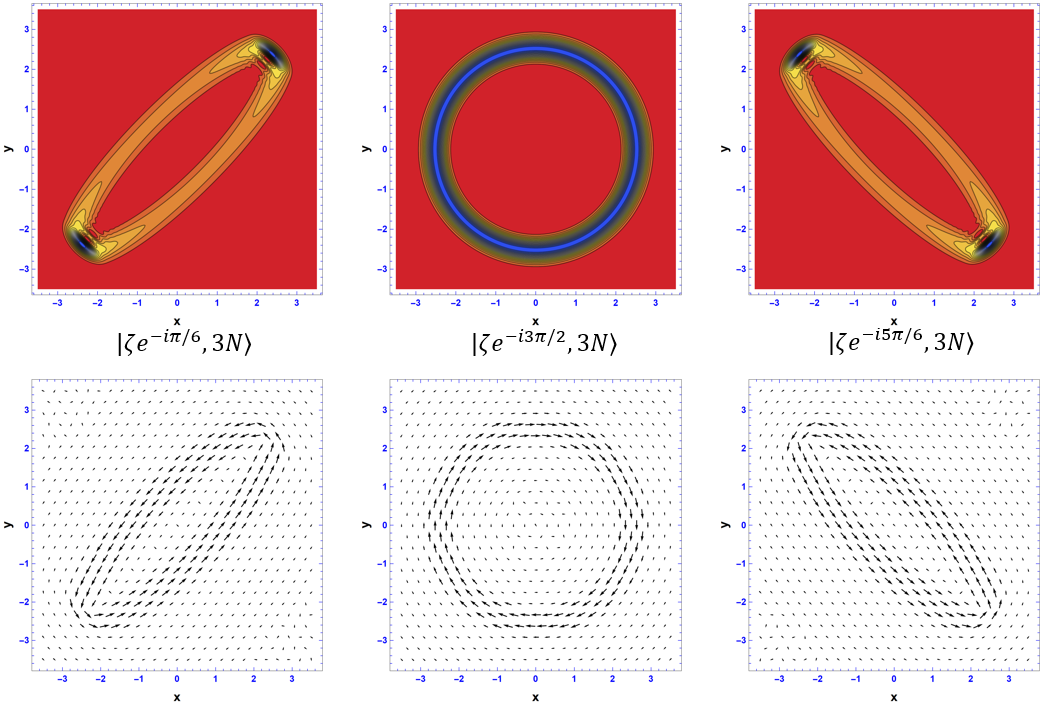}
    
    \caption{The contour plots of the probability densities and the current density plots of the fundamental isotropic quantum Lissajous state with $\alpha=1$ and $\beta=e^{i\pi/6},e^{i3\pi/2},e^{i5\pi/6}$. A coherent superposition of these probability densities with $3N$ quanta and a frequency of $3\omega_0$ makes up the 3:3 quantum Lissajous state with $\alpha=1$ and $\beta=e^{i\pi/6}$, Fig. \ref{fig:LJ-figure-N20p3q3alpha1-betaExpiPi6}.} 
    \label{fig:LJ-figure-N60p1q1alpha1-beta(ExpiPi6,Expi3Pi2,Expi5Pi6)}
\end{figure}
The figures presented in this section are the isotropic higher harmonic quantum Lissajous states. They are formed by a coherent superposition of $m$ fundamental isotropic quantum Lissajous states (SU(2) coherent states) with $mN$ quanta, frequencies of $m\omega_0$, and complex amplitudes corresponding to the $m^{th}$ roots of unity, or the roots of unity shifted by an already existing phase, $\phi$. There is $m$ extrema on both the $x-$ and $y-$axis, and $m^2$ vortices when applicable ($m$ columns and $m$ rows).
\subsection{Anisotropic Higher Harmonic Quantum Lissajous State}
The results presented in this section correspond to the theory developed on the isotropic higher harmonic quantum Lissajous states, Eq. (\ref{eq:aniso higher harmonic quantum lissajous state}), and its wavefunction, Eq. (\ref{eq:aniso higher harmonic quantum lissajous state wavefcn}). The probability density is calculated using the anisotropic higher harmonic analog to Eq. (\ref{eq:hh iso prob dens}) and the probability current density using Eq. (\ref{eq:prob current density}).
\begin{figure}[H]
    \centering
    \includegraphics[width=\textwidth]{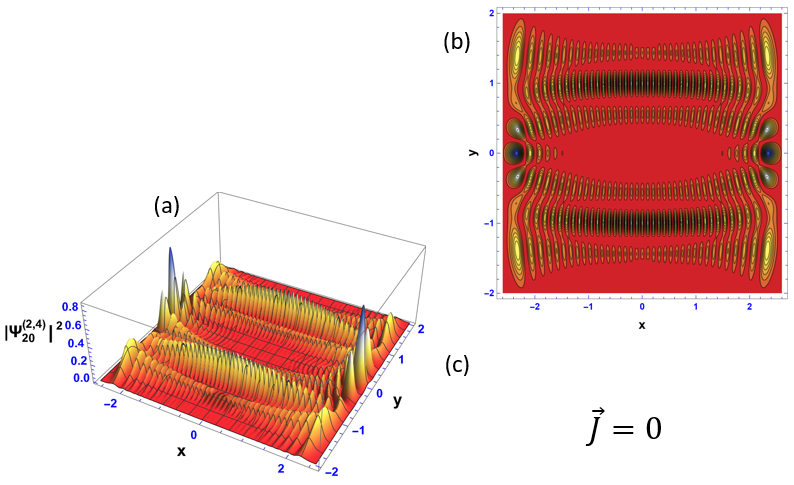}
    
    \caption{The surface (a) and contour (b) plots of the probability density of the anisotropic higher harmonic quantum Lissajous state with $p=2$, $q=4$, and $\alpha=\beta=1$. The classical Lissajous figure is that of the 1:2 classical Lissajous figure for $\phi=0$, oscillating at double the frequency of Fig. \ref{fig:CLJ-p1q2-phiVARIED}(a). The corresponding probability current density (c) vanishes and is a \textbf{\underline{static limit}} of the 2:4 quantum Lissajous states for the anisotropic higher harmonic 2DHO. The 2:4 quantum Lissajous figure is a coherent superposition of two fundamental anisotropic quantum Lissajous figures having $2N$ quanta, complex amplitudes that are the square roots of unity, and oscillating at double the frequency of Fig. \ref{fig:LJ-figure-N20p1q2alpha1-beta1}. The interference on the overlaps of the two states in superposition is similar to interference on the overlaps of Fig. \ref{fig:LJ-figure-N20p3q3alpha1-beta1}. Unlike the 2:4 classical Lissajous figures, there are two distinct extrema on the $y-$axis and four distinct extrema on the $x-$axis.} 
    \label{fig:LJ-figure-N20p2q4alpha1-beta1}
\end{figure}

\begin{figure}[H]
    \centering
    \includegraphics[width=\textwidth]{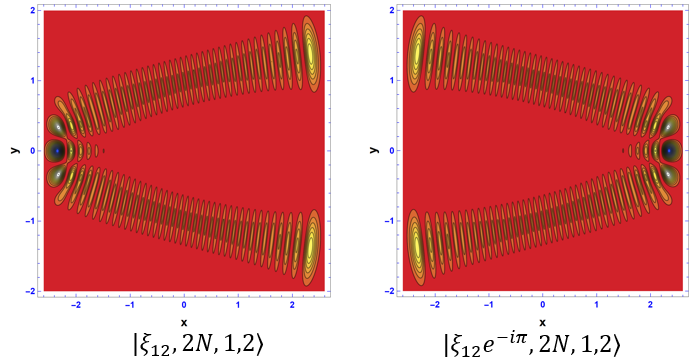}
    
    \caption{The contour plots of the probability densities and the current density plots of the fundamental anisotropic quantum Lissajous state with $p=1$, $q=2$ $\alpha=1$ and $\beta=1,e^{i\pi/2}$. $\beta^q$ gives the square root of unity, $e^{i\pi}$. A coherent superposition of these probability densities with $2N$ quanta and a frequency of $2\omega_0$ makes up the 2:4 quantum Lissajous state with $\alpha=1$ and $\beta=1$, Fig. \ref{fig:LJ-figure-N20p2q4alpha1-beta1}.} 
    \label{fig:LJ-figure-N40p1q2alpha1-beta(1,ExpiPi2)}
\end{figure}

\begin{figure}[H]
    \centering
    \includegraphics[width=\textwidth]{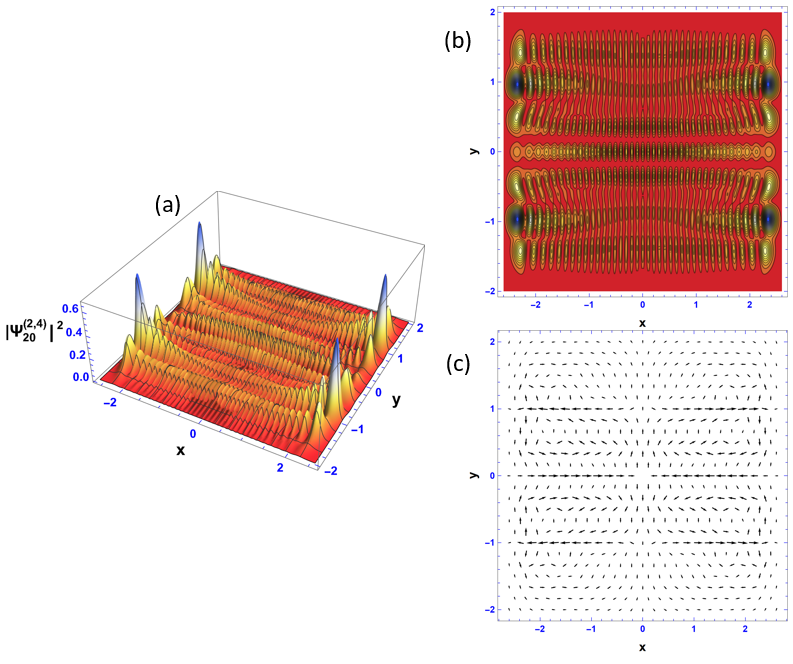}
    
    \caption{The surface (a) and contour (b) plots of the probability density of the anisotropic higher harmonic quantum Lissajous state with $p=2$, $q=4$, $\alpha=1$, and $\beta=e^{i\pi/6}$. The classical Lissajous figure is that of the 1:2 classical Lissajous figure for $\phi=\pi/6$, oscillating at double the frequency of the curves shown in Fig. \ref{fig:CLJ-p1q2-phiVARIED}. The corresponding probability current density (c) is non-zero and steady, making this a vortex state of the 2:4 quantum Lissajous states for the anisotropic higher harmonic 2DHO. Interference fringes at the overlaps of the curves in superposition exhibit the same sort of interference as that of the fundamental anisotropic cases and isotropic higher harmonic cases with non-zero probability current density. The 2:4 quantum Lissajous figure is a coherent superposition of two fundamental isotropic quantum Lissajous figures having $2N$ quanta, complex amplitudes that are the square roots of unity shifted by $\pi/6$, and oscillating at double the frequency of Figs. \ref{fig:LJ-figure-N20p1q2alpha1-beta1}-\ref{fig:LJ-figure-N20p1q2alpha1-betaExpiPi8}. Unlike the 2:4 classical Lissajous figures, there are two distinct extrema on the $y-$axis and four distinct extrema on the $x-$axis. There are eight vortices present, two columns and four rows.} 
    \label{fig:LJ-figure-N20p2q4alpha1-betaExpiPi6}
\end{figure}

\begin{figure}[H]
    \centering
    \includegraphics[width=\textwidth]{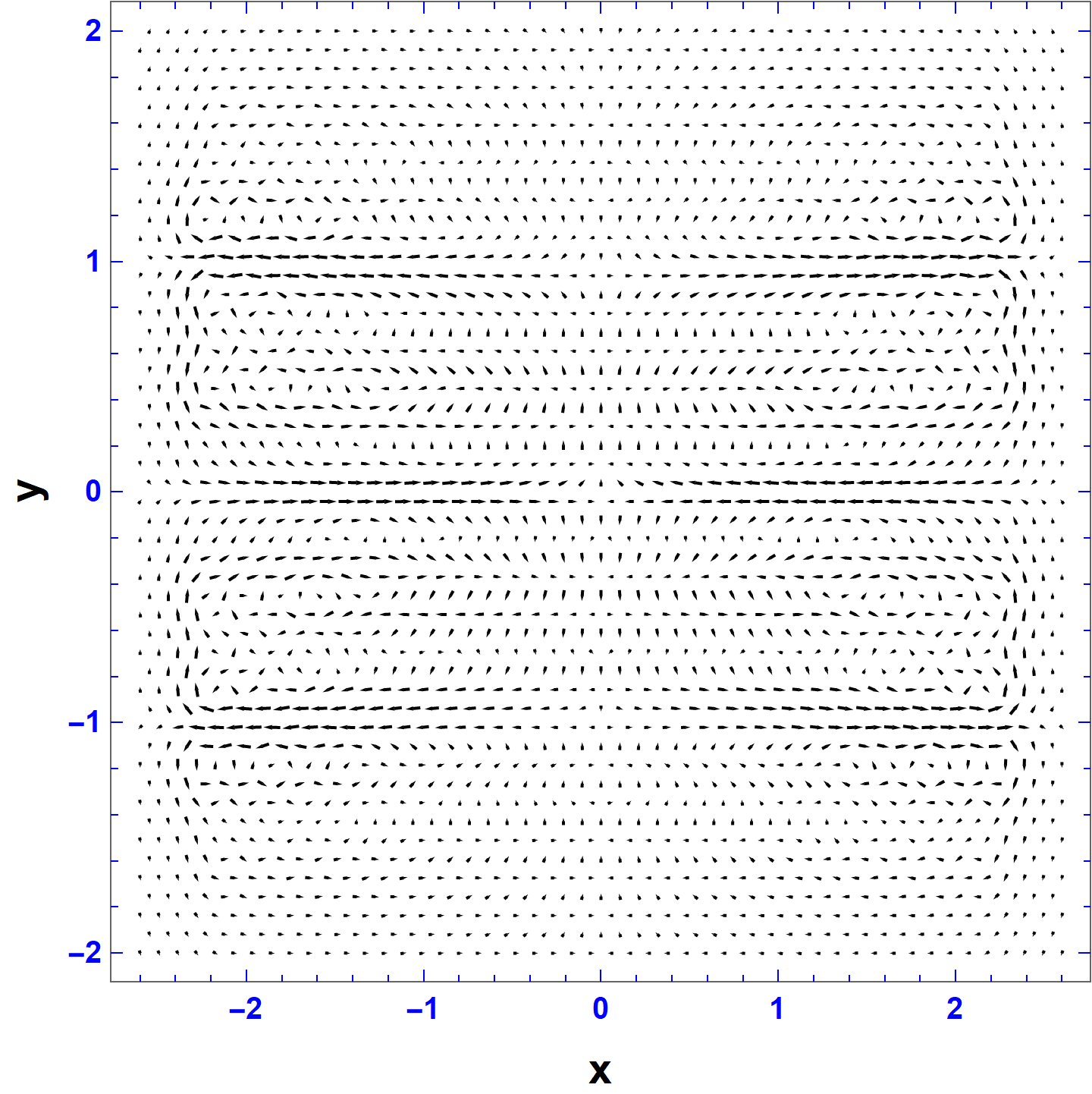}
    
    \caption{Presents a higher resolution vector plot of the probability current density, Fig. \ref{fig:LJ-figure-N20p2q4alpha1-betaExpiPi6}(c), to demonstrate a clearer picture of the vortices.} 
    \label{fig:LJ-currentdensity-N20p2q4alpha1-betaExpiPi6(vectorPoints50)}
\end{figure}

\begin{figure}[H]
    \centering
    \includegraphics[width=\textwidth]{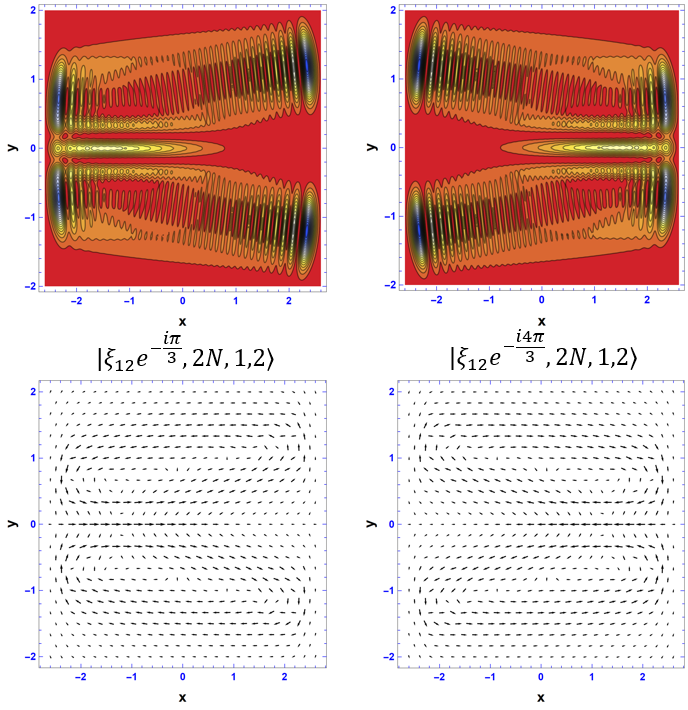}
    
    \caption{The contour plots of the probability densities and the current density plots of the fundamental anisotropic quantum Lissajous state with $p=1$, $q=2$, $\alpha=1$, and $\beta=e^{i\pi/6},e^{i2\pi/3}$. A coherent superposition of these probability densities with $2N$ quanta and a frequency of $2\omega_0$ makes up the 2:4 quantum Lissajous state with $\alpha=1$ and $\beta=e^{i\pi/6}$, Fig. \ref{fig:LJ-figure-N20p2q4alpha1-betaExpiPi6}.} 
    \label{fig:LJ-figure-N40p1q2alpha1-beta(ExpiPi6,Expi2Pi3)}
\end{figure}

\begin{figure}[H]
    \centering
    \includegraphics[width=\textwidth]{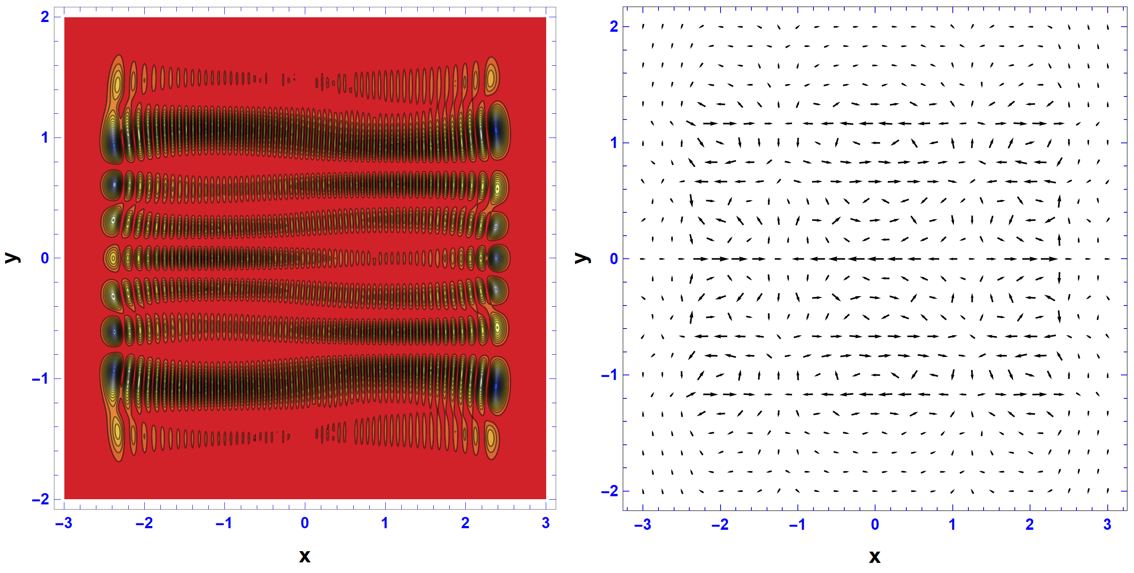}
    
    \caption{This example of an anisotropic higher harmonic quantum Lissajous state is left for the reader as an example of how to determine the Lissajous ratio of $p$ and $q$. Recall that the Lissajous ratio determines how many vortices will appear in the probability current density.} 
    \label{fig:LJ-figure-N20p3q6alpha1-betaExpiPi8}
\end{figure}

\begin{figure}[H]
    \centering
    \includegraphics[width=\textwidth]{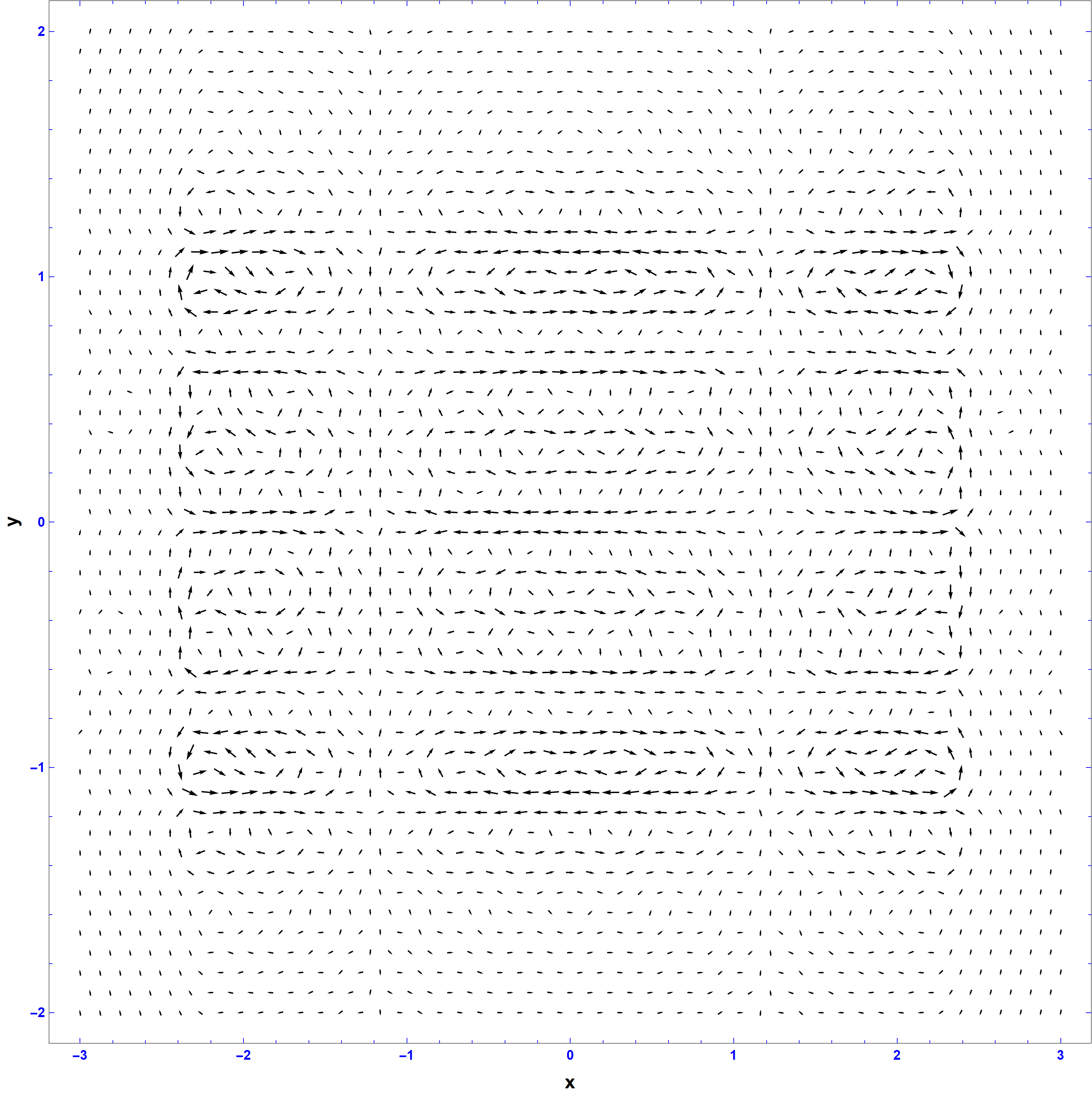}
    
    \caption{Presents a higher resolution vector plot of the probability current density in Fig. \ref{fig:LJ-figure-N20p3q6alpha1-betaExpiPi8} to demonstrate a clearer picture of the vortices.} 
    \label{fig:LJ-currentdensity-N20p3q6alpha1-betaExpiPi8(vectorPoints50)}
\end{figure}

The figures presented in this section are the isotropic higher harmonic quantum Lissajous states. It becomes difficult to visually distinguish which fundamental anisotropic states are in superposition; the 2:4 case being the easiest to discern. They are formed by a coherent superposition of $m$ fundamental isotropic quantum Lissajous states (SU(2) coherent states) with $mN$ quanta, frequencies of $m\omega_0$, and complex amplitudes corresponding to the $m^{th}$ roots of unity, or the roots of unity shifted by an already existing phase, $\phi$. There is $p$ extrema on the $y-$axis and $q$ extrema on the $x-$axis, and $qp$ vortices when applicable ($p$ columns and $q$ rows).

\section{Summary}
\label{chap5_Summary}
We have presented a theoretical framework in which quantum Lissajous figures arise ‘organically’ via projection of a two-mode ordinary coherent state onto a degenerate subspace of the 2DHO. The states resulting from this process are quantum mechanically stationary, having time-independent configuration space probability densities localized along the corresponding classical Lissajous figures. For the isotropic 2DHO, the fundamental case naturally reproduced the SU(2) coherent states; for the anisotropic 2DHO the fundamental cases \textbf{are not SU(2) coherent states} in even a meaningful “generalized” way (as some claim them to be), still the anisotropic fundamental quantum Lissajous states can be used as a non-orthogonal basis for the expansion of anisotropic higher harmonic quantum Lissajous states having the same frequency ratios. This is in direct analogy with the role of the SU(2) coherent states with respect to isotropic higher harmonic cases.

The higher harmonic quantum Lissajous states are radically different from their classical counterparts in one very important way. The classical version is simply a copy of the fundamental case with an $m$ times larger frequency, tracing the figure $m$ times faster than the relevant fundamental case (same orientation and eccentricity behaviors as the fundamental), owing to the larger amount of mechanical energy in the system. In the quantum version, a linear superposition of the fundamental states relevant of a given case naturally emerges; for the $m^{th}$ harmonic state the complex amplitudes of the states involved in the superposition are weighted by the $m^{th}$ roots of unity. We have related the amplitude factors involved to the phases of the originally projected ordinary coherent states. The superposition that arises is the signature of quantum interference in these higher harmonic states, and the evidence for this is clear in our graphical results. We have given a detailed description of the important interplay between quantum mechanical probability current density and quantum interference, especially in connection with the vortex states that arise when considering the quantum Lissajous states. We have pointed out the crucial role played by continuity with respect to the stationary nature of all quantum Lissajous states (both vortex and static). We believe that this discussion sheds light on an issue that seems to have been muddled or ignored in the existing literature on Lissajous figures in quantum mechanics.

Our work has several clear connections and possible extensions: The systematic study of roots-of-unity states for ordinary, SU(2), and SU(1,1) coherent states (we have already done a lot of this but have omitted it from this manuscript). One obvious extension of this work is the examination of quantum Lissajous states via projection in multi-mode systems (e.g., 3DHO, etc.). Another interesting avenue is the investigation of the possibility of quantum stationary states via projection for the “Kepler” problem (H-atom); this is the only other problem that might support them owing to (the Ehrenfest limit of) Bertrand’s theorem. One could also consider the projections of states other than the ordinary coherent states. Do these generally bear any relation to classical Lissajous figures? We don’t know, but this could be the basis for a provable theorem. Systematic study of the non-classical properties and interactions of all types of quantum Lissajous states could yield interesting results for quantum technologies based on integrated nano-photonics and quantum dots. Finally, finding a viable method for producing quantum Lissajous states in the lab would be an essential breakthough for their possible application to possible non-classical systems.

\section{Appendix: SU(2) Coherent States}
\label{appendix a}
\subsection{SU(2) Coherent States in the Angular Momentum Representation}
A big theme of this project involves SU(2) coherent states. We have encountered the SU(2) coherent states in a non-traditional way, by projecting the ordinary coherent state onto a degenerate subspace of the 2DHO. This appendix will summarize well known earlier derivations of SU(2) coherent states (Arecchi \textit{et al.})\cite{arecchi_atomic_1972,inomata_path_1992} in the angular momentum basis. In addition, the Schwinger Realization\cite{schwinger_angular_2015} that, relates the angular momentum operators to the raising and lowering operators of the quantum harmonic oscillator, is implemented to yield the SU(2) coherent state in the Fock state basis.  

The angular momentum basis states, just like the Fock states, can be constructed by applying the raising operator on the ground state
\begin{align}
\label{eqn:angular momentum basis state}
\begin{split}
\ket{J,M}=\frac{1}{(J+M)!}\binom{2J}{J+M}^{-1/2}\hat{J}_+^{J+M}\ket{J,-J},
\end{split}
\end{align}
where $-J\leq M\leq J$. The angular momentum representation has its own version of raising and lowering operators
\begin{align}
\label{eqn:angular momentum raising and lowering}
\begin{split}
\hat{J}_{\pm}=\hat{J}_x\pm i\hat{J}_y,
\end{split}
\end{align}
and $\hat{J}_3$ can be written in terms of raising and lowering operators
\begin{align}
\label{eqn:J3 in terms of raising and lowering}
\begin{split}
\hat{J}_{3}=\frac{1}{2}\left(\hat{J}_{+}\hat{J}_{-}-\hat{J}_{+}\hat{J}_{-}\right).
\end{split}
\end{align}
Eqs. (\ref{eqn:angular momentum raising and lowering}) and (\ref{eqn:J3 in terms of raising and lowering}) obey the commutation relations
\begin{align}
\label{eqn:angular momentum commutators}
\begin{split}
\left[\hat{J}_{\pm},\hat{J}_3\right]=\mp\hat{J}_\pm.
\end{split}
\end{align}
The Casimir operator for this algebra is $\hat{J}^2$, given by Eq. (\ref{eq:casimir operator}). It is important to recall the eigenvalue equations for $\hat{J}_3$ and $\hat{J}^2$
\begin{align}
\label{eqn:J3 eigenvalue eqn}
\begin{split}
\hat{J}_3\ket{J,M}=M\ket{J,M},
\end{split}
\end{align}
and
\begin{align}
\label{eqn:J^2 eigenvalue eqn}
\begin{split}
\hat{J}^2\ket{J,M}=J(J+1)\ket{J,M}.
\end{split}
\end{align}
Similar to how the ordinary coherent state is created by displacing the ground Fock state, the SU(2) coherent state is created by a rotational "displacement" of the angular momentum ground state. Using the normally ordered form of the rotation operator, the second equality of Eq. (\ref{eq:rotation operator}), one finds
\begin{align}
\label{eqn:su2 coherent state ang mom rep}
\begin{split}
\ket{\zeta,J}&=(1+\abs{\zeta}^2)^{-J}e^{\zeta\hat{J}_+}\ket{J,-J}\\
&=(1+\abs{\zeta}^2)^{-J}\sum_{M=-J}^J\binom{2J}{J+M}^{1/2}\zeta^{J+M}\ket{J,M}.
\end{split}
\end{align}
Eq. (\ref{eqn:su2 coherent state ang mom rep}) is the SU(2) coherent state in the angular momentum representation.
\subsection{SU(2) Coherent States in the Fock State Representation}
Using Eqs. (\ref{eq:Schwinger realization}) and (\ref{eq:J0}), we can obtain the eigenvalue equations for $\hat{J}_3$ and $\hat{J}^2$ in the Fock state basis,
\begin{align}
\label{eqn:J3 eigenvalue eqn fock state}
\begin{split}
\frac{1}{2}\left(\hat{a}_x^\dagger\hat{a}_x-\hat{a}_y^\dagger\hat{a}_y\right)\ket{K,N-K}&=\left(K-\frac{N}{2}\right)\ket{J,M},
\end{split}
\end{align}
and
\begin{align}
\label{eqn:J^2 eigenvalue eqn fock state}
\begin{split}
\frac{1}{2}\left(\hat{a}_x^\dagger\hat{a}_x+\hat{a}_y^\dagger\hat{a}_y\right)\left(\frac{1}{2}\left(\hat{a}_x^\dagger\hat{a}_x+\hat{a}_y^\dagger\hat{a}_y\right)+1\right)\ket{K,N-K}&=\frac{N}{2}\left(\frac{N}{2}+1\right)\ket{K,N-K}.
\end{split}
\end{align}
Comparing Eqs. (\ref{eqn:J3 eigenvalue eqn fock state}) and (\ref{eqn:J^2 eigenvalue eqn fock state}) with Eqs. (\ref{eqn:J3 eigenvalue eqn}) and (\ref{eqn:J^2 eigenvalue eqn}), we can see that
\begin{align}
\label{eqn:J to N relationship}
\begin{split}
J=\frac{N}{2},
\end{split}
\end{align}
and
\begin{align}
\label{eqn:J and M to K relationship}
\begin{split}
K=J+M.
\end{split}
\end{align}
A simple substitution of these into Eq. (\ref{eqn:su2 coherent state ang mom rep}) yields the SU(2) coherent state in the Fock state representation, that we found via projection of the two-mode coherent state onto a degenerate subspace of the 2DHO
\begin{align}
\label{eqn:su2 coherent state fock state rep}
\begin{split}
\ket{\zeta,N}&=(1+\abs{\zeta}^2)^{-N/2}\sum_{K=0}^N\binom{N}{K}^{1/2}\zeta^{K}\ket{K,N-K}.
\end{split}
\end{align}

\end{document}